\documentclass[]{elsarticle}
\usepackage[dvipsnames]{xcolor}

\makeatletter
\def\ps@pprintTitle{%
 \let\@oddhead\@empty
 \let\@evenhead\@empty
 \def\@oddfoot{}%
 \let\@evenfoot\@oddfoot}
\makeatother
\usepackage{amssymb}
\usepackage{mathtools}
\usepackage{subcaption}
\usepackage[section]{placeins}
\usepackage{lineno}

\usepackage[numbers]{natbib}
\usepackage{listings}
\usepackage{hyperref}
\usepackage{floatrow}
\usepackage{algorithm}
\usepackage{graphicx}
\usepackage{algpseudocode}
\usepackage{amsmath}
\usepackage[size=footnotesize]{subcaption}

\graphicspath{{figures/}}
\usepackage{cleveref}
\crefname{appendix}{}{}
\usepackage{soul}
\usepackage{geometry}
\usepackage{algorithmicx}
\usepackage{booktabs}
\usepackage{longtable}
\usepackage{siunitx}
\usepackage{stmaryrd}
\usepackage{algpseudocode}
\usepackage{algorithm}
\usepackage{adjustbox}
\usepackage{color, colortbl}
\usepackage{multirow}
\usepackage{xcolor}
\usepackage{ulem}

\usepackage{tikz}
\usetikzlibrary{positioning,shapes,arrows,chains}
\makeatletter
\let\tikz@preaction@layer=\pgfutil@empty     
\tikzset{preaction layer/.store in=\tikz@preaction@layer} 
\makeatother

\tikzstyle{rounded shadow} = [
    copy shadow={%
      preaction layer=shadow,
      fill=gray!25,
      draw=none,
      shadow xshift=0.2em,
      shadow yshift=-0.25em
   }]

\tikzstyle{test shadow} = [
    copy shadow={%
      preaction layer=shadow,
      fill=gray!25,
      draw=none,
      shadow xshift=0.1em,
      shadow yshift=-0.2em
   }]

\pgfdeclarelayer{shadow} 
\pgfsetlayers{shadow,main}

\newcommand{\rhoLENT}{$\rho$LENT}
\newcommand{\Cell}[1]{\Omega_{#1}}

\newcommand{\tnn}{t^{n+1}}
\newcommand{\Sf}{\mathbf{S}_f}
\newcommand{\Sfhat}{\hat{\mathbf{S}}_f}
\renewcommand{\v}{\mathbf{v}}
\newcommand{\x}{\mathbf{x}}
\newcommand{\n}{\mathbf{n}}

\newcommand{\rhoratio}{\rho^- / \rho^+}

\newcommand{\e}[1]{\ensuremath{\times10^{#1}}}

\usepackage[disable]{todonotes}

\usepackage[dvipsnames]{xcolor}
\definecolor{cadmiumred}{rgb}{0.89, 0.0, 0.13}
\definecolor{brandeisblue}{rgb}{0.0, 0.44, 1.0}
\colorlet{Reviewer1}{black}
\colorlet{Reviewer3}{black}
\colorlet{Reviewer4}{black}
\colorlet{Reviewer24}{black}
\colorlet{Dieter}{black}
\colorlet{Reviewer25}{black}

\allowdisplaybreaks

\let\oldequation\equation
\let\oldendequation\endequation

\renewenvironment{equation}
  {\linenomathNonumbers\oldequation}
  {\oldendequation\endlinenomath}

\let\oldalign\align
\let\oldendalign\endalign

\renewenvironment{align}
  {\linenomathNonumbers\oldalign}
  {\oldendalign\endlinenomath}

\begin{document}

\frenchspacing

%\thispagestyle{empty}
%\input{significance-and-novelties.tex}
%\clearpage

%\thispagestyle{empty}
%\input{research-highlights.tex}
%\clearpage

%\thispagestyle{empty}
%\input{credit-statement.tex}
%\clearpage

\begin{frontmatter}

%% Title, authors and addresses

%% use the tnoteref command within \title for footnotes;
%% use the tnotetext command for theassociated footnote;
%% use the fnref command within \author or \address for footnotes;
%% use the fntext command for theassociated footnote;
%% use the corref command within \author for corresponding author footnotes;
%% use the cortext command for theassociated footnote;
%% use the ead command for the email address,
%% and the form \ead[url] for the home page:
%% \title{Title\tnoteref{label1}}
%% \tnotetext[label1]{}
\author{Jun Liu}
\ead{jun.liu@tu-darmstadt.de}
\author{Tobias Tolle}
\ead{tolle@mma.tu-darmstadt.de}
\author{Dieter Bothe}
\ead{bothe@mma.tu-darmstadt.de}
\author{Tomislav Mari\'{c}\corref{corr}}
\cortext[corr]{Corresponding author}
\ead{maric@mma.tu-darmstadt.de}

\address{Mathematical Modeling and Analysis, Technische Universit\"{a}t Darmstadt}
%\fntext[label2]{Mathematical Modeling and Analysis, Technische Universit\"{a}t Darmstadt}
%% \fntext[label3]{}

\title{An unstructured finite-volume Level Set / Front Tracking method for two-phase flows with large density-ratios}
% \title{\textcolor{Reviewer1}{A consistent discretization of the single-field two-phase momentum convection term for the unstructured finite volume \\ Level Set / Front Tracking method}}

%\title{Recovering exact numerical consistency for the two-phase single-field Navier-Stokes momentum convection with high density ratios, discretized with the unstructured collocated finite volume Level Set / Front Tracking method}

%% use optional labels to link authors explicitly to addresses:
%% \author[label1,label2]{}
%% \address[label1]{}
%% \address[label2]{}

\begin{abstract}

    % \item Exact numerical consistency recovered for the two-phase momentum convection. 
    % \item Theoretical basis provided for the use of the auxiliary mass conservation equation.
    % \item Theoretical relationship between the mass flux density and the phase indicator.
    % \item Implicit discretization of the two-phase momentum convective term.
    % \item Increased numerical stability with strong surface tension forces and density-ratios.
\textbf{This is the preprint version of the published manuscript \url{https://doi.org/10.1016/j.jcp.2023.112426}: please cite the published manuscript when refering to the contents of this document.}

\textcolor{Reviewer4}{We extend the unstructured \textbf{LE}vel set / fro\textbf{NT} tracking (LENT) method \citep{maric2015lentfoam,tolle2020saample} for handling two-phase flows with strongly different densities (high-density ratios)}
by providing the theoretical basis for the numerical consistency between the mass and momentum conservation in the collocated Finite Volume discretization of the single-field two-phase Navier-Stokes equations.
% The cause of the numerical inconsistency lies in the way the cell-centered density is computed in the new time step ($\rho_c^{n+1}$). 
% Specifically, if $\rho_c^{n+1}$ is computed from the approximation of the fluid interface at $t^{n+1}$, and it is not computed by solving a mass conservation equation (or its equivalent), the two-phase momentum convection term will automatically be inconsistently discretized.
\textcolor{Reviewer24}{Our analysis provides the theoretical basis for the mass conservation equation introduced by \citet{ghods2013} and used in \cite{nangia2019,raessi2012,orazzo2017,zuzio2020,chenadec2013vof}}. %\textcolor{Dieter}{but for Volume of Fluid method.} 
% The face-centered mass flux density we base on the fundamental principle of mass conservation, used to model the single-field density, contrary to the use of different weighted averages of cell-centered single-field densities and alternative reconstructions of the mass flux density by other contemporary methods.
%
\textcolor{Reviewer24}{We use a mass flux that is consistent with mass conservation in the implicit Finite Volume discretization of the two-phase momentum convection term, \textcolor{Reviewer4}{and solve the single-field Navier-Stokes equations with our SAAMPLE segregated solution algorithm \citep{tolle2020saample}}.}
The proposed \rhoLENT{} method recovers exact numerical stability for the two-phase momentum advection of a spherical droplet with density ratios  $\rhoratio \in [1, 10^4]$. Numerical stability is demonstrated for in terms of the relative $L_\infty$ velocity error norm, for \textcolor{Reviewer24}{density-ratios in the range of $[1,10^4]$}, dynamic viscosity-ratios in the range of $[1, 10^4]$ and very strong surface tension forces, for challenging mercury/air and water/air fluid pairings. \textcolor{Reviewer24}{In addition, the solver performs well in cases characterized by strong interaction between two phases, i.e., oscillating droplets and rising bubbles.}
%The mercury/air combination is a replacement for molten metal droplets moving through air, but with a known surface tension coefficient.} 
The proposed \rhoLENT{} method\footnote{The implementation in OpenFOAM is publicly available at \url{https://gitlab.com/leia-methods/lent/-/tree/2022-02-rhoLENT-R1} \citep{LENTcode}.} is applicable to any other two-phase flow simulation method that discretizes the single-field two-phase Navier-Stokes Equations using the collocated unstructured Finite Volume Method but does not solve an advection equation for the phase indicator using a flux-based approach, \textcolor{Reviewer4}{by adding the proposed geometrical approximation of the mass flux and the auxiliary mass conservation equation to the solution algorithm}.
\end{abstract}

%%Graphical abstract
%\begin{graphicalabstract}
%\includegraphics{grabs}
%\end{graphicalabstract}

%%Research highlights
%\begin{highlights}
%\item Research highlight 1
%\item Research highlight 2
%\end{highlights}

\begin{keyword}
%% keywords here, in the form: keyword \sep keyword

%% PACS codes here, in the form: \PACS code \sep code

%% MSC codes here, in the form: \MSC code \sep code
%% or \MSC[2008] code \sep code (2000 is the default)

level set \sep front tracking \sep finite volume \sep unstructured mesh \sep high density ratios
\end{keyword}

\end{frontmatter}

%\linenumbers

%% main text
\section{Introduction}
\label{sec:intro}

A variety of natural and industrial two-phase flow processes involve gas/liquid flows, characterized by density ratios $\rhoratio\ge 10^3$\footnote{
In this publication $\rho^-$ denotes the density of the denser fluid, so that 
$\rho^- \geq \rho^+$ and $\rhoratio \geq 1$ holds. 
}, such as the atomization of fuel jets \citep{Li2016}, sloshing tank \citep{godderidge2009investigation}, mold filing \citep{soukane2006application}, water flooding \citep{gao2011numerical}. Large density ratios at the fluid interface cause severe challenges for numerical simulations \citep{desjardins2010}. \textcolor{Reviewer4}{For segregated solvers, the \textcolor{Reviewer24}{discrete} pressure Poisson equation} becomes ill-conditioned  \textcolor{Reviewer24}{if density is cell-centered, since its abrupt change across the interface between the two fluids can lead to a large variation in the matrix coefficients. Additionally,} spurious numerical errors in the solution of the momentum equation accumulate because of inconsistencies between mass and momentum advection. 
%\citet{ghods2013} point out that a standard two-phase solver utilizes the phase-indicator function to compute the mass flux, while the momentum flux is obtained by solving a non-conservative form of the momentum transport equations.
\citet{ghods2013} point out that for level set methods mass and momentum are typically transported in different, inconsistent ways. While mass is transported by a solution of the level set equation, momentum is obtained
from solving a non-conservative form of a momentum balance equation.
Hence, a large non-physical change in the momentum can be generated by a small error in the interface position when the density ratio is high. \citet{nangia2019} state that the abrupt change in density often introduces notable shear at the interface and adds difficulties in the discretization of governing momentum equations at the interface, which further leads to higher stiffness of the linear equation system. 

Many researchers have addressed these problems, and some indicated further that because of the sizeable numerical error resulting from high-density ratios, some flow algorithms or solvers can only be used to solve low density-ratio cases with $\rhoratio \in [1,10]$ \cite{raessi2012}. However, in engineering applications, density ratios usually range from $10^2$ to $10^3$, and even $10^4$ for molten metals or water-water vapor systems. Hence, a solution algorithm with the ability to handle a broader range of density ratio problems is required to simulate real-world engineering problems.     

A pioneering attempt to alleviate numerical instability of the VOF method caused by  high-density ratios was made by \citet{rudman1998}. \citet{rudman1998} has used a sub-mesh with a doubled mesh resolution for advecting volume fractions, compared to the mesh used for the momentum and pressure equations. 
The goal of this two mesh approach was the reduction of small errors in the discrete momentum that cause large errors in the velocity. However, an additional higher mesh resolution for the volume fractions requires a discrete divergence free velocity on the finer mesh. Furthermore, using an additional mesh for the volume fractions increases the computational costs significantly, and it is not applicable to general unstructured meshes. \citet{rudman1998} demonstrates qualitatively a reduction of parasitic currents for the stationary droplet case with $\rhoratio = 100$, and improved results for more complex cases. Another important finding of \citet{rudman1998} is the role of the densities used in the mass flux and the momentum flux in ensuring numerical consistency of the two-phase momentum advection.

\citet{bussmann2002} extended the work of \citet{rudman1998} for the unstructured collocated finite volume method. \citet{bussmann2002} employ the conservative form for the momentum convection. At first, the momentum advection is solved separately, using an explicit Euler time integration scheme. \citet{bussmann2002} use the unstructured unsplit Volume-of-Fluid method of \citet{Rider1998}, which enables the simplification of the numerical consistency requirement for the density and momentum equations. Specifically, the solution of the volume fraction equation results in phase-specific volumes at face centers. Those phase-specific volumes are then used to compute the volume fractions at face centers. These volume fractions are used by \citet{bussmann2002}, together with a simple average of cell densities, and velocities calculated by the least squares reconstruction technique, to compute the momentum fluxes at face centers. Since the velocity is continuous at the interface, the least squares approximation is acceptable. However, calculating face-centered densities by an average does not yield numerical stability in all cases. Contrary to \citet{rudman1998}, \citet{bussmann2002} do not require an additional finer mesh. They do, however, limit the solution to first-order accuracy in time and introduce the CFL condition by solving the momentum advection equation explicitly. \citet{bussmann2002} introduce the important case of a translating droplet in a quiescent ambient fluid. This test case can be used to demonstrate numerical consistency in the momentum transport. Their solutions show accurate results for high density ratios, especially considering the fact that even the unsplit VOF method distorts the interface during the translation \citep{ccerne2002numerical}. However, for $\ \rhoratio \in [1, 10^2]$, the constant translation velocity is modified by the solution of the pressure and momentum equations, which implies a remaining numerical inconsistency in this approach.

\citet{sussman2007sharp} employ the CLSVOF method \cite{sussman2000} for obtaining a robust and stable solution for the density ratio of $1000$ by extrapolating the liquid velocities into the gas domain. The interface is advected using the extrapolated liquid velocity field only.

\citet{raessi2012} propose a 2D staggered discretization of conservative single-field form of two-phase Navier-Stokes equations for handling high density ratios. Like \citet{bussmann2002} did, \citet{raessi2012} first solve the momentum advection equation, using second-order (or higher) explicit integration schemes, and upwinding for the velocity near the interface. The density used in the momentum convective term is computed as a weighted combination of signed distances from the old and the new time step. For the partially submerged line segments bounding 2D rectangular cells, intersection between the mesh and the zero level set (iso-surface) is performed using the marching cubes algorithm. \citet{raessi2012} point out that there is still an inconsistency between the face-centered density and the momentum transport, as the Level Set equation remains decoupled / inconsistent with the momentum transport. The verification of numerical stability was done using the translating droplet case from \citet{bussmann2002}, and results demonstrate qualitative improvement for the density ratio $\rhoratio = 10^6$. Other density ratios have not been verified. A viscous oscillating droplet case demonstrates quantitative improvement in terms of the improved amplitude decay rate, compared to non-conservative form of the momentum equation.

\citet{lechenadec2013} extend their forward/backward Lagrangian tracking and Eulerian remapping VOF method \citep{lechenadec2013} for handling high density ratios. Equivalent to volume fractions in \citep{lechenadec2013}, the density and the momentum are advected in the Lagrangian forward/backward tracking step by observing the control volume as a material volume and moving the mesh forward / backward with the flow velocity. While the content of material volumes does not change on the continuum level, this condition cannot be discretely ensured and is a source of conservation errors. In the Eulerian re-mapping step, physical properties are transferred from the Lagrangian to the Eulerian mesh, and the geometrical intersections between the PLIC interface on the forward/backward image of the mesh, and the background mesh, are another source of volume conservation errors. Ensuring numerical consistency further requires the transfer of velocities located at the center of mass. Since the velocities associated with the cell centroids are used, an inconsistency is introduced. Qualitative results show  significant improvements for the stationary droplet with $\rhoratio = 10^9$, and quantitative improvement is shown for the standing wave by \citet{prosperetti1981motion} with $\rhoratio = 850$.

\citet{ghods2013} have developed a Consistent Rescaled momentum transport (CRMT) method. The CRMT method discretizes the conservative form of the single-field Navier-Stokes equations using a collocated unstructured Finite Volume method. To increase the numerical stability for high density ratio, CRMT solves \textcolor{Reviewer24}{what we call an "auxiliary"} mass conservation equation using a mass flux either by upwinding the face-centered density in the interface cells and their face-neighbors (defined by a volume fraction tolerance), or by averaging the densities elsewhere. The same discretization scheme used for the face-centered density is also applied to the mass flux in the convective term of the momentum equation. A difference is therefore introduced in the mass flux of the continuity equation and the mass flux in the convective term of the momentum equation when upwinding is used, because the upwinded face-centered density in the continuity equation uses the face-centered velocity, while the upwinded mass flux in the momentum equation includes both the upwind velocity and density. We show that any difference in the discretization of the mass flux to be a source of numerical inconsistency for the two-phase momentum advection. Like \citet{bussmann2002}, the explicit discretization of the momentum convective term introduces the CFL condition, limiting the time step for convection-dominated multiphase flows, where high density ratios play a major role. Using upwind schemes makes the discretization first-order accurate. The droplet translation case \cite{bussmann2002}, with $\rhoratio = 10^6$, is compared in terms of the droplet shape, that remains stable. Other density ratios are not reported for this verification case. It is our opinion, that the droplet shape errors may result from the interface advection scheme\footnote{The Level Set and VoF methods do not exactly preserve the shape of a translating droplet.}, and should be generally substituted by the $L_\infty$ norm of the velocity error to demonstrate numerical consistency.

\textcolor{Reviewer4}{\citet{vaudor2014consistent} base their approach on a CLSVOF code from \citet{Menard2007} and \citet{aniszewski2014volume}}, which can switch between LS-based and VOF-based mode to calculate momentum fluxes. They \textcolor{Reviewer24}{\citep{vaudor2014consistent}} chose the VOF-based momentum fluxes calculation mode and implemented the framework of Rudman's method \citep{rudman1998} but with more accurate interpolation schemes for velocities and velocity gradients on faces of staggered meshes to ensure consistency. This method is developed in two-dimensions and exploits two sets of meshes. To provide a more widely applicable method, \citet{vaudor2017} advanced the method in their more recent study. In contrast to the previous work \cite{vaudor2014consistent}, the LS method tracks the interface, while the VOF method is utilized to update density. They exploited the identical scheme to discretize conservative convective term in mass and momentum equation. In addition, the mass flux is also identical in both discretized equations. A new strategy that leverages half cell-faces' and half cells' quantities of volume fraction and density to couple staggered mass cells and momentum cells is introduced to avoid the need for a refined mesh in the original method by \citet{rudman1998}. A prominent feature of this new method is that it can be used to simulate three-dimensional applications. Besides, comparing with the method from \citet{rudman1998}, the new method shows relatively low computational cost when simulating the same 2D application. 
 
\citet{owkes2017} presented a three-dimensional, unsplit, second-order semi-Lagrangian VOF scheme that conserves mass and momentum and ensures consistency between the mass (volume fraction) and momentum fluxes. The volume fractions are geometrically transported near the fluid interface using the method from \citep{Owkes2014}. As in \citep{rudman1998}, \citet{owkes2017} introduce an additional refined mesh for the calculation of semi-Lagrangian fluxes. The motivation for the refined mesh is to enforce the consistency between semi-Lagrangian mass and momentum fluxes, similar to \citet{rudman1998}. Results confirm mass and momentum conservation, and stability of the momentum convection. The method proposed  by \citet{owkes2017} relies on the staggered variable arrangement and this, together with the use of the additional finer mesh, makes this approach inapplicable to unstructured finite volume meshes.

\citet{orazzo2017}, similarly to \citet{rudman1998}, resolve the volume fraction function on twice finer sub-cells and update density from the volume fraction. After that, they update face-centered density on mass cells by averaging density on sub-cells, and then evaluate the mass flux on the faces of standard staggered momentum cells. These density and mass flux values are used to initialize and calculate interim momentum and velocity during the prediction step. \citet{zuzio2020} made no changes and applied Orazzo's method \cite{orazzo2017}. Besides, they further verified and validated this method with \textcolor{Reviewer24}{more complex} cases, \textcolor{Reviewer24}{e.g., liquid jet in cross-flow}. \citet{yang2021robust} notice that the high-density ratio has a profound effect on robustly simulating two-phase flows at high Reynolds numbers. To mitigate the problem, they adopt the consistent framework from \citet{nangia2019} and replace the interface-capturing method in \cite{nangia2019}, which is standard LS, with CLSVOF method \cite{sussman2000} to ensure mass conservation.
 
\citet{patel2017} employ the method of \citet{ghods2013}, a high-resolution scheme called Cubic Upwind Interpolation (CUI) for the convective terms of momentum and volume fraction transport equations, and the solution of a momentum equation in the face-normal direction. The face-normal momentum equation leads to a combined collocated/staggered variable arrangement, that requires the use of nonlinear solvers, as this equation is a non-linear algebraic equation. \citet{patel2017} demonstrate the balanced nature of their discretization for the stationary droplet using exact curvature and density ratios $\rhoratio \in [10,1000]$. Numerical stability is demonstrated with reduced parasitic currents when the curvature is approximated numerically for $\rhoratio = 10, \text{We}=1$. For the verification test case of the two-phase momentum advection problem, $\rhoratio = 10^6$ is used without surface tension and viscous forces and qualitative results show slight deformations of the interface shape, the $L_\infty$ norm of the velocity error is not reported. With enabled surface tension and viscous forces and exact curvature prescribed, and density ratios $\rhoratio=1,1000$, the velocity error in the $L_\infty$ norm lies within $[10^{-3},10^{-2}]$.
 
\citet{manik2018}, similarly to \cite{patel2017}, attempt to enforce numerical consistency by applying the similar discretization scheme on the conservative form of the volume fraction advection equation and the momentum conservation equation. \citet{manik2018} are using a collocated unstructured Finite Volume method for the equation discretization and the CUBISTA scheme (\citet{alves2003convergent}) to discretize convective terms. The verification of the numerical consistency for the two-phase momentum advection is done using the droplet translation case of \citet{bussmann2002} and density ratios $\rhoratio=10^3,10^6$, that demonstrates qualitative improvement compared to a naive discretization of the momentum convective term with the upwind method. The qualitative evaluation is based on the shape of the droplet, given by the $0.5$ iso-surface of the volume fraction. Although the proposed method demonstrates improvement w.r.t. an obviously inconsistent approach, some shape deformation is still visible, so one can conclude that $L_\infty(\mathbf{v})\ne 0$ and some non-zero velocities are still generated. 
 
A recent second-order accurate LS method is proposed by \citet{nangia2019}, extending the work from \citet{ghods2013} that is first-order accurate. Similar to the method proposed by \citet{ghods2013}, an additional mass conservation equation is solved, and the identical mass flux is used for both mass and momentum transport. Two techniques are employed: one is the third-order accurate Koren's limited CUI, which is modified to consistently discretize the convective term of both mass and momentum equation. This scheme satisfies the convection-boundedness criterion (CBC) and is total variation diminishing (TVD). The second technique is the solution of an update equation for the face-centred densities. In this step, a third-order accurate strong stability preserving Runge-Kutta (SSP-RK3) scheme is used for time integration. The update is performed in every fix-point iteration, and the updated face-centered density is then employed to solve the discretized momentum equation.

\citet{zuzio2020} also follow \citet{ghods2013} by solving an auxiliary continuity equation for increasing the numerical consistency in discretizing the two-phase momentum convection term. Their Consistent Mass-Momentum (CMOM) transport method utilizes a staggered Cartesian variable arrangement and utilizes the two-phase incompressible Navier-Stokes equations in the conservative form, solved using \textcolor{Reviewer4}{Chorin's projection method} \citep{Chorin1967} together with the CLSVOF method for tracking the fluid interface. The solution of the auxiliary density equation requires the evaluation of staggered (face-centered) densities, by constructing staggered control volumes, and evaluating the densities using sub-grid quadtree (octree in 3D) refinement and intersection with the PLIC interfaces. This aspect of CMOM shows the importance of evaluating the densities at face-centeres that are required for the solution of the auxiliary continuity equation. Momentum flux reconstruction scales the fluxed phase-specific volume from the VOF method. Finally, the two-phase momentum is advected in the staggered cells, and scaled with the corresponding density to obtain velocity components in all spatial directions. \citet{zuzio2020} demonstrate significant improvements in numerical stability in a very detailed way, reporting shape, position and kinetic energy errors for canonical verification and validation cases. The kinetic energy for the dense translating droplet \cite{bussmann2002} with a density ratio of $10^6$ is reported, and CMOM recovers a numerically stable solution.

\citet{Arrufat2021} consider the conservative form of the advection equation of a discontinuous property to enforce numerical consistency of the advected two-phase momentum, using face averages that are derived by integrating the advection equation in space and time. Since the discontinuity of the property introduced by the interface complicates the evaluation of the face averages, two additional equations are introduced, one for each phase. The method is derived for the MAC staggered variable arrangement. Results demonstrate a numerically stable droplet shape when it is advected with a constant velocity, however, the authors consider this case to only test the consistency of the implementation and not the numerical consistency of the method - we consider it important for both - so the results are not quantified in terms of kinetic energy or $L_\infty$ velocity errors. Still, the method shows significant improvements for realistic multiphase flows with high density ratios.

The high-density ratio is also challenging for other numerical methods for two-phase flows, like the phase-field and lattice Boltzmann. The corresponding surveys are beyond the scope of this work, more details can be found in \cite{ding2007diffuse,wang2015,huang2020,inamuro2004lattice,lee2005stable,zheng2006lattice}.
%The high density ratio problem is not that pronounced for Front Tracking methods \citep{Shin2011} because the marker field (phase-indicator) is not as sharp as in the unstructured Volume-of-Fluid method \citep{Maric2020} and the unstructured Level Set / Front Tracking method \citep{maric2015lentfoam,tolle2020saample}.\todo{Same beginning of the sentence.}
Contrary to the numerical two-phase methods mentioned so far, the difficulties with
high density ratios are far less pronounced for Front Tracking methods \citep{Shin2011} because the marker field (phase-indicator) is not as sharp as in the unstructured Volume-of-Fluid method \citep{Maric2020} and the unstructured Level Set / Front Tracking method \citep{maric2015lentfoam,tolle2020saample}.

The methods of \citet{bussmann2002,ghods2013,patel2017,manik2018} utilize the unstructured Finite Volume equation discretization, other above-mentioned methods utilize a staggered variable arrangement that is not applicable to unstructured meshes.
\textcolor{Reviewer3}{Compared to contemporary collocated Finite Volume methods, our proposed \rhoLENT{} method achieves the numerical consistency in the two-phase momentum advection exactly.}
We derive the requirement for the auxiliary mass conservation equation introduced by \citet{ghods2013}, and derive the requirement for the face-centered (flux) density from the mass conservation principle. 
Compared to a similar observation by \cite{Arrufat2021}, we avoid the integration in time that complicates the evaluation of face-centered quantities, as demonstrated in detail below.
Although hybrid Level Set / Front Tracking LENT method \cite{maric2015lentfoam} is used for interface capturing, the \rhoLENT{} solution algorithm can be used with other interface capturing methods, where there is a discrepancy in the evaluation of the collocated density. 

\textcolor{Reviewer3}{We utilize a collocated unstructured Finite Volume discretization \textcolor{Reviewer24}{because it is ideal for geometrically complex domains}. \textcolor{Reviewer24}{At its core, the proposed} unstructured collocated finite-volume \rhoLENT{} \textbf{LE}vel set / fro\textbf{NT} tracking method \textcolor{Reviewer24}{geometrically approximates the face-centered density in the mass flux} and implicitly discretizes the two-phase momentum convective term, \textcolor{Reviewer24}{thus avoiding the interpolation of face-centered densities and the CFL stability criterion introduced in  \citep{bussmann2002}}}. %The \rhoLENT{} method solves the single-field Navier-Stokes equations in a segregated way, using the SAAMPLE algorithm \citep{tolle2020saample}.}

\section{Mathematical model}
\label{sec:mathematical-model}

\begin{figure}[!htb]
    \centering
    \def\svgwidth{0.6\textwidth}
    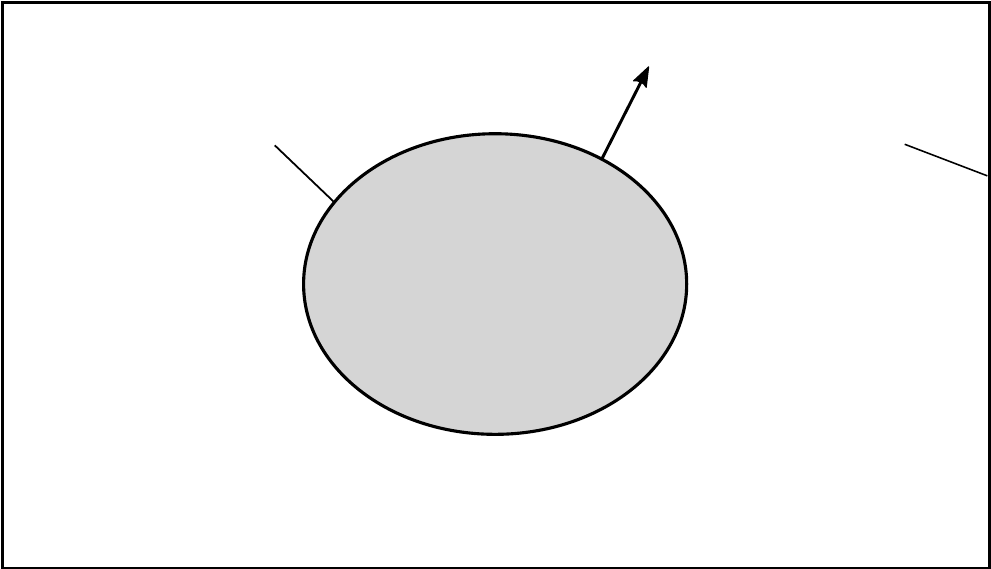\notag
    \caption{\textcolor{Reviewer24}{The domain $\Omega$, split by the fluid interface $\Sigma(t)$ into two sub-domains $\Omega^\pm$}.}
    \label{fig:drop-gas}
\end{figure}

As shown in \cref{fig:drop-gas}, the overall flow domain $\Omega\subset \mathbb{R}^3$ is filled with two phases which occupy subdomains $\Omega^-(t)$ and $ \Omega^+(t)$, such that $\Omega = \Omega^+(t) \cup \Omega^-(t) \cup \Sigma(t)$. The unit normal vector $\boldsymbol{n}_\Sigma$ of the interface $\Sigma(t)$ is oriented outwards for the subdomain $\Omega^-(t)$. These two phases have different material properties that change sharply across the interface $\Sigma(t)$, separating the two subdomains. To identify the phase at a particular location $\mathbf{x}$ and time $t$, the phase indicator is utilized and defined as  
\begin{equation}
  \chi(\x,t) := 
    \begin{cases}
      1, & \x \in \Omega^-(t), \\ 
      0, & \x \in \Omega^+(t) . 
    \end{cases}
  \label{eq:indicator}
\end{equation}
In this work, a single-field formulation of governing equations is used to model \textcolor{Reviewer24}{incompressible} two-phase flows \textcolor{Reviewer24}{without phase-change}. Constant density and dynamic viscosity of the two phases, namely $\rho^-$, $\rho^+$ and $\mu^-$, $\mu^+$ are combined into single-fields using the phase indicator:
\textcolor{Reviewer24}{
\begin{align}
    \rho(\x,t) &= \chi(\x,t) \rho^- + (1-\chi(\x,t)) \rho^+ 
        = (\rho^- -  \rho^+)\chi(\x,t) + \rho^+, \label{eq:rhoindicator} \\
    \mu(\x,t) &= \chi(\x,t) \mu^- + (1-\chi(\x,t)) \mu^+ 
        = (\mu^- - \mu^+) \chi(\x,t) + \mu^+,
\end{align}}
and used in the Navier-Stokes equations in the single-field formulation model the incompressible two-phase flow, 
\textcolor{Reviewer25}{
\begin{align}
    \nabla\cdot\v &= 0 \label{eq:volume-transport}, \\
    %\partial_t \rho + \nabla \cdot (\rho \v) &= 0 \label{eq:mass-transport},\\
    \partial_t(\rho \v)+\nabla\cdot(\rho \v \otimes \v)& = -\nabla P + \rho\mathbf{g} +\nabla\cdot\left(\mu\left(\nabla\v + (\nabla\v)^T\right)\right) + \mathbf{f}_\Sigma. 
    %\partial_t(\rho \v)+\nabla\cdot(\rho \v \otimes \v-\mathbf{T}) &=  -\nabla \rho (\mathbf{g}\cdot \mathbf{x}) + \mathbf{f}_\Sigma,
    \label{eq:momentum-orig}
\end{align}}
% \textcolor{Dieter}{
% We define the stress tensor $\mathbf{T}$ as}  
% %
% \textcolor{Dieter}{
% \begin{equation}
%     % \mathbf{T}=-\left(p+\frac{2}{3}\mu\nabla\cdot\v\right)\mathbf{I} + \mu\left(\nabla\v + (\nabla\v)^T\right).
%     \mathbf{T}=-p\mathbf{I} + \mu\left(\nabla\v + (\nabla\v)^T\right).
% \end{equation}}
%
% \textcolor{Reviewer24}{The pressure $P$ in the momentum equation \ref{eq:momentum-orig} is split into the dynamic pressure $p$ and the hydrostatic pressure
% \begin{equation}
%      P = p + \rho \mathbf{g} \cdot \mathbf{x},
% \end{equation}
% with $\x$ as the position vector giving $\nabla \x = I$, resulting in}
\textcolor{Reviewer25}{With $\x$ as the position vector and constant gravitational acceleration $\mathbf{g}$, we can write
\begin{equation}
    \nabla(\rho\mathbf{g}\cdot \mathbf{x}) 
    = (\mathbf{g}\cdot\mathbf{x})\nabla \rho + 
      \rho \mathbf{g}\cdot \nabla \mathbf{x} 
    = (\mathbf{g}\cdot\mathbf{x})\nabla \rho + 
      \rho \mathbf{g},
      \label{eq:gradrhogx}
\end{equation}
since $\nabla \mathbf{x} = \mathbf{I}$ ($\mathbf{I}$ is the identity matrix). We rearrange $-\nabla P + \rho\mathbf g$ on the r.h.s. of \cref{eq:momentum-orig} using \cref{eq:gradrhogx} 
\begin{equation}
    -\nabla P + \rho \mathbf{g} 
    = -\nabla P + \nabla(\rho \mathbf{g}\cdot \mathbf{x}) - (\mathbf{g} \cdot \mathbf{x})\nabla \rho
    = -\nabla(P - \rho \mathbf{g}\cdot \mathbf{x})
    - (\mathbf{g} \cdot \mathbf{x})\nabla \rho,
\end{equation} and defining the dynamic pressure as $p:=P - \rho\mathbf{g}\cdot \mathbf{x}$ results in}
\textcolor{Reviewer24}{
\begin{equation}
     \partial_t(\rho \v)+\nabla\cdot(\rho \v \otimes \v) = -\nabla p -  (\mathbf{g}\cdot\mathbf{x}) \nabla \rho + \nabla\cdot\left(\mu\left(\nabla\v + (\nabla\v)^T\right)\right) + \mathbf{f}_\Sigma. 
     \label{eq:momentum-transport}
\end{equation}}
% In the standard model, i.e.\ with a constant surface tension and without phase change, 
\textcolor{Reviewer24}{The surface tension force per unit volume $\mathbf{f}_\Sigma$ exerts a force on the interface $\Sigma(t)$} and is modeled using a CSF (Continuum Surface Force) model \citep{brackbill1992continuum}
\begin{equation}
    \mathbf{f}_\Sigma = \sigma \kappa \mathbf{n}_\Sigma\delta_\Sigma
    \label{eq:csf-model}
\end{equation}
\textcolor{Reviewer24}{with a constant surface tension coefficient $\sigma$}, \textcolor{Reviewer24}{$\kappa$ as twice the local mean curvature of $\Sigma(t)$}, and $\delta_\Sigma$ as the interface Dirac distribution.
% Note that in case of variable surface tension coefficient and phase change, \cref{eq:csf-model,eq:volume-transport} would need to be modified.

% When only the advection of the two-phase momentum is considered, all source terms on the r.h.s. of Eq. \ref{eq: momentum-transport} are disregarded, and the momentum transport equation obtains the following form: 

% \begin{equation}
%     \partial_t(\rho \v)+\nabla\cdot(\rho \v \otimes \v)=0.
%     \label{eq:momentum-advection}
% \end{equation}

% With such a sharp interface model, the fluid flow is usually described by a continuum mechanical model. An approximative solution of this model requires a disjoint decomposition of the solution domain into non-overlapping volumes, so-called \emph{cells} $\Cell{c}$, denoted by
% %
% \begin{equation}
%     \Omega_h = \{ \Cell{c} \}_{c \in C}, \quad \text{ such that } \Omega = \cup_{c \in C} \Cell{c}.
%     \label{eqn:omegah}
% \end{equation}
% %

\section{A solution algorithm for two-phase flows with high density ratios using the collocated unstructured Finite Volume method}
\label{sec:numerical-method-and-implementation}

% \int_{\Omega_c} \nabla \mu \, dV = (\nu^- - \nu^+) 
% \int_{\Omega_c} \nabla \chi \, dV = (\nu^- - \nu^+)\int_{\partial \Omega_c} \chi \mathbf{n} \, dS = (\nu^- - \nu^+) \frac{|S_f|}{|S_f|}\int_{\partial \Omega_c} \chi \mathbf{n} \, dS = (\nu^- - \nu^+)|S_f| \sum_f \alpha_f \mathbf{\hat{S}}_f = (\nu^- - \nu^+) \sum_f \alpha_f \mathbf{S}_f

% \sum_f \mu_f \mathbf{S}_f 
% = \sum_f [(\nu^- - \nu^+) \alpha_f + \nu^+]  \mathbf{S}_f
% = (\nu^- - \nu^+) \sum_f \alpha_f \mathbf{S}_f + \nu^+ \sum_f \mathbf{S}_f

The volume fraction $\alpha (\x,t)$ is defined as \textcolor{Reviewer24}{a volumetric average of the phase indicator}\textcolor{Reviewer24}{ $\chi(\x,t)$} over a fixed control volume $\Cell{c}$, i.e. 
\begin{equation}
  \alpha_c(t) := \frac{1}{|\Cell{c}|}\int_{\Cell{c}} \chi(\x, t) \, dV, 
  \label{eq:volfracdef}
\end{equation}
\textcolor{Reviewer24}{and in the equation discretization using the unstructured Finite Volume method, $\Omega_c$ corresponds to the finite volume (mesh cell), generating a discrete field of volume fractions $\{\alpha_c\}_{c \in C}$. For simplicity, we denote both $\{\alpha_c\}_{c \in C}$ and $\alpha_c$ with $\alpha_c$ and use the plural "volume fractions" when discussing $\{\alpha_c\}_{c \in C}$ and singular "volume fraction" when discussing $\alpha_c$ in the text below}.
%
% The control volume $\Omega_c$ is equivalent to the finite volume in the discrete setting.
%
\textcolor{Reviewer24}{The volume fractions $\alpha_c(t)$ are approximated from the geometrical approximation of the fluid interface $\Sigma(t)$, the so-called Front $\tilde{\Sigma}(t)\approx \Sigma(t)$, that is advected using the flow velocity $\v(t)$ by unstructured finite-volume Level Set / Front Tracking (LENT) method \citep{maric2015lentfoam}. The LENT method geometrically computes  signed distances $\psi(\x,t)$, positive in the direction of $\n_\Sigma$ (cf. \cref{fig:drop-gas}), details are given in \citep{maric2015lentfoam,tolle2020saample,Tolle2022}.}

The \textcolor{Reviewer24}{collocated unstructured Finite Volume method \cite{Jasak1996,Hirsch1997,Moukalled2016}}, which is implemented in the OpenFOAM open-source software \cite{Jasak2007,Jasak2009,OFprimer}\textcolor{Reviewer24}{,} associates with the centroid of the \textcolor{Reviewer24}{finite} volume $\Omega_c$ the density
\begin{equation}
    \label{eq:mixed_rho_eqn}
    \rho_c(t) = (\rho^- - \rho^+) \alpha_c(t) + \rho^+,
\end{equation}
and the viscosity
\begin{equation}
    \mu_c(t) = (\mu^- - \mu^+) \alpha_c(t) + \mu^+.
    \label{eq:mixed_mu_eqn}
\end{equation}
\Cref{eq:mixed_rho_eqn,eq:mixed_mu_eqn} are \textcolor{Reviewer24}{ used in a collocated FV discretization of} \cref{eq:momentum-transport}.
The surface tension force given by \cref{eq:csf-model} is \textcolor{Reviewer24}{approximated} as
\begin{equation}
    \mathbf{f}_{\Sigma,c} \approx \sigma \kappa_c \nabla \alpha_c,
\end{equation}
where \textcolor{Reviewer24}{cell-centered curvature $\kappa_c$ }\textcolor{Reviewer4}{is calculated using $\nabla\cdot(\frac{\nabla \psi}{|\nabla \psi|})$, \textcolor{Reviewer24}{and $\psi$ is the geometrically computed distance from the Front \citep{maric2015lentfoam}}. The cell curvature $\kappa_c$ is propagated as a constant in the interface-normal direction using a so-called spherical correction. Details on curvature calculation are given in \cite{tolle2020saample}. The term $\nabla\alpha_c$ denotes a discrete finite volume gradient of
the volume fraction used to approximate the interface Dirac distribution}.

\subsection{The unstructured Finite Volume Hybrid Level Set / Front Tracking method}
Hybrid multiphase flow simulation methods combine the sub-algorithms of the Front Tracking, Level Set, or Volume-of-Fluid methods to achieve better overall results. The structured Hybrid Level Set / Front Tracking method (\cite{shin2002modeling,shin2005accurate,ceniceros2010robust,Shin2011,shin2017solver}) has demonstrated remarkable capabilities for simulating a wide range of multiphase flows. 
The unstructured Level Set / Front Tracking method - the LENT method \cite{maric2015lentfoam,tolle2020saample,Tolle2022} - shows promising computational efficiency and accuracy for surface tension driven flows on unstructured meshes. 

\textcolor{Reviewer24}{However, LENT method in its existing form cannot handle two-phase flows with strongly different densities.  
 Cell-centered volume fractions  $\alpha_c(t)$ are computed from signed distances $\psi_c(t)$, that are computed geometrically from the Front $\tilde{\Sigma}(T)\approx \Sigma(t)$: a triangular surface mesh that approximates the fluid interface $\Sigma(t)$. This geometrical calculation of $\alpha_c$ from $\tilde{\Sigma}(t)$ and, subsequently, the calculation of $\rho_c$ from $\alpha_c$ by \cref{eq:mixed_rho_eqn}, together with the interpolation of the face-centered density in the mass flux of the discretized convective term from \cref{eq:momentum-transport}, introduces an inconsistency that we describe in detail and address below.}

%%%%%%%%%%%%%%%%%%%%%%%%%%%%%%%%%%%%%%%%%%%%%%%%%%%%%%%
\subsection{Numerical consistency of the single-field conservative two-phase momentum convection term}
\label{sec:momconsistency}

\textcolor{Reviewer24}{As outlined in the introduction section \ref{sec:intro}, many authors have addressed numerical instabilities in various discretizations and two-phase flow methods arising from high-density ratios. Here, we provide a detailed analysis of the inconsistencies that lead to numerical instabilities by studying the relationships between mass conservation, phase indicator function conservation, and momentum convection. It turns out that the conservative formulation of conservation equations permits us to precisely define equalities that must hold in the mathematical model and its discretization to achieve consistency in the equation system and prevent numerical instabilities.}

\citet{bussmann2002} were the first to consider the problem of numerical consistency of the two-phase momentum convective term in the setting of the collocated unstructured Finite Volume method. We expand on their work by improving the accuracy of the face-centered density evaluation and employing a solution algorithm that allows for an implicit discretization of the convective term, thus removing the CFL condition. \textcolor{Reviewer24}{We discretize the two-phase momentum convection term from \cref{eq:momentum-transport} using the collocated unstructured \textcolor{Reviewer24}{finite volume method (FVM)} \citep{Jasak1996, maric2014openfoam, moukalled2016finite} as 
\begin{equation}
    \int_{\Omega_c} \nabla \cdot( \rho \v \otimes \v) \, dV
    = \int_{\partial \Omega_c} ( \rho \v \otimes \v) \cdot \n \, ds
    = \sum_{f \in F_c} \rho_f F_f \v_f + O_{\rho\v,con}(h^2).
    \label{eq:momterm}
\end{equation}
 \textcolor{Reviewer24}{The second-order discretization error is denoted as $O_{\rho\v,con}(h^2)$, with the ${\rho\mathbf{v}}$ subscript indicating the equation ($\rho\mathbf{v}$ for momentum eq. \ref{eq:momentum-transport}), ${con}$ subscript the convective term of the equation, and $h$ the discretization length.}The convective term in \cref{eq:momterm} has been linearized with respect to the solution variable $\mathbf{v}$ in order to obtain a linear equation system. Here, $\rho_f$ represents the face-center density, and $F_f$ represents the linearized volumetric flux.
 %The convective term has been linearized with respect to $\mathbf{v}$ in \cref{eq:momterm} with $\mathbf{v}$ as the solution variable of the resulting linear equation system, $\rho_f$ as the linearized face-centered density, and $F_f$ as the linearized volumetric flux. 
 Details on the flux linearization and temporal integration are given in \cref{sec:density update}, here we first focus on the spatial discretization. }

\textcolor{Reviewer24}{The discretization \ref{eq:momterm} requires a mass flux $\rho_f F_f$. %The incompressibility of both phases relates volume conservation \cref{eq:volume-transport} to mass conservation.
\textcolor{Reviewer25}{The volume fraction conservation and the conservation of mass are equivalent if both phases are incompressible.
% However, collocated FV methods that solve single-field Navier-Stokes equations, but are not inherently volume conservative, introduce an inconsistency between \cref{eq:volume-transport} and 
To show this,} we write the mass conservation equation in conservative form \textcolor{Reviewer25}{in a fixed (time-independent) control volume $\Omega_c \ne \Omega_c(t)$} as  
\begin{equation}
    \partial_t \int_{\Omega_c} \rho \, dV =  - \int_{\partial\Omega_c} \rho\v\cdot\n \, dS. 
    \label{eq:mass-transport}
\end{equation}
Applying \cref{eq:rhoindicator} and \cref{eq:volume-transport} to \cref{eq:mass-transport} leads to 
\textcolor{Reviewer25}{
\begin{equation}
    (\rho^- - \rho^+)\partial_t \int_{\Omega_c} \chi \, dV =  -(\rho^- - \rho^+) \int_{\partial \Omega_c} \chi \v \cdot \n \, dS - \rho^+ \int_{\partial_{\Omega_c}} \v \cdot n \, dS, 
    \label{eq:chiscaled}
\end{equation}
with $\int_{\partial_{\Omega_c}} \v \cdot n \, dS = \int_{\Omega_c} \nabla \cdot \v \, dV = 0$ because of \cref{eq:volume-transport}.} Dividing \cref{eq:chiscaled} by $|\Omega_c|$, and using the volume fraction definition \ref{eq:volfracdef} leads to 
\begin{equation}
     (\rho^- - \rho^+) \partial_t \alpha_c(t) =   -(\rho^- - \rho^+) \frac{1}{|\Omega_c|}\int_{\partial \Omega_c} \chi \v \cdot \n \, dS. 
    \label{eq:alphac-scaled}
\end{equation}
%
%Dividing \cref{eq:mass-transport} with $|\Omega_c|$ gives
\textcolor{Reviewer24}{The \cref{eq:mass-transport} implies}
\begin{equation}
    \partial_t \rho_c(t) =  - \frac{1}{|\Omega_c|} \int_{\partial \Omega_c} \rho\v\cdot\n \, dS. 
    \label{eq:rhoc-transport}
\end{equation}
An important equality arises from \cref{eq:mass-transport,eq:alphac-scaled,eq:rhoc-transport}, namely
\begin{align}
    \partial_t \rho_c(t)  
    =  -\frac{1}{|\Omega_c|} \int_{\Omega_c} \rho\v\cdot\n \, dV
    =  (\rho^- - \rho^+) \partial_t \alpha_c(t)    
    =  -(\rho^- - \rho^+) \frac{1}{|\Omega_c|}\int_{\partial \Omega_c} \chi \v \cdot \n \, dS. \label{eq:consistency}
\end{align}
Selecting $\partial_t \rho_c(t) = (\rho^- - \rho^+) \partial_t \alpha_c(t)$ from \cref{eq:consistency}, and integrating over the time interval $[t^n, t^{n+1}]$ leads to
\begin{equation}
    \rho_c^{n+1} = (\rho^- - \rho^+)\alpha_c^{n+1} + \rho_c^n - (\rho^- -\rho^+)\alpha_c^n,
    \label{eq:rhocinterim}
\end{equation}
and applying \cref{eq:mixed_rho_eqn} at $t^n$ to \cref{eq:rhocinterim} leads to 
\begin{equation}
    \rho_c^{n+1} = (\rho^- - \rho^+)\alpha_c^{n+1} + \rho^+,
\end{equation}
which is \cref{eq:mixed_rho_eqn} at $t^{n+1}$. Note that the time integration is exact because of the fundamental theorem of calculus. 
The equality $\partial_t \rho_c(t) = (\rho^- - \rho^+) \partial_t \alpha_c(t)$ from \cref{eq:consistency} can lead to a false conclusion of consistency of the two-phase momentum convection. If other equalities from \cref{eq:consistency} are not upheld, e.g. when the method that advects the phase indicator $\alpha_c$ does not rely on phase-specific fluxes (cf. \citep{Maric2020} for a recent review), inconsistencies arise.
}

\textcolor{Reviewer24}{Since unstructured finite volumes are bounded by faces $S_f$, we can rewrite  
\begin{linenomathNonumbers}
\begin{equation*}
    \frac{1}{|\Omega_c|} \int_{\Omega_c} \rho\v\cdot\n \, dV
    = (\rho^- - \rho^+) \frac{1}{|\Omega_c|}\int_{\partial \Omega_c} \chi \v \cdot \n \, dS 
\end{equation*}
\end{linenomathNonumbers}
from \cref{eq:consistency} as 
\begin{equation}
    \sum_{f \in F_c} \int_{S_f} \rho\v\cdot\n \, dS
    = (\rho^- - \rho^+) \sum_{f \in F_c} \int_{S_f}  \chi \v \cdot \n \, dS,
\end{equation}
and discretize it further using second-order-accurate face-averages, resulting in  
\begin{equation}
   \begin{aligned}
    \sum_{f \in F_c} \rho_f F_f + O_{\rho, con}(h^2) & 
    = 
     \sum_{f \in F_c} (\rho^- - \rho^+) \frac{F_f}{|\Sf|} \int_{S_f}  \chi dS + O_{\alpha, con}(h^2) \\
     & = \sum_{f \in F_c} (\rho^- - \rho^+) F_f \alpha_f + O_{\alpha, con}(h^2), 
    \end{aligned}
    \label{eq:fluxequality}
\end{equation}
with $\phi_f := \frac{1}{|S_f|}\int_{S_f} \phi \, dS$ defining the face-average associated to the centroid of each face $S_f$. \textcolor{Reviewer24}{In \cref{eq:fluxequality},} $F_f:= \v_f \cdot \Sf$ is the linearized volumetric flux given by the velocity $\v$ from the discretized single-field Navier-Stokes equations \cref{eq:volume-transport,eq:momentum-transport}.
\Cref{eq:fluxequality} \textcolor{Reviewer24}{reveals an important fact}: %sets an important requirement. 
the mass flux $\rho_f F_f$ - necessary for the discretization of the two-phase momentum convective term \ref{eq:momterm} - must be linearly proportional to the phase-specific volumetric flux $F_f \alpha_f$ used to advect the phase indicator $\alpha_c$, with  $(\rho^- - \rho^+)$ as the proportionality coefficient. 
%Any two-phase flow simulation method that utilizes a flux-based approach to solve the advection equation for the volume fractions ensures the fulfillment of this consistency condition.
This consistency is not ensured by any two-phase flow simulation method that does not solve an advection equation for the volume fractions using a flux-based discretization method.}

\textcolor{Reviewer24}{We extend the unstructured Level Set / Front Tracking LENT method \citep{maric2015lentfoam,tolle2020saample,Tolle2022} to ensure that the condition from \cref{eq:fluxequality} is upheld. Before describing the numerical method in detail, we discuss a verification case of a droplet advected in a constant velocity field.}

\textcolor{Reviewer24}{\subsubsection{Verification case: droplet translating with constant velocity}}

\textcolor{Reviewer24}{We consider the Euler explicit collocated unstructured FV discretization  of \cref{eq:mass-transport}, \textcolor{Reviewer24}{i.e.,} 
\begin{equation}
        % \frac{\rho_c^{n+1}-\rho_c^{n}}{\Delta t} + \frac{1}{|\Omega_{c}|}\sum_{f \in F_c}\rho_{f}^{n}F_{f}^{n} = 0, \\
        \rho_c^{n+1} = \rho_c^{n}- \frac{\Delta t}{|\Omega_{c}|}\sum_{f \in F_c}\rho_{f}^{n}F_{f}^{n}. \label{eq:density-update-conti}
\end{equation}}
%
% where $F_f$ is the volumetric flux at the face defined by the index $f$ from the set of all indexes of cell-faces $F_c$ that belong to $\Omega_c$, defined as
% %
% \begin{equation}
%     F_f := \v_f \cdot\Sf,
%     \label{eq:fluxdef}
% \end{equation}
% %
% with $\v_f^n$ as the face-centered velocity and $\Sf:=|S_f|\Sfhat$ as the face area-normal vector, with $|S_f|$ denoting the area of the face $S_f$. 
%
The two-phase momentum advection is modeled using \cref{eq:momentum-transport} with a prescribed initial constant velocity and without forces on the r.h.s, namely
\begin{equation}
    \partial_t(\rho \v)+\nabla\cdot(\rho \v \otimes \v)=0.
    \label{eq:momentum-advection}
\end{equation}
Without forces on the r.h.s. of \cref{eq:momentum-advection}, the initial constant velocity should remain spatially constant.
Therefore, a numerically consistent unstructured collocated FVM discretization of the two-phase momentum convection equation (\cref{eq:momentum-advection}) must ensure that no artificial acceleration or deceleration occurs.
For example, just like \cref{eq:density-update-conti}, the Euler explicit discretization of \cref{eq:momentum-advection} is
\begin{equation}
    \rho_c^{n+1}\v_c^{n+1} = \rho_c^{n}\v_c^{n} - \frac{\Delta t}{|\Omega_{c}|}\sum_{f \in F_c}\rho_{f}^{n}F_{f}^{n}\v_f^{n}.
    \label{eq:momentumeuler}
\end{equation}
Given a consistent discretization, the velocity field remains spatially constant, so
\begin{equation}
    \v_f^n=\v_c^n,
    \label{eq:vfeqvc}
\end{equation}
which is, of course, ensured for the initial spatially constant velocity $(\v_f^0 = \v_c^0)$. \Cref{eq:vfeqvc}, applied to \cref{eq:momentumeuler}, results in
\begin{equation}
    \rho_c^{n+1}\v_c^{n+1}=\v_c^n\left(\rho_c^n-\frac{\Delta t}{|\Omega_{c}|}\sum_{f \in F_c}\rho_{f}^{n}F_{f}^{n}\right),
\end{equation}
and dividing by $\rho_c^{n+1}$ finally gives
\begin{equation}
    \v_c^{n+1}=\dfrac{\v_c^n\left(\rho_c^n-\dfrac{\Delta t}{|\Omega_{c}|}\sum_{f \in F_c}\rho_{f}^{n}F_{f}^{n}\right)}{\rho_c^{n+1}}.
    \label{eq:momentum-advection-v-update}
\end{equation}
As there are no forces on the r.h.s. of \cref{eq:momentum-advection}, the velocity should not be changed simply by advecting the two-phase momentum, i.e.
\begin{equation}
    \v_c^{n+1}=\v_c^n,
    \label{eq:velocity-no-change}
\end{equation}
and this condition is ensured in \cref{eq:momentum-advection-v-update} if
\begin{equation}
    \dfrac{\rho_c^n-\dfrac{\Delta t}{|\Omega_{c}|}\sum_{f \in F_c}\rho_{f}^{n}F_{f}^{n}}{\rho_c^{n+1}}=1,
    \label{eq:momcondition}
\end{equation}
which is equivalent to \cref{eq:density-update-conti}: the Euler explicit discretization of the mass conservation equation. Consequently, a numerically consistent discretization of the momentum convection equation \textcolor{Reviewer24}{ in this verification case} requires the new cell-centered density $\rho_c^{n+1}$ to be computed by solving a mass conservation equation. 
%This observation justifies theoretically the use of the auxiliary mass conservation equation, introduced by \citet{ghods2013}, and used by \cite{nangia2019,raessi2012,orazzo2017,zuzio2020,chenadec2013vof}.

\textcolor{Reviewer1}{Modern unstructured geometric flux-based Volume-of-Fluid \textcolor{Reviewer24}{(VOF)} methods (\cite{Ivey2017,owkes2017,Maric2018,Scheufler2019}, see \cite{maric2020unstructured} for a recent review) potentially ensure this property, since they solve the conservative formulation of the volume fraction advection equation \citep{Maric2020} for $\alpha_c^{n+1}$ by computing phase-specific fluxed volumes, \textcolor{Reviewer24}{scale the phase-specific fluxed volumes to compute the mass flux}, and use the cell-centered volume fraction $\alpha_c^{n+1}$ to compute $\rho_c^{n+1}$ with \cref{eq:mixed_rho_eqn}. However, the temporal discretization scheme used in the momentum equation for the convective term must be consistent with the integration of the fluxed phase-specific volumes, used to obtain $\alpha_c^{n+1}$. \textcolor{Reviewer24}{Even if the mass flux can be computed by scaling the phase-specific fluxed volumes with $\delta t$, any difference between the temporal integration schemes used for the volume fraction and momentum equations, or any flux limiting in the momentum equation, cause inconsistencies.} Additionally, the $\alpha_c^{n+1} \in [0,1]$ must hold near machine epsilon. Any correction to $\alpha_c^{n+1}$ performed after the numerical solution of the volume fraction advection equation, that bounds $\alpha_c^{n+1}$ within $[0,1]$, results in a discrepancy between $\rho_c^{n+1}$ computed using the mass flux that gives unbounded $\alpha_c^{n+1}$, and the $\rho_c^{n+1}$ computed from the a-posteriori bounded $\alpha_c^{n+1}$ using \cref{eq:mixed_rho_eqn}.} 

It is important to note that if the pressure gradient is included on the r.h.s of \cref{eq:momentum-advection}, any error in $\v_c^{n+1}$ will result in non-zero source terms on the r.h.s. of the resulting pressure equation, in the $p-\v$ coupling algorithm. Since the pressure gradient enforces $\nabla\cdot \v =0$ ($\sum_{f \in F_c} F_f = 0$ on the discrete level), this results in artificial velocities similar to parasitic currents caused by the surface tension force. 

% However, numerical methods such as Front Tracking, Level Set, and their hybrids, update the cell-centered density $\rho_c^{n+1}$ from the approximated fluid interface that is not advected by solving the volume fraction advection equation.
\citet{bussmann2002} \textcolor{Reviewer24}{have utilized} the consistency of the Volume-of-Fluid method \textcolor{Reviewer24}{and the availability of phase-specific volumetric fluxes in the VOF method} to first solve \cref{eq:momentum-advection} explicitly in the first step, followed by the second step that includes volume and surface forces. The approach from \citet{bussmann2002} \textcolor{Reviewer24}{cannot be applied without modifications to the Level Set method, the Front Tracking method, their hybrids, or any other collocated FV two-phase flow simulation method that does not rely on phase-specific volumetric fluxes to discretely advect volume fractions. If the phase-specific volumetric fluxes (or volumes) \textcolor{Reviewer24}{are not calculated by the method}, they cannot be used to construct a consistent mass flux.} The solution algorithm for high density ratios that we propose avoids the CFL condition imposed by \citet{bussmann2002} and increases the accuracy of the face-centered density $\rho_f$ required by the mass flux, and it is applicable to any multiphase flow simulation method that utilizes the single-field formulation of the Navier-Stokes equations.
\hfill\\

\subsection{A semi-implicit solution algorithm for high-density ratios}\label{sec:density update}

\begin{figure}[!htb]
    \centering
    \def\svgwidth{0.3\textwidth}
    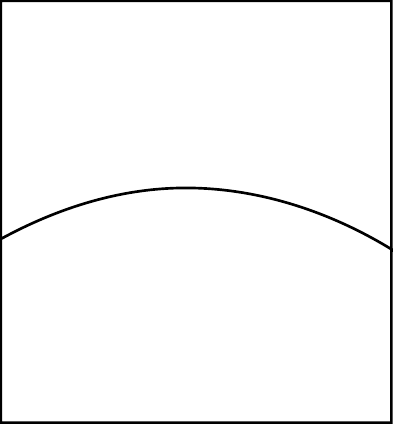
    \caption{A two-phase fixed control volume $\Omega_c$ separated by the interface $\Sigma(t)$.}
    \label{fig:alpha_f_model}
\end{figure}

\Cref{sec:momconsistency} provides the formal reasoning behind solving the mass conservation equation (or its equivalent) for $\rho_c^{n+1}$.
Since \citet{ghods2013} introduced \textcolor{Reviewer24}{what we call an} "auxiliary" mass conservation equation, other researchers have adopted this approach, with the main difference in the way the face-centered (mass flux) density $\rho_f$ is evaluated both in the discretized mass conservation equation (\cref{eq:mass-transport}) and the discretized momentum equation (\cref{eq:momentum-transport}). 

The condition given by \cref{eq:momcondition}, derived from \cref{eq:momentum-advection-v-update,eq:velocity-no-change} can be fulfilled only if the same face-centered (mass flux) density is used when discretizing the auxiliary mass conservation and momentum equations.
Going one step further, the volumetric flux $F_f$ must also be the same in the discretized auxiliary mass conservation and momentum equations.
Put together, the mass flux in the auxiliary discretized mass conservation equation must be equal to the mass flux in the discretized momentum conservation equation: this is the requirement for the mass flux consistency, mentioned throughout the literature.

It is relevant to point out that the same model for the single-field density given by \cref{eq:mixed_rho_eqn} is used throughout the literature.
The basis of this model is mass conservation, and this fundamental principle further leads to an interesting conclusion regarding the evaluation of the face-centered (mass flux) density $\rho_f$ in the discretized mass and momentum conservation equations.
The face centered density is evaluated differently throughout scientific publications reviewed in \cref{sec:intro}, and here we show that there is a strict relationship between the phase indicator and the face centered density $\rho_f$.

Consider the fixed control volume $\Omega_c$ in \cref{fig:alpha_f_model}, that is separated by the fluid interface $\Sigma(t)$ into two parts, occupied by fluids $\Omega^\mp(t)$.
The single-field density model given by \cref{eq:rhoindicator} is adopted in every publication reviewed in \cref{sec:intro}, an in the rest of the scientific literature on two-phase flow simulations.
The mass conservation principle together with the single-field density model (\cref{eq:rhoindicator}) give
\begin{equation}
    \frac{d}{dt}\int_{\Omega_c} \rho \, dV = 
    -\int_{\partial \Omega_c} \rho \v \cdot \n dS = 
    -\int_{\partial \Omega_c} 
       [\rho^- \chi + \rho^+(1 - \chi)] \v \cdot \n \, dS. 
    \label{eq:rhoanalyticflux}
\end{equation}
The equality of surface integrals in \cref{eq:rhoanalyticflux},
\begin{linenomathNonumbers}
\begin{equation*}
    \int_{\partial \Omega_c} \rho \v \cdot \n dS = 
        \int_{\partial \Omega_c} [\rho^- \chi + \rho^+(1 - \chi)] \v \cdot \n \, dS,
\end{equation*}
\end{linenomathNonumbers}
demonstrates that the mass flux of the single-field density over $\partial \Omega_c$ is determined by the constant densities $\rho^\mp$ and the phase indicator given by \cref{eq:indicator}, if \cref{eq:rhoindicator} is used to model the single-field density. 
In other words, the single-field density at $\partial \Omega_c$ should be computed using the phase indicator as done on the r.h.s. of \cref{eq:rhoanalyticflux}, otherwise the mass conservation of the single-field density model given by \cref{eq:rhoindicator} will not be upheld.
This relevant condition transfers to the discrete level, leading to an interesting consequence for the computation of the face-centered (mass flux) density, that has so far been computed in many ways throughout the literature.
 
Specifically, when the surface integrals in \cref{eq:rhoanalyticflux} are discretized using the unstructured collocated finite volume method, 
\begin{equation}
    \begin{aligned}
        \sum_{f \in F_c} \rho_f F_f 
        & = 
        \sum_{f \in F_c} 
        \left[
            \rho^- \left(\int_{S_f} \chi \, dS\right) \v_f \cdot \Sfhat
            + \rho^+ \left(\int_{S_f} dS\right) \v_f \cdot \Sfhat - 
            \rho^+ \left(\int_{S_f} \chi \, dS\right)\v_f \cdot \Sfhat
        \right] \\ 
        & = 
        \sum_{f \in F_c} 
        \left[
            \rho^- \frac{\|\Sf\|}{\|\Sf\|} \left(\int_{S_f} \chi \, dS\right) \v_f \cdot \Sfhat
            + \rho^+ \left(\int_{S_f} dS\right) \v_f \cdot \Sfhat \right. \\
            & \left. - \rho^+ \frac{\|\Sf\|}{\|\Sf\|} \left(\int_{S_f} \chi \, dS\right)\v_f \cdot \Sfhat
        \right] \\
        & = 
        \sum_{f \in F_c} 
            \left[
                \rho^- \alpha_f 
                + \rho^+ (1 - \alpha_f)
            \right] F_f, 
    \end{aligned}
    \label{eq:rhodiscreteflux}
\end{equation}
where 
\begin{equation}
    \alpha_f := \frac{1}{|S_f|}\int_{S_f} \chi \, dS 
    \equiv \dfrac{|\Omega^-(t) \cap S_f|}{|S_f|}
\end{equation} 
is the \emph{area fraction} of the face $S_f \subset \partial \Omega_c$, i.e.\ the ratio of the area of $S_f$ submerged in $\Omega^-(t)$, and the total face-area $|S_f|$. 
Further, $\|\Sf\| \equiv |S_f|$, and $F_f$ is the volumetric flux in \cref{eq:rhodiscreteflux}.

An important consequence of \cref{eq:rhodiscreteflux} is the requirement for the evaluation of the face-centered (mass flux) density, necessary for ensuring the numerical consistency of the single-field two-phase momentum convection. 
\Cref{eq:rhodiscreteflux} requires all methods\footnote{All two-phase flow simulation methods encountered by the authors use \cref{eq:mixed_rho_eqn}.} that define $\rho$ using \cref{eq:mixed_rho_eqn} to either compute $\rho_f$ using the area fractions or $\int_{S_f}\chi \, dS$ from \cref{eq:rhodiscreteflux}, or to achieve this equivalently when computing $\rho_c^{n+1}$ from the advected volume fractions $\alpha_c^{n+1}$, which is possible for the flux-based VOF methods \cite{maric2020unstructured}.

Another important realization is that \cref{eq:rhodiscreteflux} is valid at any time $t$ - which is very relevant for the semi-implicit discretization developed within the \rhoLENT{} method, that applies \cref{eq:rhodiscreteflux} at $t^{n+1}$.  

\textcolor{Reviewer24}{Any simulation method that relies on the collocated unstructured FV discretization of single-field two-phase Navier-Stokes equations, but does not advect the phase indicator by solving an advection equation using phase-specific volumetric fluxes, does not provide the phase-specific volumetric fluxes for the approximation of the mass fluxes needed to ensure the consistency of the two-phase momentum transport. This, however, does not infer that \cref{eq:rhodiscreteflux} cannot be applied.}
The idea of using an auxiliary mass conservation equation introduced by \cite{ghods2013}, made into a formal requirement by \cref{eq:momentum-advection-v-update,eq:velocity-no-change}, allows the use of \cref{eq:rhodiscreteflux}: $\alpha_f$ can be computed regardless of the approximation of the fluid interface $\Sigma(t)$ and the method used to advect it. 

\begin{figure}[!htb]
    \centering
    \subfloat[Interface $\Sigma$ at $t^n$ and $t^{n+1}$ and the respective $\Omega^-(t^n)$ and $\Omega^-(t^{n+1})$ in gray color, used to compute $\alpha_c^{n}$ and $\alpha_c^{n+1}$, that are further used to compute $\rho_c^{n}$ and $\rho_c^{n+1}$ in an inconsistent way.]
    {
      \def\svgwidth{0.8\textwidth}
      \graphicspath{{figures/}}
      %% Creator: Inkscape inkscape 0.92.4, www.inkscape.org
%% PDF/EPS/PS + LaTeX output extension by Johan Engelen, 2010
%% Accompanies image file 'rho_update_FT.pdf' (pdf, eps, ps)
%%
%% To include the image in your LaTeX document, write
%%   \input{<filename>.pdf_tex}
%%  instead of
%%   \includegraphics{<filename>.pdf}
%% To scale the image, write
%%   \def\svgwidth{<desired width>}
%%   \input{<filename>.pdf_tex}
%%  instead of
%%   \includegraphics[width=<desired width>]{<filename>.pdf}
%%
%% Images with a different path to the parent latex file can
%% be accessed with the `import' package (which may need to be
%% installed) using
%%   \usepackage{import}
%% in the preamble, and then including the image with
%%   \import{<path to file>}{<filename>.pdf_tex}
%% Alternatively, one can specify
%%   \graphicspath{{<path to file>/}}
%% 
%% For more information, please see info/svg-inkscape on CTAN:
%%   http://tug.ctan.org/tex-archive/info/svg-inkscape
%%
\begingroup%
  \makeatletter%
  \providecommand\color[2][]{%
    \errmessage{(Inkscape) Color is used for the text in Inkscape, but the package 'color.sty' is not loaded}%
    \renewcommand\color[2][]{}%
  }%
  \providecommand\transparent[1]{%
    \errmessage{(Inkscape) Transparency is used (non-zero) for the text in Inkscape, but the package 'transparent.sty' is not loaded}%
    \renewcommand\transparent[1]{}%
  }%
  \providecommand\rotatebox[2]{#2}%
  \newcommand*\fsize{\dimexpr\f@size pt\relax}%
  \newcommand*\lineheight[1]{\fontsize{\fsize}{#1\fsize}\selectfont}%
  \ifx\svgwidth\undefined%
    \setlength{\unitlength}{557.44982394bp}%
    \ifx\svgscale\undefined%
      \relax%
    \else%
      \setlength{\unitlength}{\unitlength * \real{\svgscale}}%
    \fi%
  \else%
    \setlength{\unitlength}{\svgwidth}%
  \fi%
  \global\let\svgwidth\undefined%
  \global\let\svgscale\undefined%
  \makeatother%
  \begin{picture}(1,0.42559762)%
    \lineheight{1}%
    \setlength\tabcolsep{0pt}%
    \put(0,0){\includegraphics[width=\unitlength,page=1]{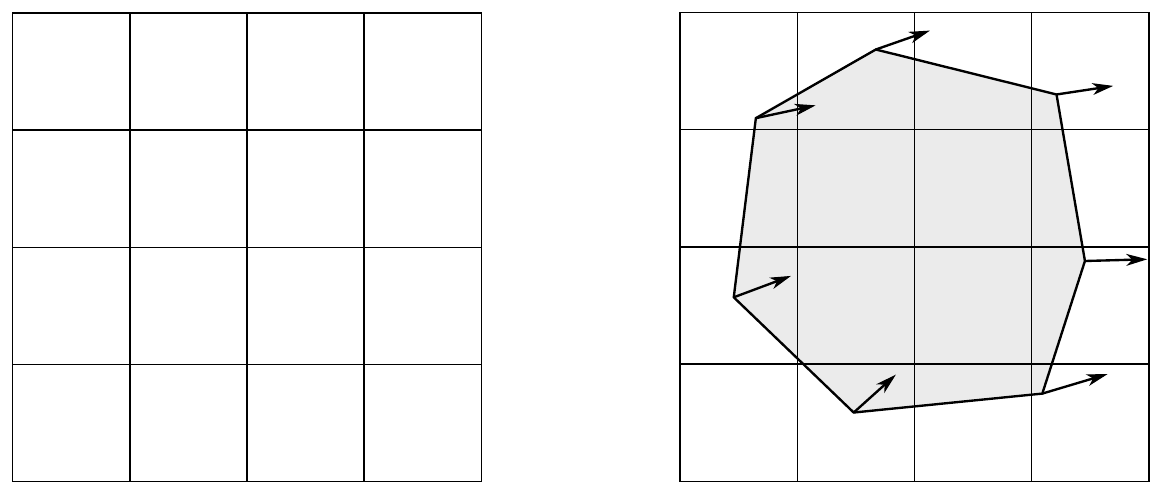}}%
    \put(0.23942012,0.02943337){\color[rgb]{0,0,0}\makebox(0,0)[lt]{\lineheight{1.25}\smash{\begin{tabular}[t]{l}$\Sigma (t^n)$\end{tabular}}}}%
    \put(0.77013764,0.02671519){\color[rgb]{0,0,0}\makebox(0,0)[lt]{\lineheight{1.25}\smash{\begin{tabular}[t]{l}$\Sigma (t^{n+1})$\end{tabular}}}}%
    \put(0,0){\includegraphics[width=\unitlength,page=2]{rho_update_FT.pdf}}%
  \end{picture}%
\endgroup%
  
      \label{subfig:rho_update_FT}
    } \\
    \subfloat[Interface at $\Sigma(t^{n+1})$ used to compute $\alpha_f^{n+1}$, then $\rho_f^{n+1}$ and finally $\rho_c^{n+1}$ in a consistent way, by solving a mass conservation equation.]
    {
      \def\svgwidth{0.35\textwidth}
      %% Creator: Inkscape inkscape 0.92.4, www.inkscape.org
%% PDF/EPS/PS + LaTeX output extension by Johan Engelen, 2010
%% Accompanies image file 'rho_update_Eqn.pdf' (pdf, eps, ps)
%%
%% To include the image in your LaTeX document, write
%%   \input{<filename>.pdf_tex}
%%  instead of
%%   \includegraphics{<filename>.pdf}
%% To scale the image, write
%%   \def\svgwidth{<desired width>}
%%   \input{<filename>.pdf_tex}
%%  instead of
%%   \includegraphics[width=<desired width>]{<filename>.pdf}
%%
%% Images with a different path to the parent latex file can
%% be accessed with the `import' package (which may need to be
%% installed) using
%%   \usepackage{import}
%% in the preamble, and then including the image with
%%   \import{<path to file>}{<filename>.pdf_tex}
%% Alternatively, one can specify
%%   \graphicspath{{<path to file>/}}
%% 
%% For more information, please see info/svg-inkscape on CTAN:
%%   http://tug.ctan.org/tex-archive/info/svg-inkscape
%%
\begingroup%
  \makeatletter%
  \providecommand\color[2][]{%
    \errmessage{(Inkscape) Color is used for the text in Inkscape, but the package 'color.sty' is not loaded}%
    \renewcommand\color[2][]{}%
  }%
  \providecommand\transparent[1]{%
    \errmessage{(Inkscape) Transparency is used (non-zero) for the text in Inkscape, but the package 'transparent.sty' is not loaded}%
    \renewcommand\transparent[1]{}%
  }%
  \providecommand\rotatebox[2]{#2}%
  \newcommand*\fsize{\dimexpr\f@size pt\relax}%
  \newcommand*\lineheight[1]{\fontsize{\fsize}{#1\fsize}\selectfont}%
  \ifx\svgwidth\undefined%
    \setlength{\unitlength}{237.0885876bp}%
    \ifx\svgscale\undefined%
      \relax%
    \else%
      \setlength{\unitlength}{\unitlength * \real{\svgscale}}%
    \fi%
  \else%
    \setlength{\unitlength}{\svgwidth}%
  \fi%
  \global\let\svgwidth\undefined%
  \global\let\svgscale\undefined%
  \makeatother%
  \begin{picture}(1,1.00000009)%
    \lineheight{1}%
    \setlength\tabcolsep{0pt}%
    \put(0,0){\includegraphics[width=\unitlength,page=1]{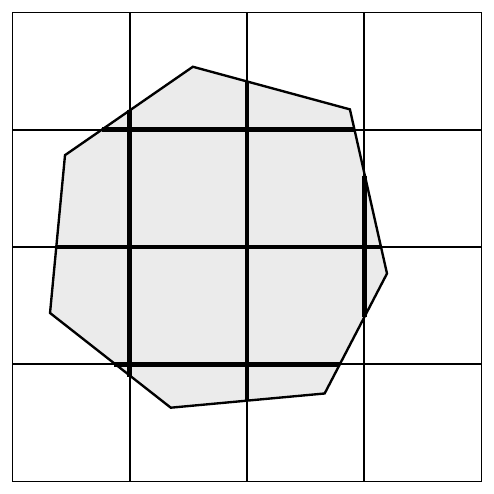}}%
    \put(0.34296113,0.60506934){\color[rgb]{0,0,0}\makebox(0,0)[lt]{\lineheight{1.25}\smash{\begin{tabular}[t]{l}$\rho_f F_f$\end{tabular}}}}%
    \put(0.56665702,0.88073484){\color[rgb]{0,0,0}\makebox(0,0)[lt]{\lineheight{1.25}\smash{\begin{tabular}[t]{l}$\alpha_f$\end{tabular}}}}%
    \put(0,0){\includegraphics[width=\unitlength,page=2]{rho_update_Eqn.pdf}}%
  \end{picture}%
\endgroup%
  
      \label{subfig:rho_update_Eqn}
    } 
   \caption{Updating the face-centered (mass flux) density in the \rhoLENT{} method.}
    \label{fig:rho_update}
\end{figure}

Similar to other contemporary methods, the \rhoLENT{} method also first advects the interface using the velocity from the previous time step as shown in the left image of \cref{subfig:rho_update_FT}, resulting in the new position of the interface shown in the right image in \cref{subfig:rho_update_FT}, that is then used to geometrically calculate the face-centered density $\rho_f^{n+1}$, by calculating area fractions $\alpha_f^{n+1}$ from the interface approximation, as shown in \cref{subfig:rho_update_Eqn}. 
The face-centered density $\rho_f^{n+1}$ and the volumetric flux $F_f^o$ are then used to update the cell-centered density $\rho_c^{n+1}$ by solving a mass conservation equation.
The index $o$ in the volumetric flux refers to the linearization of the convective term in the momentum equation. 
The same mass flux $\rho_{f}^{n+1}F_{f}^{o}$ is used in the implicitly discretized momentum conservation equation. The pressure-velocity coupling algorithm iterates the linearized volumetric flux $F_f^o$ to $F_f^{n+1}$. Finally, the cell-centered velocity $\mathbf{v}_c^{n+1}$ is obtained, which is used to evolve the fluid interface in the next time step, from $t^{n+1}$ to $t^{n+2}$. 
At this point, the numerically consistent cell-centered density $\rho_c^{n+1}$ has served its purpose and is reset \textcolor{Reviewer24}{according to} \cref{eq:mixed_rho_eqn}, \textcolor{Reviewer24}{using $\alpha_c$ approximated from signed distances \citep{tolle2020saample}}, to make it consistent again with the fluid interface approximation.

Any two-phase flow simulation method has the possibility to compute the face-centered density $\rho_f(t)$ from the interface approximation in some way.
The \rhoLENT{} method computes the face-centered density $\rho_f(t)$ by computing the face area fraction $\alpha_f(t)$ (short: area fraction) of the face $S_f$, submerged in the phase $\Omega^-(t)$. 
The calculation of $\alpha_f$ uses signed distances available in the unstructured LENT \citep{maric2015lentfoam,tolle2020saample}. 
%This is contrary to other contemporary methods \cite{} that interpolate cell-centered densities $\rho_c$ to the face center, which causes instabilities when the density field changes abruptly.
\textcolor{Reviewer24}{Any two-phase flow simulation method that utilises the collocated FV method for discretizing single-field Navier-Stokes equations can be adapted as described above to} geometrically approximate the area fraction $\alpha_f$ \textcolor{Reviewer24}{thus avoiding erroneous} interpolation of \textcolor{Reviewer24}{fields} that abruptly change in the interface-normal direction.
 
In the original Front-Tracking method, the density is updated utilizing the new position of marker points (the approximated interface) \cite{muradoglu2008front}. 
After the velocity field in the current step is computed, the position of marker points in the new time step can be updated immediately by 
\begin{equation}
    \x_p^{n+1} = \x_p^{n} + \Delta t \, \v_p^n,
    \label{eq:markerpointupdate}
\end{equation}
where $\x_p$,$\v_p$ indicate the position and interpolated velocity of marker points respectively, and $\Delta t$ is the time step length. \textcolor{Reviewer1}{The advection of marker points along Lagrangian trajectories eventually corrupts the triangular mesh, leading to discrepancies in the ratios of triangular angles and areas and self-intersections of the triangular mesh. The original Front Tracking method \cite{Tryggvason2001} deals with this by redistributing marker points based on quality criteria imposed on the triangular mesh, which involves manipulating the connectivity of the triangular mesh.}

\textcolor{Reviewer1}{Contrary to original Front Tracking \cite{Tryggvason2001}, the LENT method reuses the principles from LCRM / LFRM methods \cite{shin2002modeling,shin2005accurate,Shin2011,shin2017solver} and reconstructs the interface using an iso-surface reconstruction algorithm. The iso-surface reconstruction does not add/delete marker points locally by changing the connectivity of the triangular surface mesh; it reconstructs the entire interface in the solution domain as an iso-surface. Following the strategy from LCRM / LFRM, the physics of the problem determines the iso-surface reconstruction frequency. The LENT method uses the marching tetrahedra \cite{Treece1999} algorithm to enable the iso-surface reconstruction on unstructured meshes.}
%However, the marching tetrahedra algorithm introduces many triangles per cell (even with regularization), causing instabilities in front tracking. We are developing an alternative iso-surface reconstruction that relies on a higher-order signed-distance interpolation and results in a favorable ratio of triangle-to-cell length scales.}

\begin{figure}[!htb]
    \centering
    \def\svgwidth{0.7\textwidth}
    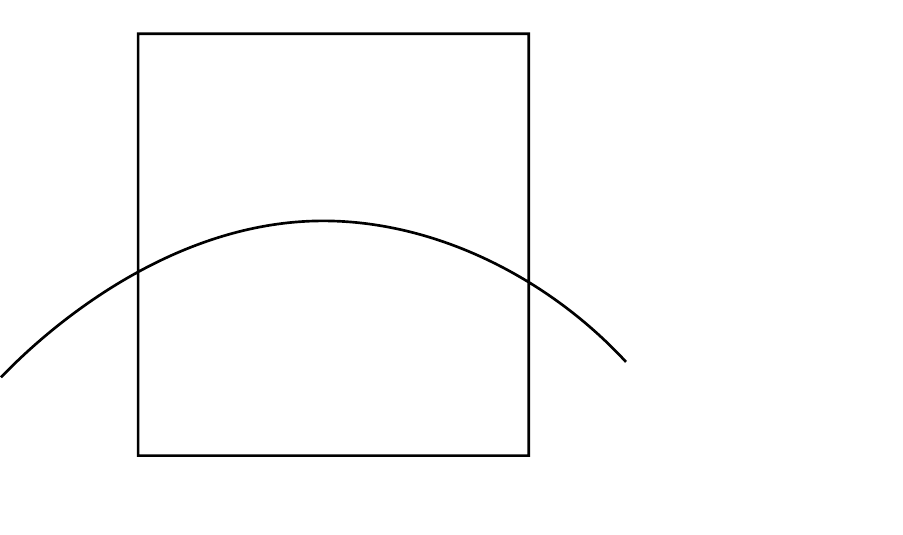
    \caption{Computing area fractions from signed distances in the method.}
    \label{fig:alpha_f}
\end{figure}

\textcolor{Reviewer1}{Once the marker points are advected and redistributed}, the cell density is updated depending on $\x_p^{n+1}$, namely
\begin{equation}
    \rho^{n+1} = \rho (\x_p^{n+1}).
\end{equation}
The face-centered density used for the mass flux is then interpolated \textcolor{Reviewer24}{by the LENT method} from densities of two adjacent cells. 
Contrary to LENT, the face-centered density is updated by \rhoLENT{} using the phase indicator approximated at each cell-face \textcolor{Reviewer24}{ by an area fraction}. 
A 2D interface is depicted in \cref{subfig:rho_update_Eqn}, where $\alpha_f^{n+1}$ is the area fraction at $t^{n+1}$: the ratio of the cell-face area submerged in the phase $\tilde{\Omega}^-(t^{n+1}) \approx \Omega^-(t^{n+1})$, and the total face area $|S_f|$. 
More precisely, the area fraction $\alpha_f^{n+1}$ is computed by the \rhoLENT{} method using a second-order accurate approximation from signed distances \citep{detrixhe2016level}, used in \cite{tolle2020saample} to approximate the volume fraction $\alpha_c$ (see \cref{eq:volfracdef}).
The Level Set component of the LENT method \cite{maric2015lentfoam} calculates signed distances from the triangular surface mesh that approximates the interface $\tilde{\Sigma}(\tnn)\approx\Sigma(\tnn):=\partial \tilde{\Omega}^-(\tnn)$. 
With the narrow band approach from \cite{maric2015lentfoam}, the signed distances can be computed efficiently at any point in a close vicinity of $\tilde{\Sigma}(t)$.  
The original LENT method \cite{maric2015lentfoam} computes signed distances at cell-centers and cell corner-points, and the proposed \rhoLENT{} additionally computes signed distances at face centers.
Each face $S_f$ is triangulated using its centroid $\x_f$, as shown in \cref{fig:alpha_f}. 
The face centroid $\x_f$, together with the two successive cell-corner points that belong to the face $S_f$, $\x_{f,i}$, $\x_{f,i+1}$, forms a triangle $(\x_f, \x_{f,i}, \x_{f,i+1})$. 
Face-triangles may be partially submerged in the phase $\tilde{\Omega}^-(t^{n+1})$, in which case the submerged area of the triangle is computed using the nearest signed distances to $\tilde{\Sigma}(t^{n+1})$ from the triangle points $(\x_f, \x_{f,i}, \x_{f,i+1})$, namely $(\psi_f, \psi_{f,i}, \psi_{f,i+1})$, as shown in \cref{fig:alpha_f}. 
The second-order approximation developed in \citep{detrixhe2016level} is used here for computing the area fraction of a triangle submerged in $\tilde{\Omega}^-(t^{n+1})$. 
Any other second-order method can be applied. 
For example, a linear interpolation of signed distances along the edges of the triangle may be used equivalently, or a geometrical intersection between $\tilde{\Omega}^-(t^{n+1})$ and the triangle. 
The total submerged area of the face $S_f$ is then the sum of the submerged areas of face-triangles 
\textcolor{Reviewer24}{\begin{equation}
    A_f^{n+1} :=|\Omega^-(t^{n+1}) \cap S_f| = \sum_{k \in T_f} |\Omega^-(t^{n+1}) \cap T_k|,
\end{equation}}
where $T_f$ is the set of indexes of the triangles in the triangulation of the face $S_f$. 
As mentioned above, \textcolor{Reviewer24}{any other two-phase flow simulation method that discretizes single-field Navier-Stokes equations but does not utilise phase-specific fluxes} can be adapted to compute \textcolor{Reviewer24}{$|\Omega^-(t^{n+1}) \cap T_k|$}.

The area fraction $\alpha_f^{n+1}$ is then computed as 
\begin{equation}
    \alpha_f^{n+1} := \dfrac{|\tilde{\Omega}^-(t^{n+1}) \cap S_f |}{|S_f|} = \dfrac{A_f}{|S_f|}, 
\end{equation}
as shown in \cref{fig:alpha_f}. Once the area fraction $\alpha_f^{n+1}$ is approximated, it is used to compute the face-centered densities required by \cref{eq:rhodiscreteflux}, namely

\begin{equation}
    \rho^{n+1}_f = \alpha_f^{n+1}\rho^-+(1-\alpha_f^{n+1})\rho^+, 
    \label{eq:face-density-update}
\end{equation}
at the new time step, because the interface has been advected forward in time to $\tnn$ with the available velocity $\v^n$. The discretized continuity equation (\cref{eq:density-update-conti}) then attains the form 
\begin{equation}
\textcolor{Reviewer24}{
    \rho^{o+1}_c = \rho^n_c + \frac{\Delta t}{|V_{\Cell{c}}|}\sum_f \rho^{n+1}_f F^o_f, \quad F^o_f = \v_f^o \cdot \Sf.}
    \label{eq:rho_update_1}
\end{equation}
It is important to note that, although $\rho_f^{n+1}$ appears in \cref{eq:rho_update_1}, \rhoLENT{} does not use an implicit discretization for \cref{eq:rho_update_1}: $\rho_f^{n+1}$ is geometrically computed from the fluid interface approximation $\tilde{\Sigma}^{n+1}$, so \cref{eq:rho_update_1} is solved exactly.  
\textcolor{Reviewer24}{The exact (iterative) evaluation of cell-center density at loop $o+1$, i.e. $\rho_c^{o+1}$ from \cref{eq:rho_update_1}, alongside \cref{eq:momentum-advection-v-update}, further infers the possibility of \emph{exact numerical consistency} for the discretized convective term in the single-field momentum equation, which is in fact achieved and supported by the results.} 

In addition to density, the viscosity is updated utilizing the area fraction $\alpha_f$. Note that there is no need to calculate the cell-centered viscosity for the unstructured FVM discretization, only the face-centered viscosity is updated as follows
\begin{equation}
    \mu_f^{n+1} = \alpha_f^{n+1}\rho^-\nu^-+(1-\alpha_f^{n+1})\rho^+\nu^+.
    \label{eq:face-viscosity-update}
\end{equation}

The non-linearity of the convective term in the momentum equation \cref{eq:momentum-transport}, namely $\rho \v \v$, is usually linearized when solving the single-field Navier-Stokes equations using the unstructured Finite Volume method. 
The convective term is discretized as 
\begin{equation}
    \int_{\Cell{c}}\nabla \cdot (\rho \v \v) dV \approx \sum_{f \in F_c} \rho^{n+1}_f F^o_f \v^{n+1}_f.
    \label{eq:rholentflux}
\end{equation}
\textcolor{Reviewer24}{Numerical consistency imposed by \cref{eq:consistency} does not depend on the implicit or explicit discretization: the proportionality between the mass flux and the phase-specific flux, and the equivalence of the mass flux in the mass conservation equation and the momentum transport equation must both hold at any time, and in any iteration of the solution algorithm.}
Therefore, the requirement given by \cref{eq:momentum-advection-v-update,eq:velocity-no-change}, is valid for an implicit discretization as well.

The volumetric flux $F^o_f$ is initialized to $F^n_f$ and iterated within the  SAAMPLE \cite{tolle2020saample} pressure-velocity coupling algorithm loop until $o = n+1$ is reached. \textcolor{Reviewer1}{The \rhoLENT{} algorithm is outlined in \cref{alg:rholent} and it extends the SAAMPLE algorithm \cite{tolle2020saample}. It is relevant to note that $F_f^o$ is iterated} from $F_f^n$ to $F_f^{n+1}$ and $p^o$ \textcolor{Reviewer1}{is solved for} from $p^n$ to $p^{n+1}$ such that the discrete incompressibility condition $\sum_{f \in F_c} F_f^{n+1}$ is ensured. \textcolor{Reviewer24}{}

\begin{center}
\begin{algorithm}[H]
    \centering
    \caption{The \rhoLENT{} solution algorithm.}
    \label{alg:rholent}
    {\small
    \centering
    \begin{algorithmic}[1]
        \While{simulation time $\le$ end time}
            \State Advect the interface to $\tilde{\Sigma}^{n+1}$. \Comment \cite{maric2015lentfoam}
            \State Compute the signed-distance field $\psi^{n+1}$ from $\tilde{\Sigma}^{n+1}$ at $\x_c, \x_f, \x_p$ in the narrow-band. \Comment \cite{maric2015lentfoam}
            \State Compute $\alpha_c^{n+1}$ from $\psi_c^{n+1},\psi_p^{n+1}$. \Comment \cite{tolle2020saample}
            \State Compute the area fraction $\alpha_f^{n+1}$ from the signed distance fields $\psi_f^{n+1}, \psi_p^{n+1}$. \Comment \Cref{fig:alpha_f}
            \State Compute the face-centered densities $\rho_f^{n+1}$ using $\alpha_f^{n+1}$. \Comment \Cref{eq:face-density-update}
            \textcolor{Reviewer24}{\While{$F_f^o$ does not converge or $o < o_{max}$}
                \State Solve the continuity equation using $\rho_f^{n+1}F_f^o$ for cell-centered densities $\rho_c^{o+1}$. \Comment \Cref{eq:rho_update_1}
                \While{$r > \text{tol}_{ls}$ and $i < i_{max}$}
                    \State Use $\rho_c^{o+1}$ and $\rho_f^{n+1}F_f^{o}$ in $p-\mathbf{v}$ coupling to compute $\v_c^{i+1}, F_f^{i+1}$. \Comment \cite{tolle2020saample} and \cref{eq:rholentflux}
                \EndWhile
            \EndWhile  }
            \State Make $\rho_c^{n+1}$ consistent with $\tilde{\Sigma}^{n+1}$, i.e.\ $\rho_c^{n+1}=\alpha_c^{n+1}\rho^-+(1-\alpha_c^{n+1})\rho^+$.
            \State Make $\mu_c^{n+1}$ consistent with $\tilde{\Sigma}^{n+1}$, i.e. $\mu_c^{n+1}:=\alpha_c^{n+1}\rho^-\nu^-+(1-\alpha_c^{n+1})\rho^+\nu^+$.
        \EndWhile
    \end{algorithmic}
    }
\end{algorithm}
\end{center}

\textcolor{Reviewer1}{The $p-\mathbf{v}$ coupling - mentioned in the step $8$ in \cref{alg:rholent} - requires some further explanation. The semi-implicit discretization (with the convective term linearized as an explicit mass-flux and implicit velocity) of the single-field momentum equation using the implicit collocated unstructured finite volume method \cite{maric2014openfoam,moukalled2016finite}, results in} 
\textcolor{Reviewer1}{\begin{equation}
    a_c \mathbf{v}_c^{n+1} + \sum_{k \in N_c} a_k \mathbf{v}_k^{n+1} = -(\nabla p)_c^{n+1} - [(\nabla \rho)^{n+1}\cdot (\mathbf{g}\cdot \mathbf{x})]_c + (\mathbf{f}_\Sigma)_c^{n+1},
    \label{eq:momentumdiscrete}
\end{equation}}\noindent\textcolor{Reviewer1}{where $N_c$ is the index-set of cells that are face-adjacent to cell $\Omega_c$, and the total pressure is expressed using the dynamic and the hydrostatic pressure. The diagonal coefficient $a_{c}$ corresponds to the cell $\Omega_c$, and $k$ denotes the coefficients contributed from cells that are face-adjacent to $\Omega_c$. \textcolor{Reviewer24}{We discretize the surface tension force $(\mathbf{f}_\Sigma)_c^{n+1})$ using the semi-implicit model from \citep{tolle2020saample}.}}

% An interesting aspect of the semi-implicit discretization in \cref{eq:momentumdiscrete}, is a potentially more stable discretization of the surface tension force term $(\mathbf{f}_\sigma)_c$ at the new time step $t^{n+1}$. Namely, the Front Tracking \cite{Tryggvason2001} in the LENT method \cite{maric2015lentfoam} (also in LCRM \cite{shin2002modeling}, and LCRM \cite{Shin2011}) discretizes \cref{eq:markerpointupdate} in a Lagrangian reference frame. Once the new interface is obtained as $\{\mathbf{x}_p^{n+1}\}_{p \in P}$ in a Lagrangian reference frame, the new interface is mapped onto an Eulerian signed-distance field. Since the Lagrangian advection is unconditionally stable, the mapping is geometrical and therefore bounded.
% and $\mathbf{f}_\sigma^{n+1}:=\mathbf{f}_\sigma^{n+1}(\{\mathbf{x}_p^{n+1}\}_{p \in P})$, $\mathbf{f}_\sigma^{n+1}$ does not introduce a CFL condition, contrary to other Eulerian two-phase flow methods. This is a potential benefit that will be investigated in the future, as it may allow for larger time steps, than those limited by hydrodynamic stability condition \cite{Popinet2018} or CFL. 
% Here, the Euler explicit method has been used in \cref{eq:markerpointupdate} for simplicity; other explicit higher-order temporal discretizations can be used instead of \cref{eq:markerpointupdate}.} 

\textcolor{Reviewer1}{Equation \cref{eq:momentumdiscrete} is also discretized semi-implicitly, because of the linearized convective term, that contributes the volumetric flux $F_f$ to the $a_{c,k}$ coefficients in \cref{eq:momentumdiscrete}. Linearizing the convective term introduces a need for iteration. Iterations are also introduced by splitting \cref{eq:momentumdiscrete} into two equations: one for $\mathbf{v}_c^{n+1}$, and another for $\mathbf{p}_c^{n+1}$. Dividing the equation \cref{eq:momentumdiscrete} with $a_c$ and applying the discrete divergence $\nabla_c \cdot$, results in the pressure equation}   
\textcolor{Reviewer24}{\begin{equation}
    \begin{aligned}
        \sum_{f \in F_c} \left(\frac{1}{a_c}\right)^o(\nabla p)_f^{i+1} \cdot \mathbf{S}_f 
        = & 
   %  \sum_{f \in F_c} \left(\frac{1}{a_c}\right)^l [\mathbf{H}(\mathbf{v}^o)]_f \cdot \mathbf{S}_f + 
   %      \sum_{f\in F_c} \left(\frac{1}{a_c}\right)^l [(\nabla \rho)^{n+1} \cdot (\mathbf{g}\cdot \mathbf{h})]_f \cdot \mathbf{S}_f + \\
   %      & \sum_{f\in F_c} \left(\frac{1}{a_c}\right)^l (\mathbf{f}_\Sigma)_f^{n+1} \cdot \mathbf{S}_f, \\
   % = & 
    \sum_{f \in F_c} \left(\frac{1}{a_c}\right)^o [\mathbf{H}(\mathbf{v}^i)]_f \cdot \mathbf{S}_f + 
        \sum_{f\in F_c} \left(\frac{1}{a_c}\right)^o [(\nabla \rho)^{i} \cdot (\mathbf{g}\cdot \mathbf{x})]_f \cdot \mathbf{S}_f + \\
        & \sum_{f\in F_c} \left(\frac{1}{a_c}\right)^o \sigma \kappa_f^{n+1} (\nabla \alpha)_f^{i} \cdot \mathbf{S}_f,    
    \end{aligned}
    \label{eq:pressurediscrete}
\end{equation}}\textcolor{Reviewer24}{where we use the CSF model \citep{brackbill1992continuum} to model the surface tension force $(\mathbf{f}_\Sigma)_f \approx \sigma \kappa_f (\nabla \alpha)_f$. The discrete divergence-free condition imposed on $\mathbf{v}_c^{n+1}$ in \cref{eq:momentumdiscrete} results in the divergence-free volumetric flux} 
\textcolor{Reviewer1}{\begin{equation}
    \sum_{f \in F_c} F_f^{o} = 0,
\end{equation}}\textcolor{Reviewer1}{used as the control variable for the convergence of outer iterations $o$ by the SAAMPLE algorithm \cite{tolle2020saample}. The outer iterations $o$ are used for linearizing the volumetric flux as described above and contribute the volumetric flux to the coefficients $a_{c,k}$, from \cref{eq:momentumdiscrete}, while $\mathbf{H}(\mathbf{v})$ in \cref{eq:pressurediscrete} is the contribution of convection and diffusion operators from face-adjacent cells in \cref{eq:momentumdiscrete}. Note that $(\nabla \rho)_{c,f}^{n+1}$ and \textcolor{Reviewer24}{the implicit part of} $(\mathbf{f}_\sigma^{n+1})_{f,c}$ are known at $t^{n+1}$ from $\mathbf{f}_\sigma^{n+1}:=\mathbf{f}_\sigma^{n+1}(\{\mathbf{x}_p^{n+1}\}_{p \in P})$, and \cref{eq:rho_update_1}.} 

\textcolor{Reviewer1}{This segregated solution for $(p_c^{n+1}, \mathbf{v}_c^{n+1})$ is standard in the context of collocated unstructured finite volume method \cite{moukalled2016finite}: the inner iterations and the assembly of the pressure equation originates from the PISO algorithm \cite{issa1986solution}, the outer iterations originate from the SIMPLE algorithm \cite{patankar1972}, and the tolerance-based control of outer iterations is described in detail in \cite{tolle2020saample}. In addition, the implementations of the LENT method \cite{maric2015lentfoam}, the SAAMPLE algorithm \cite{tolle2020saample} and the \rhoLENT{} method are publicly available \textcolor{Reviewer24}{\citep{LENTcode}}. This description, the details on the tolerance-based outer iteration control in \cite{tolle2020saample}, and the publicly available implementation in OpenFOAM, provide sufficient information for an interested reader willing to understand or further extend the methodology.}

\begin{figure}[!htb]
    \footnotesize
    \hspace{5em}
    \begin{tikzpicture}[%
        >=triangle 60,              % Nice arrows; your taste may be different
        start chain=going below,    % General flow is top-to-bottom
        node distance=3.8mm and 50mm, % Global setup of box spacing
        every join/.style={norm}    % Default linetype for connecting boxes
        ]
        \tikzset{
          base/.style={draw, on chain, on grid, align=center},
          rounded/.style={base, rounded corners, text width=16em}, 
          test/.style={base, diamond, aspect=1}, 
          arrow/.style={->, draw}
        }
        \node [rounded] (p0) {Calc. search distances \citep{maric2015lentfoam}};
        \node [test]    (p1) {};
        \node [rounded] (p2) {$t^{n+1} = t^n + \Delta t$};
        \node [rounded, dashed, thick] (p3) {Reconstruct then evolve front \citep{maric2015lentfoam}};
        \node [rounded] (p4) {Calculate signed distances \citep{maric2015lentfoam}};
        \node [rounded] (p5) {Calculate phase indicator\\$\alpha_{c}^{n+1}$ \citep{tolle2020saample}};
        \node [rounded, dashed, thick] (p6) {Update cell face density $\rho_f^{n+1}$ (\cref{eq:face-density-update})};
        \node [test]    (p7) {};
        \node [rounded] (p8) {Calc. mass flux $m_f^o = \rho_f^{n+1} F_f^o$ using $\rho_f^{n+1}$ from \cref{eq:face-density-update}};
        \node [rounded, dashed,thick] (p9) {Update mixture density $\rho_c^{o+1}$ from \cref{eq:rho_update_1}};
        \node [rounded] (p10) {Calc. predicted velocity by solving momentum predictor equation}; %\\ $\mathbf{v}^I, \, I = 0$, from \textbf{ref}, discretized with $F^K$};
        \node [rounded] (p11) {Initialize pressure residual norm $r$};
        \node [test]    (p12) {};
        \node [rounded] (p13) {Solve pressure equation};%\\ $p^I$ from $\mathbf{v}^I$ using \textbf{ref}};
        \node [rounded] (p14) {Update residual norm $r$};
        \node [rounded] (p15) {Update flux to obtain $F_f^{i+1}$};%Flux update \\$F_f^{I+1} = \left(\frac{1}{a_p}\right)_f H(\mathbf{v}^I)_f - \textbf{D}^{\mathbf{v}}_f\nabla{p^I}_f$};
        \node [rounded] (p16) {Reconstruct velocity $\mathbf{v}_c^{i+1}$};%Velocity reconstruction\\$\mathbf{v}^{I+1}$ from \textbf{ref}};
        \node [rounded] (p17) {END};
        \node [rounded, text width=14em, dashed, thick] (p18) [right=of p4] { Update mixture properties $\mu_c^{n+1}, \rho_c^{n+1}$ (\cref{eq:mixed_mu_eqn,eq:mixed_rho_eqn})};

        % Straight paths
        \draw[->] (p0)  --  (p1);
        \draw[->] (p1)  -- node[anchor=west] {$t < t_{END}$} (p2);
        \draw[->] (p2)  --  (p3);
        \draw[->] (p3)  --  (p4);
        \draw[->] (p4)  --  (p5);
        \draw[->] (p5)  --  (p6);
        \draw[->] (p6)  --  (p7);
        \draw[->] (p7)  -- node[anchor=west] {$F_f^o$ does not converge or $o < o_{max}$} (p8);%{$conv(F_f^K) != 0$ (\textbf{ref}) or $K < K_{max}$} (p8);
        \draw[->] (p8)  --  (p9);
        \draw[->] (p9)  --  (p10);
        \draw[->] (p10)  -- (p11);
        \draw[->] (p11)  -- (p12);
        \draw[->] (p12) -- node[anchor=west] {$r > \text{tol}_{ls}$ and $i < i_{max}$} (p13);
        \draw[->] (p13) --  (p14);
        \draw[->] (p14) --  (p15);
        \draw[->] (p15) --  (p16);

        % Broken paths
        \draw[->] (p16.east) -| ++(8mm,0) |- (p12.east);
        \draw[->] (p12) -| ++(-40mm,0) |- (p7);
        \draw[->] (p7) -| (p18.south);
        \draw[->] (p18.north) |- (p1);
        \draw[->] (p1) -| ++(-55mm,0) |- (p17.west);

    \end{tikzpicture}

    \caption{\textcolor{Reviewer24}{Flowchart of the \rhoLENT{} method. The dashed blocks denote the new and modified elements of the SAAMPLE method \citep{tolle2020saample}. The indices $o_{max}$ and $i_{max}$ in the flowchart indicate the maximal iteration numbers for the outer and inner loop, respectively, while $tol_{ls}$ denotes the prescribed linear solver tolerance.}}
    \label{fig:flowchart}
\end{figure}

\textcolor{Reviewer3}{\subsection{Volume correction method}
\begin{figure}[H]
    \centering
    \def\svgwidth{0.6\textwidth}
    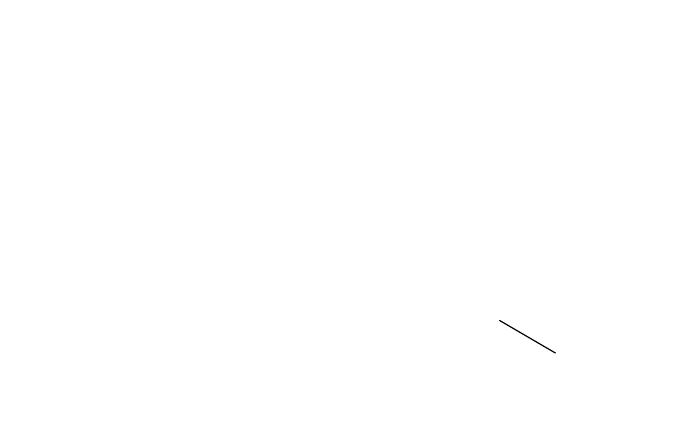\notag
    \caption{\textcolor{Reviewer3}{The volume correction method: iso-value compensates the volume-change.}}
    \label{fig:volume-correction}
\end{figure}
Level Set / Front Tracking methods are Lagrangian / Eulerian methods that kinematically evolve the fluid interface without utilising fluxes through control-volume boundaries and are therefore inherently not mass(volume)-conservative. Rising bubble and oscillating droplet simulations are presented in the results section to demonstrate the benefits of the \rhoLENT{} method for stronger momentum interaction between fluid phases that have strongly different densities. Mass conservation is crucial for accurately \textcolor{Reviewer24}{simulating} rising bubbles \textcolor{Reviewer25}{ (\citet{Singh_2007,Hua_2007,Hua_2008,Pivello_2014})}. In particular, \citet{Hua_2007} conducted comparative analyses which revealed that mass conservation carries equivalent importance to mesh resolution and domain size in influencing the accuracy. 
 The Front reconstruction in the LENT method \citep{maric2015lentfoam,tolle2020saample} uses marching tetrahedrons with linear interpolation of the iso-surface root-points \citet{Treece1999}, that causes volume loss. To demonstrate the benefits of the proposed \rhoLENT{} method for handling high density ratios with stronger interface deformation and momentum exchange, we ensure volume conservation using by extending/contracting the Front with a modified iso-value.}
 
\textcolor{Reviewer3}{
Figure \cref{fig:volume-correction} depicts the volume correction at time step $n$, where $\tilde{\Sigma}_0$ denotes the Front at the time step $n$. The value $\psi_0=0$ is the iso-value used to reconstruct the $\tilde{\Sigma}_0$ at $t^n$. From $\tilde{\Sigma}_0$ that contains volume-conservation errors, we compute the corrected Front $\tilde{\Sigma}_1$, as $\tilde{\Sigma}_0$ extended in the normal direction by $h$. \textcolor{Reviewer24}{We consider volume loss, with no loss of generality in the case of volume gain.} For sufficiently small $h$, for any $\x \in \tilde{\Sigma}_0$, we define $\x':= \x + h\n_{\Sigma}(\x), \x' \in \tilde{\Sigma}_1$. The linear Taylor-series approximation
\begin{equation}
    \psi(\x') \doteq \psi(\x) + \nabla \psi(\textcolor{Reviewer24}{\x}) \cdot h \n_{\Sigma}(\x),
\end{equation}
with $\psi(\x) = 0$ by the definition of an iso-surface $\forall \x \in \tilde{\Sigma}_0$, results in 
\textcolor{Reviewer24}{\begin{equation}
     \psi(\x') \doteq h \nabla \psi(\x) \cdot  \n_{\Sigma}(\x).
\end{equation}}
Since Level Set / Front Tracking ensures $\nabla \psi(\x)=\n_\Sigma$ \textcolor{Reviewer24}{and thus $ \nabla \psi(\x) \cdot  \n_{\Sigma}(\x)=1$} by geometrically re-distancing $\psi$ from the reconstructed Front $\tilde{\Sigma}^n_0$,   
\begin{equation}
     \psi(\x') = h.
\end{equation}
The height $h$ is expressed from the change in volume between $\tilde{\Sigma}_0$ and $\tilde{\Sigma}_1$ 
\begin{equation}
\begin{aligned}
      \int_{\tilde{\Sigma}_0} h \, dS &= \textcolor{Reviewer25}{V_{\textit{target}} -V_{\textit{ini}}}, \\
       h &= \frac{\textcolor{Reviewer25}{V_{\textit{target}} - V_{\textit{ini}}}}{\int_{\tilde{\Sigma}_0} 1 \, dS},
       \label{eq:hcontinuous}
\end{aligned}
\end{equation}
Volume \textcolor{Reviewer25}{$V_{\textit{target}}$} is \textcolor{Reviewer25}{the target volume and} known before the reconstruction, and we aim to recover \textcolor{Reviewer25}{$V_{\textit{target}}$}'s  corresponding front $\tilde{\Sigma}_1$. The volume \textcolor{Reviewer25}{$V_{\textit{ini}}$} is computed from the \textcolor{Reviewer25}{initially} reconstructed $\tilde{\Sigma}_{0}$. If volume loss really occurred, then \textcolor{Reviewer25}{$V_{\textit{target}} > V_{\textit{ini}}$}, and $h>0$, so we extend $\tilde{\Sigma}_0$ in the direction of $\mathbf{n}_\Sigma$ by reconstrucing an iso-surface $\tilde{\Sigma}_1 = \{ \x' : \psi(\x')=h\}$. However, if volume gain occured, \textcolor{Reviewer25}{$V_{\textit{target}} < V_{\textit{ini}}$}, so $h<0$, and reconstrucing an iso-surface $\tilde{\Sigma}_1 = \{ \x' : \psi(\x')=h\}$ shrinks $\tilde{\Sigma}_0$ in the normal direction.}

The volume \textcolor{Reviewer25}{$V_{\textit{ini}}$} is computed geometrically \citep{Tolle2022} \textcolor{Reviewer24}{as}
\begin{equation}
    \textcolor{Reviewer25}{V_{\textit{ini}}}=\frac{1}{3}\left|\sum_{e=1}^{N_{\tilde{\Sigma}_0}}\mathbf{x}_e\cdot\mathbf{S}_e\right|,
    \label{eq:volume-calc}
\end{equation}
where $N_{\tilde{\Sigma}_0}$ is the number of triangles in $\tilde{\Sigma}_0$, $\mathbf{x}_e$ is a centroid, and $\mathbf{S}_e$ the area-normal vector of the $e$-th triangle in $\tilde{\Sigma}_0$. At reconstruction time step $t^n$, the front is first reconstructed using the iso-value $\psi_0 = 0$, as shown in \cref{fig:volume-correction}. The volume $V^n(\psi_0=0)$ is then calculated w.r.t \cref{eq:volume-calc}. Since the loss/gain of volume between successive reconstructions is small ($\mathcal{O}(10^{-4})$), the compensated extension/contraction is uniformly distributed across the Front. The iso-value adjustment given by \cref{eq:hcontinuous} is then discretized as
\begin{equation}
    h = \frac{V(\tilde{\Sigma}_1)-\textcolor{Reviewer25}{V_{\textit{ini}}}}{\sum_{e=1}^{N_{\tilde{\Sigma}_0}} |\mathbf{S}_e|},
\end{equation}
in which $|\mathbf{S}_e|$ denotes the area of the $e$-th triangle. Reconstruction with the new iso-value $\psi_1 = h^n$ generates a volume-conserved Front, as illustrated by the solid line on the right in \cref{fig:volume-correction}.

\section{Verification and validation}
\label{sec:validations-results}

\textcolor{Reviewer1}{The hybrid Level Set / Front Tracking method is not strictly volume conservative, and volume errors arise from three sources. First, the iso-surface reconstruction - that handles the topological changes of the fluid interface - introduces volume errors by interpolating the level-set function. This error source can be reduced using higher-order level set function interpolation. Second, the Front Tracking method approximates the fluid interface as a surface triangulation and advects the interface in a co-moving reference frame by displacing the triangulation points along Lagrangian trajectories. The volume errors introduced by Front Tracking are reducible significantly by a second (or higher)-order temporal integration of the Lagrangian displacements. The third source of volume conservation errors is the phase-indicator model: we approximate volume fractions from signed distances stored at cell centers and cell-corner points \cite{tolle2020saample}; however, we are investigating a more accurate geometrical intersection between the Front and the volume mesh \cite{Tolle2022}. The volume conservation of the hybrid Level Set / Front Tracking method depends on the physics of the problem. For the verification problems, the \rhoLENT{} method recovers very low maximal relative volume conservation errors of $5.13 \cdot 10^{-4}$ for the coarsest resolution of only $6$ cells per droplet diameter and $5.49\cdot10^{-5}$ for the finest resolution of $26$ cells per droplet diameter. The volume conservation errors of such small magnitude have no effect on the numerical stability of the two-phase momentum convection term, so their detailed visualization is omitted for brevity.}\\

\textcolor{Reviewer1}{\noindent Secondary data presented in this section in the form of diagrams and tables \citep{rhoLENTdata}, the snapshot of the LENT implementation used in this manuscript \citep{rhoLENTsoftware}, and the active development repository of the LENT method as an OpenFOAM module \citep{rhoLENTrepo} are publicly available.}

\subsection{Time step size}
 
The time step size limit due to the CFL condition is given by
 \begin{equation}
    \Delta t \leq \Delta t_{CFL} = \dfrac{h}{U},
 \end{equation}
where $h$ is cell length and $U$ is a characteristic velocity. In the cases, $h$ is \textcolor{Reviewer24}{the minimum cell size}, while $U$ is equal to magnitude of the ambient flow velocity vector, i.e. $U=|\v_a|=1$.
Another restriction for the time step size
%is derived from CFL condition and applied to
arises from
the propagation of capillary waves on interfaces between two fluids. This time step constraint is firstly introduced by \citet{brackbill1992continuum}, and 
afterwards revised by \citet{denner2015numerical}. It has the form
\begin{equation}
    \Delta t \leq \Delta t_{cw} = \sqrt{\dfrac{(\rho_d+\rho_a)h^3}{2\pi\sigma}},
\end{equation}
in which $\rho_d$ and $\rho_a$ are density of droplet and ambient fluid, respectively, $\sigma$ is the surface tension coefficient. In the case setup procedure, the method devised by \citet{tolle2020saample} is followed, i.e., using a compare function
\begin{equation}
    \Delta t = \min\,(k_{cw}\Delta t_{cw},\ k_{CFL}\Delta t_{CFL})
\end{equation}
where $k_{cw}$ and $k_{CFL}$ are arbitrary scale factors between 0 and 1. In the
following, $k_{cw}=0.5$ and $k_{CFL}=0.2$ are used.

\subsection{Translating droplet}\label{sec:popinet_cases}
\begin{figure}[!htb]
    \centering
    \def\svgwidth{0.8\textwidth}
    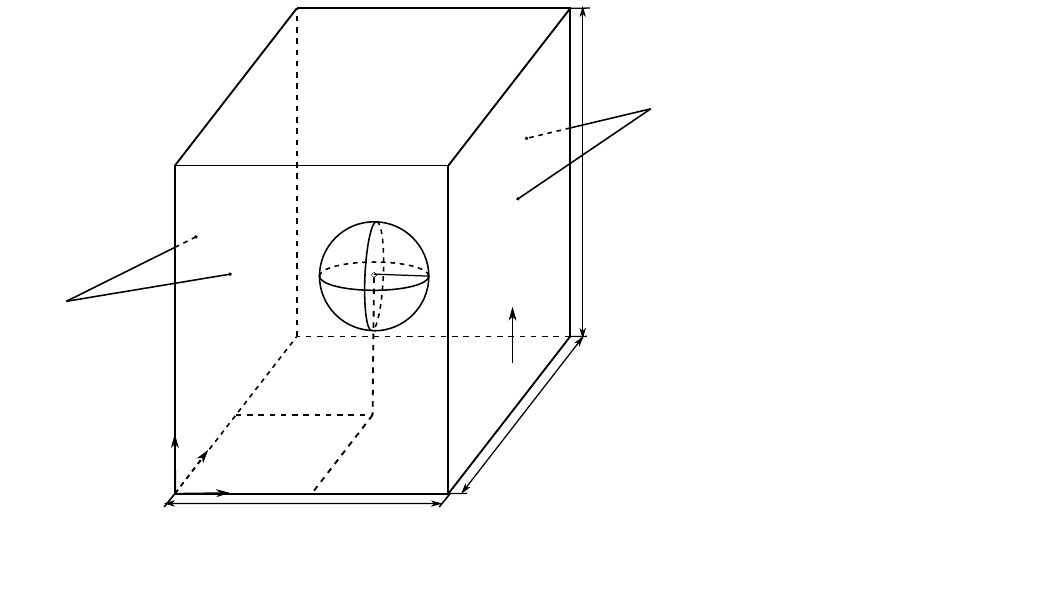
    \caption{Translating droplet case setup.}
    \label{fig:translating-droplet}
\end{figure}
Following the setup of \citet{popinet2009accurate}, a sphere of radius $R=0.2$ translates in a rectangular domain having side lengths $L_x=L_y=5R, L_z =6R$. The initial position of the sphere's centroid is $C_x=C_y=0.5,C_z=0.4$. One corner of the rectangular domain \textcolor{Reviewer24}{locates} in the origin as shown in \cref{fig:translating-droplet}. 
The boundary conditions of the rectangular domain are set as follows: $\nabla \v=0$ and $p=0$ for the outlet, $\v=\v_a$ and \textcolor{Reviewer24}{ zero gradient $\nabla p =0$ for the pressure} at the mantle and the inlet. 
The initial conditions for internal field is set to $p(t_0)=0$ and $\v(t_0)= \v_a$. 
The end time of simulation is set to $t_{end} = 0.41\text{~s}$, which corresponds to a droplet displacement of one diameter.  

 \begin{table}[]
 \begin{adjustbox}{width=1\textwidth}
\small
\begingroup
\renewcommand{\arraystretch}{2}
\begin{tabular}{llccccl}
\hline 
\rowcolor[HTML]{EFEFEF}                             & \multicolumn{5}{c}{Parameters range}                                                                                         \\ 
\rowcolor[HTML]{EFEFEF}                        & \multicolumn{1}{c}{\textcolor{Reviewer24}{Momentum equation}}      & Density ratio     & Resolution     & Kinematic viscosity                       & Surface tension coefficient \\  \cline{1-6} 
\multicolumn{1}{c}{Group $1$}  & \textcolor{Reviewer24}{$\partial_t(\rho \v)+\nabla\cdot(\rho \v \otimes \v) = -\nabla p$}     & $(1, 10^2, 10^3, 10^4)$ & $(16, 32, 64)$       & $0$                                    & $0$                               \\
\multicolumn{1}{c}{Group $2$} & \textcolor{Reviewer24}{$\begin{aligned}
    \partial_t(\rho \v)  +\nabla\cdot(\rho \v \otimes \v) & = 
      -\nabla p - (\mathbf{g}\cdot\mathbf{x}) \nabla \rho\\ &  
      + \nabla\cdot\mu\left(\nabla\v + (\nabla\v)^T\right) + \mathbf{f}_\Sigma
\end{aligned}$}    & \textcolor{Reviewer24}{$(1, 10^2, 10^3, 10^4)$}   & $(16, 32, 64)$       & $(0.057735, 0.018257, 0.0057735, 0.0)$ & $1$                               \\\cline{1-6}
\end{tabular}
\endgroup
\end{adjustbox}
\caption{\textcolor{Reviewer24}{The parameters range of the case group $1$ and the case group $2$.}}
\label{table:parameter_range} 
\end{table}

\begin{figure}[!htb]
      \centering
      \includegraphics[width=.4\textwidth]{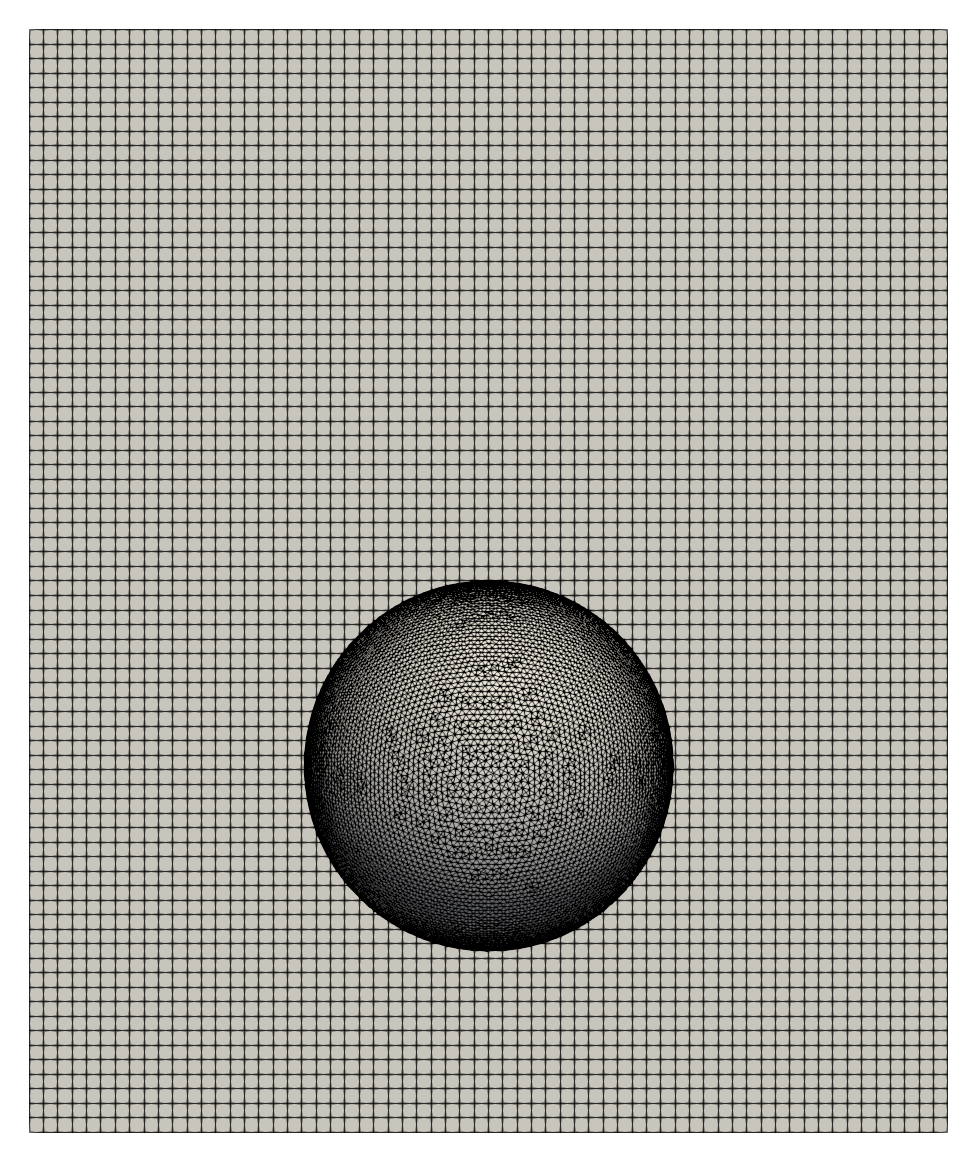}
    \caption{Half section of mesh $N=64$, droplet at initial position.}
    \label{fig:translating-droplet-mesh}
\end{figure}

Two groups of cases are tested to verify the \rhoLENT{} method, \textcolor{Reviewer4}{their parameters are listed in \cref{table:parameter_range}}. For the first group, only the advection of momentum \textcolor{Reviewer24}{and pressure term are} considered, and the ambient flow has a constant density $\rho_a=1$, while the density of the droplet $\rho_d$ varies between $(1, 10^2, 10^3, 10^4)$, resulting in four density ratios. 
Three mesh resolutions $N \in (16, 32, 64)$ are tested. For each mesh resolution $N$, the domain is discretized equidistantly into $1.2N^3$ hexahedral cells, as shown in \cref{fig:translating-droplet-mesh}. 
The exact solution is given by $\v_c^{n+1} = \v_c^n = \v_c(t_0) = \v_a$ and can be used to verify the numerically consistent discretization of the single-field conservative two-phase momentum convection. 

Viscosity and surface tension forces are included in the second test case group. \textcolor{Reviewer24}{ The same range of density ratios is simulated, $\rhoratio \in (1, 10^2, 10^3, 10^4)$. }
%The kinematic viscosity of ambient fluid is kept identical to the droplet viscosity in each case,  $\nu \in (0.057735, 0.018257, 0.0057735, 0.0)$.
The same kinematic viscosity is used for the ambient and the droplet phase, namely $\nu \in (0.057735, 0.018257, 0.0057735, 0.0)$.

The surface tension coefficient is constant $\sigma =1$. 

\subsubsection{Droplet translation without viscosity and surface tension forces}\label{sec:case_1_pure_advection}

\begin{figure}[!htb]
     \begin{subfigure}{.475\textwidth}
      \centering
      \includegraphics[width=\linewidth]{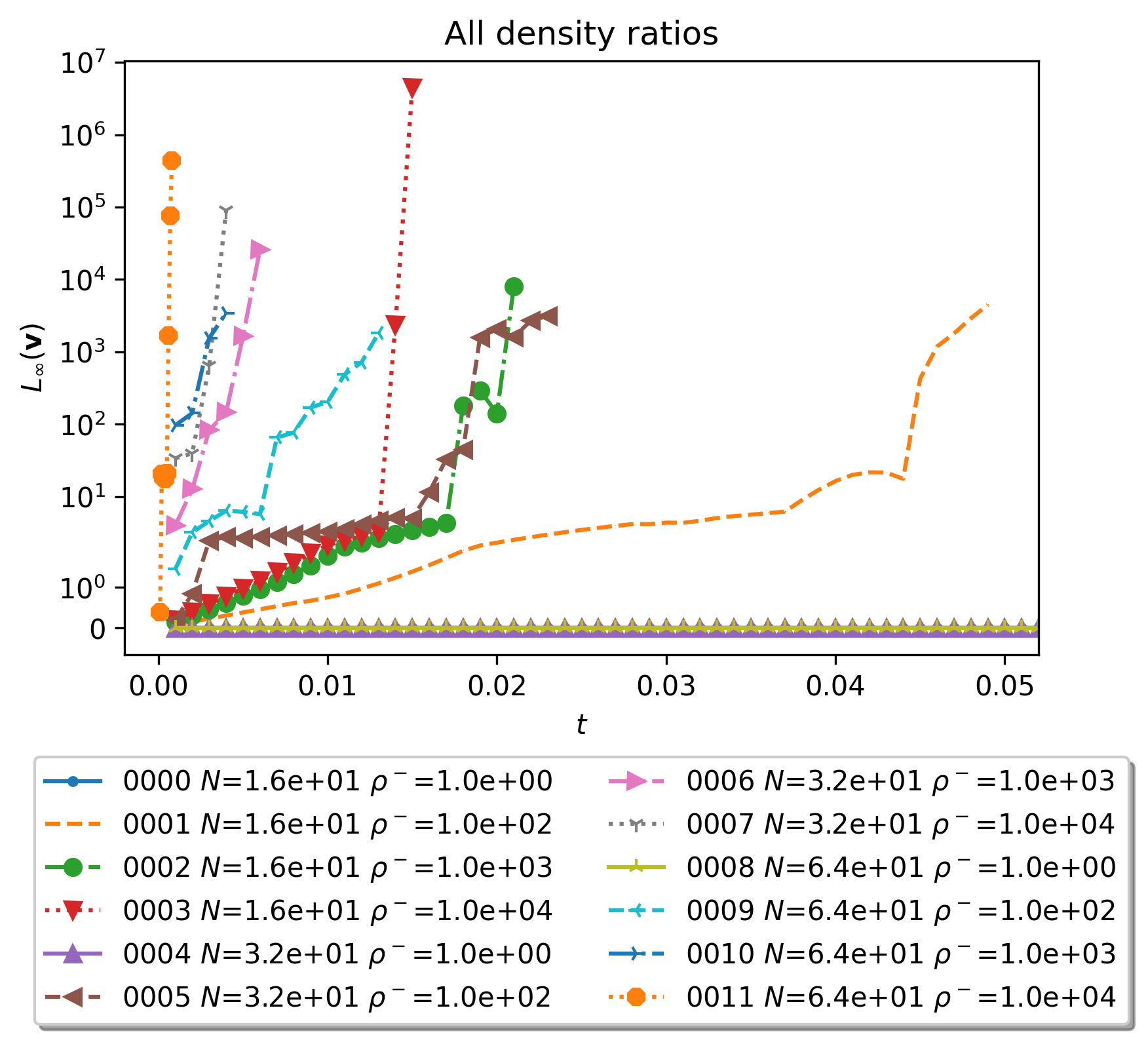}
      \caption{\textcolor{Reviewer25}{SAAMPLE method.}}
      \label{fig:old-inconsistent-method}
     \end{subfigure}
     \hfill
     \begin{subfigure}{.49\textwidth}
      \centering
      \includegraphics[width=\linewidth]{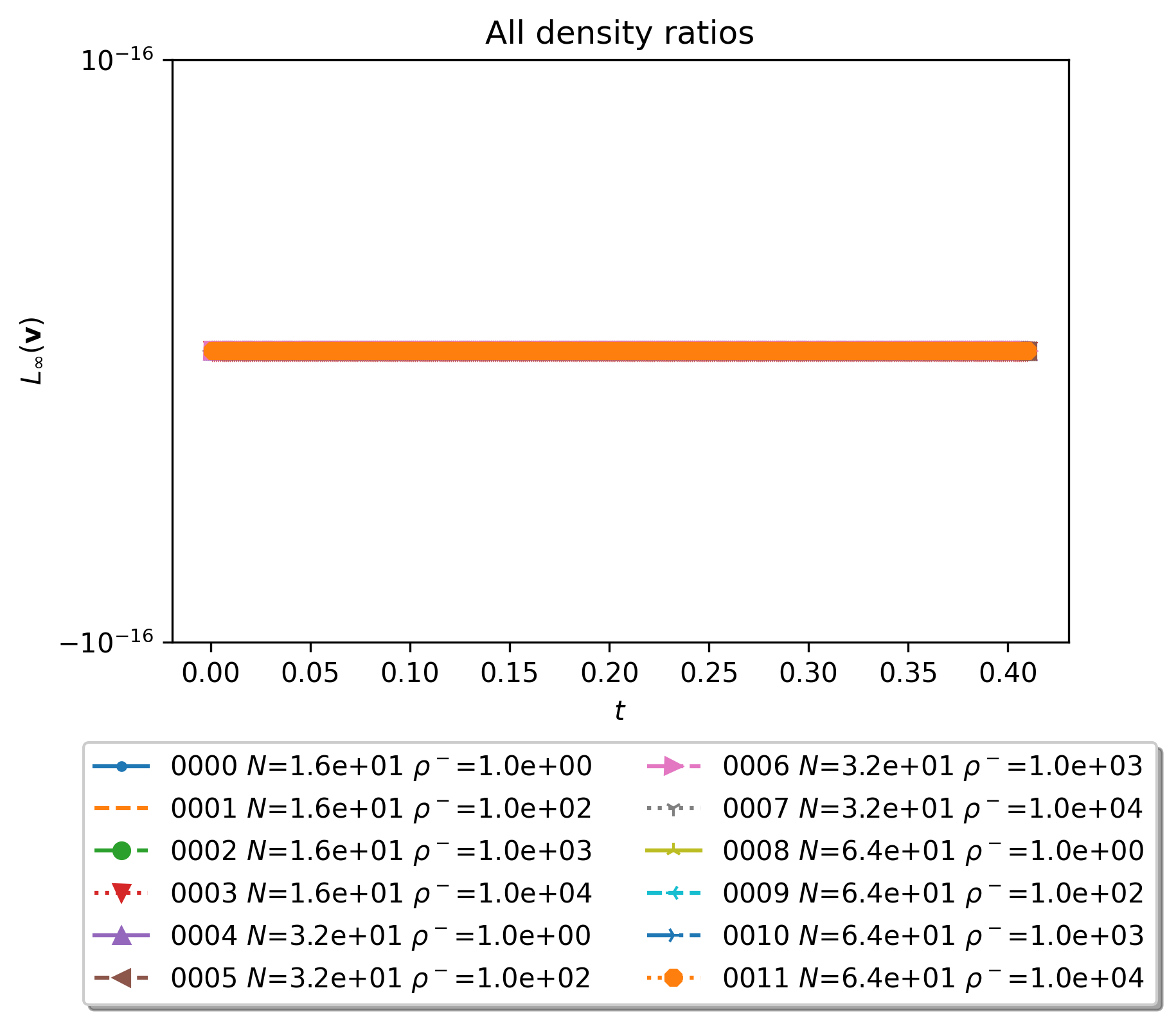}
      \caption{\textcolor{Reviewer24}{\rhoLENT{} method.}}
      \label{fig:new-consistent-method}
     \end{subfigure}    
    \caption{\textcolor{Reviewer24}{Temporal evolution of velocity error norm $L_{\infty}(\v)$: the left figure depicts the results from SAAMPLE algorithm, the right shows the results from \rhoLENT{} method.}}
    \label{fig:linf-error-norm}
\end{figure}
 
\begin{figure}[!htb]
  \centering
  %\includegraphics[width=0.5\textwidth]{figures/noRhoEqn_noForces_U.png}
  %\caption{Deterioration of solution with the inconsistent method. Example: $N=64$, $\rho^-=10^4$,  $t=0.0008s$.}
  \includegraphics[width=0.6\textwidth]{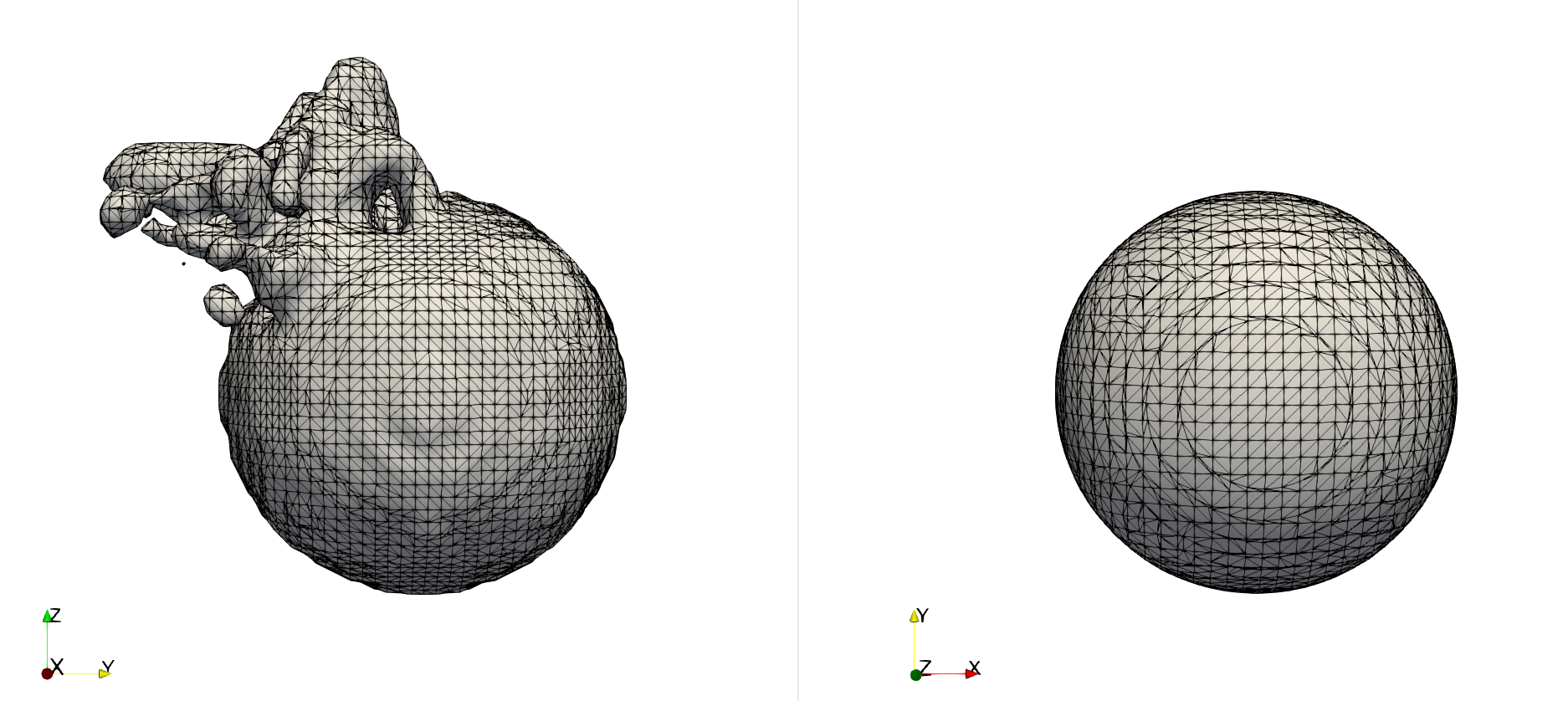}
  \caption{Comparison of the strong interface deformation with \textcolor{Reviewer24}{SAAMPLE} method (left) and the numerically consistent interface shape of the \rhoLENT{} method. Parameters: $N=64$, $\rhoratio=10^4$,  $t=0.0008s$.}
  \label{fig:U-old-inconsistent-method}
\end{figure}

When the momentum is transported only by advection, no forces are exerted on the droplet body and surface. As a result, the velocity field in the overall domain should remain spatially constant and equal to $\v_a = (0,0,1)$. The maximum norm $L_\infty$ is employed to measure how much the numerical velocity deviates from the analytical one, i.e.,\  
\textcolor{Reviewer4}{\begin{equation}
    L_{\infty}(\mathbf{v})= \max_i\left(\frac{\|\mathbf{v}_i-\mathbf{v}_{a}\|}{\|\mathbf{v}_{a}\|}\right),
    \label{eq:linfty}
\end{equation}}
where $\mathbf{v}_i$ denotes velocity of \textcolor{Reviewer24}{all cells}. \textcolor{Reviewer24}{The previous SAAMPLE method \citep{tolle2020saample}} can cause large nonphysical interface deformations leading to a complete deterioration of the solution, visible for a verification configuration in the left image in \cref{fig:U-old-inconsistent-method}. The deterioration is amplified by the $p-\v$ coupling algorithm that will calculate a pressure field $p$ that enforces $\nabla \cdot \v =0$. This, in turn, causes artificial acceleration in all cells where $\v_c^{n+1} \ne \v_a$. The consistent \rhoLENT{} method ensures the shape of the droplet is preserved, as shown on the right image in \cref{fig:U-old-inconsistent-method}.

The \cref{fig:old-inconsistent-method} contains the velocity error calculated with the old, inconsistent method. Every line in the diagram is labeled by the number of the case, mesh resolution $N$, and droplet density $\rho^-$. The default ambient density is $1$. Thus, the $\rho^-$ also represents the density ratio. As shown in \cref{fig:old-inconsistent-method}, all cases with a density ratio higher than $1$, namely $\rho^- > 1$, diverge and stop at early stage. Cases with a very high density ratio of $10^4$ (e.g., case $0011$ and $0003$) fail catastrophically. \textcolor{Reviewer25}{The complete results are shown in \cref{legend:densityRatioInfluence_fullTime}, in \cref{sec:appendix-a-imageWithLegend}.}
 %, the cases with the density ratio $\rho^-=1$ run successfully until end time $t_{end}= 0.41s$. A
 
When \rhoLENT{} is used, as shown in \cref{fig:new-consistent-method}, the velocity error remains exactly $0$ in all cases. This means that the interface velocity remains consistent with the ambient flow and is unaffected by the mesh resolution and density ratio. The results demonstrate the exact recovery of numerical consistency for the advection of the two-phase momentum, using \textcolor{Reviewer24}{an implicitly discretized momentum term in a conservative formulation of single-field two-phase Navier-Stokes equations}. 
%
%\begin{figure}[!htb]
%     \subcaptionbox{Old inconsistent method: momentum predictor on. \label{fig:momentumPredictor-on-old-inconsistent-method}}{\includegraphics[width=.475\textwidth]{figures/Trans_densityRatioInfluence_noRhoEquation_withMomentumPredictor_limitedLinearV0.5.png}}
%     \subcaptionbox{Old inconsistent method:  momentum predictor off. \label{fig:momentumPredictor-off-old-inconsistent-method}}{\includegraphics[width=.51\textwidth]{figures/Trans_densityRatioInfluence_noRhoEquation_noMomentumPredictor_limitedLinearV0.5.png}}
%     \caption{Temporal evolution of velocity error norm with and without momentum predictor: pure advection.}
%     \label{fig:momentumPredictor-old-inconsistent-method}
%\end{figure}
%

\subsubsection{Droplet translation with viscosity and surface tension forces}\label{sec:popinet_cases_nu_sigma}
\begin{figure}[!htb]
     \subcaptionbox{\textcolor{Reviewer24}{SAAMPLE}: interface stable only for cases with density ratio $\rhoratio = 1$ \label{fig:fullforces-old-inconsistent-method}}{\includegraphics[width=.48\textwidth]{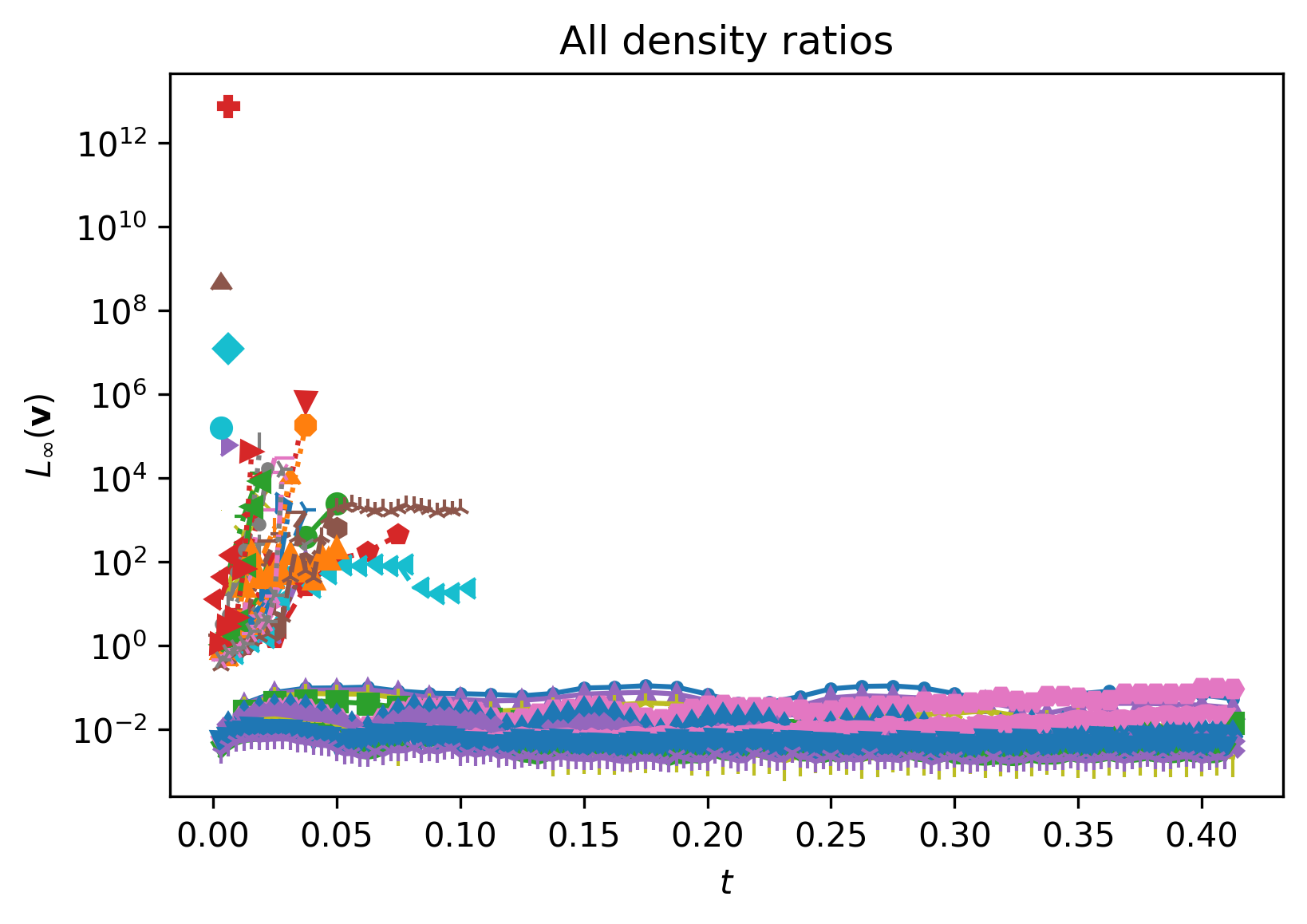}}
\subcaptionbox{\textcolor{Reviewer24}{\rhoLENT{}}: interface stable for density ratios \textcolor{Reviewer24}{$\rhoratio \in (1,10^2, 10^3, 10^4)$}  \label{fig:fullforces-new-consistent-method}}{\includegraphics[width=.48\textwidth]{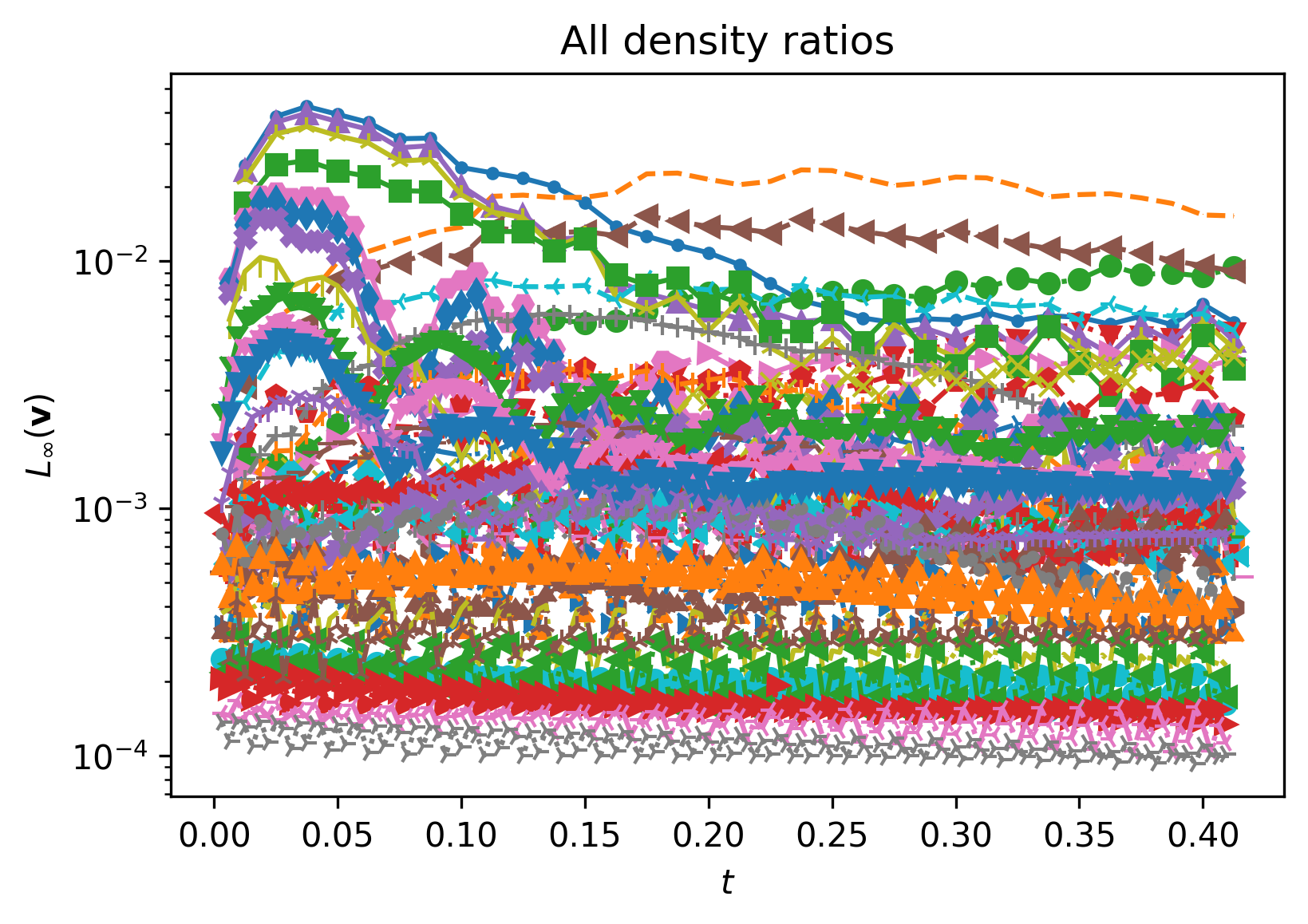}}
    \caption{Temporal evolution of velocity error norm $L_{\infty}(\v)$ for the viscous flow with surface tension forces: the left diagram depicts the results from the \textcolor{Reviewer24}{SAAMPLE} method, and the right diagram contains the results from the \rhoLENT{} method. The legends of these diagrams are large, and the full information is available in Appendix A: \cref{legend:no_with_LV0_withlegend} for \cref{fig:fullforces-old-inconsistent-method}, \cref{legend:with_with_LV0_withlegend} for \cref{fig:fullforces-new-consistent-method}.}
\label{fig:fullforces-Linf-error-norm}
\end{figure}
Here, viscous and capillary forces are taken into account when solving the momentum equation. Since SAAMPLE is a well-balanced algorithm \citep{tolle2020saample} \textcolor{Reviewer4} {- SAAMPLE balances the discrete surface tension force exactly with the pressure gradient when constant curvature is used, using the same discretization for the pressure gradient and the surface-normal gradient of the volume fraction  \citep{popinet2009accurate}. The force-balance is maintained also if the curvature is exactly calculated and propagated as a constant in the interface-normal direction. For numerically approximated curvature, the balance is obtained on a dissipation timescale with respect to initial perturbations. The translating droplet test case combines the force-balance requirement in the droplet's frame of reference, with the requirement for numerical consistency of the two-phase momentum advection.}  In the absence of gravity, such a droplet does not accelerate or decelerate. The temporal evolution of $L_\infty$ is shown in \cref{fig:fullforces-Linf-error-norm}. The inconsistent method remains stable only for $\rhoratio = 1$. For the results of all other cases, i.e., with $\rhoratio >1$, the velocity error increases exponentially, and the simulations crash. In contrast, as depicted in \cref{fig:fullforces-new-consistent-method}, the \rhoLENT{} demonstrates numerically stable results for all tested density ratios. Additional numerical errors are introduced compared with two-phase momentum advection, specifically when approximating the curvature \cite{tolle2020saample}. \textcolor{Reviewer4}{The approximation of curvature in \cite{tolle2020saample} recovers accurate $L_2$ norms of the curvature errors for a sphere, in the range $[10^{-4}, 10^{-3|}]$ for discretization lengths in the range $[128^{-1}, 16^{-1}]$ in the unit-box solution domain. Because of the numerically approximated curvature,} $L_\infty$ cannot exactly be equal to zero, as shown in \cref{fig:new-consistent-method}. However, as seen in \cref{fig:fullforces-new-consistent-method}, the final $L_\infty$ error given by \cref{eq:linfty} $10^{-4}$ and $10^{-2}$, which is acceptable.
%As in the pure advection cases, the choice of the flux-limiter and the use of the momentum predictor do not affect the results.
%
\subsubsection{Translating sub-millimeter droplet with realistic physical properties}\label{sec:small-droplet-cases}
\begin{table}[H]
\begin{adjustbox}{width=1\textwidth}
\small
\begingroup
\renewcommand{\arraystretch}{1.1}
\begin{tabular}{cccccl}
\hline
\rowcolor[HTML]{EFEFEF}
materials/properties (\SI{25}{\celsius}) & density (\si{\kilogram\per\cubic\metre}) & kinematic viscosity (\si{\square\metre\per\second}) & surface tension (\si{\newton\per\metre}) & density ratio \\
\hline
air                                                 & $1.1839$                                                    & $1.562\e{-5}$                                                             & $---$                                            & $---$                                   \citep{adamson1997physical}\\
water                                               & $997.05$                                                    & $8.926\e{-7}$                                                             & $0.07213$ (in air)                               & $842.17$ (in air)                       \citep{adamson1997physical}\\
mercury                                             & $13.5336\e{3}$                                                & $1.133\e{-7}$                                                             & $0.4855$ (in air)                                & $11431.37$(in air)                      \citep{adamson1997physical}\\
silicone oil (cSt $10$)                               & $0.934\e{3}$                                                  & $1.088\e{-5}$                                                             & $0.0201$ (in air)                                & $788.92$(in air)                       \citep{zivojnovic2012silicone} \\
silicone oil (cSt $50$)                               & $0.96\e{3}$                                                   & $5\e{-5}$                                                                 & $0.032$ (in water)                               & $0.96$ (in water) \citep{trinh1982large} \\
 \hline
\end{tabular}
\endgroup
\end{adjustbox}
\caption{Realistic fluid properties are combined into four tests: water droplet/air ambient, mercury droplet/air ambient, silicone oil droplet/air ambient, silicone oil droplet/water ambient.}
\label{table:physical-properties} 
\end{table}
\textcolor{Reviewer24}{
%In the aforementioned cases, which were configured based on the work by \citet{popinet2009accurate}, the droplet diameter is specified as $0.4$ without any unit. In our simulations presented in \cref{sec:case_1_pure_advection,sec:popinet_cases_nu_sigma}, we directly employ this value, which inherently corresponds to a diameter of $0.4m$  in the OpenFOAM framework. It should be noted that a diameter of $0.4m$ is significantly larger than the typical size of natural droplets. 
The physical properties including densities, viscosities, and surface tension coefficients in the widely used translated droplet case from \citet{popinet2009accurate} are not related to physical two-phase flows systems. We have adapted the case and used small droplet dimensions to challenge the method in terms of surface tension force approximation for capillary problems, and used real-world fluid pairings with challenging density ratios.}

\Cref{table:physical-properties} contains the physical properties used for the test-case configuration of the translating sub-millimeter droplet with realistic physical properties. In terms of size, a spherical droplet of radius $R=$ \SI{0.25}{\milli\metre} is translating a distance of three diameters with velocity \SI[per-mode=symbol]{0.01}{\metre\per\second} in $z$-direction of the rectangular solution domain ($L_x=L_y=5R,L_z=10R$). The initial centroid position of the droplet is $(2.5R,2.5R,2R)$. Surface tension and viscous forces are not considered for this
setup.
\begin{figure}[!htb]
    \subcaptionbox{Silicone oil droplet in water, density ratio 0.96 }{\includegraphics[width=.48\linewidth]{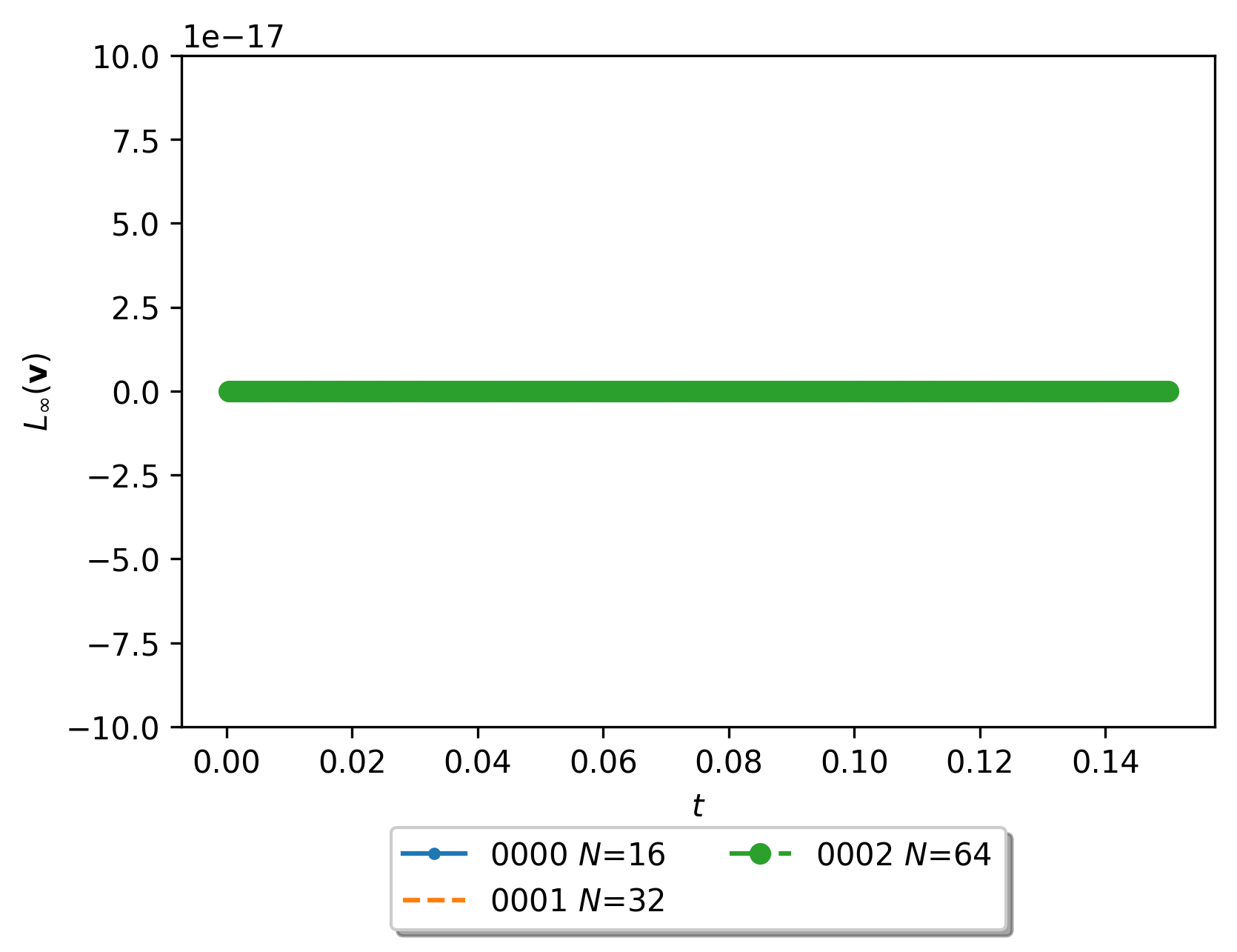}} 
    \subcaptionbox{Silicone oil droplet in air, density ratio 788.92}{\includegraphics[width=.48\linewidth]{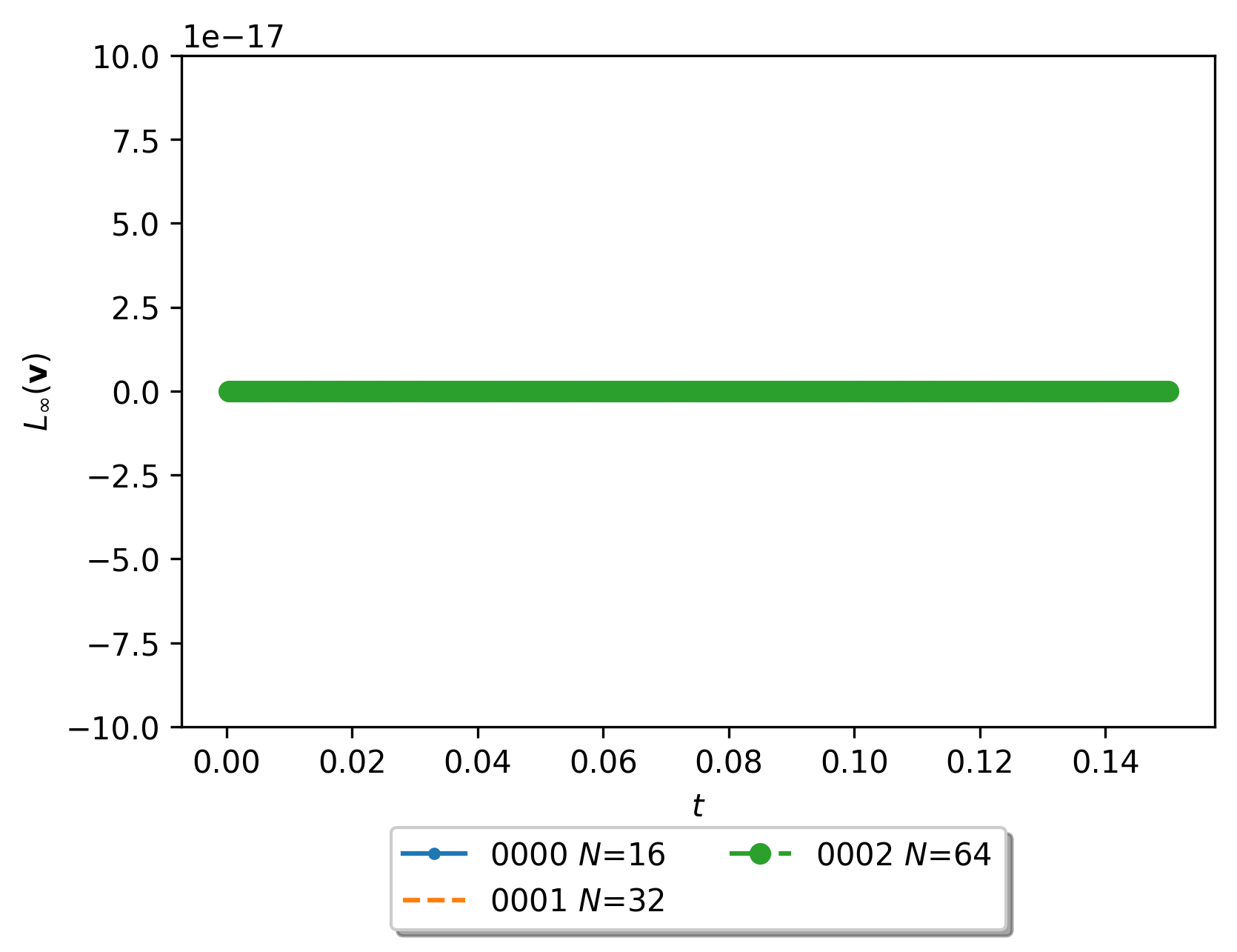}}
    \subcaptionbox{Water droplet in air, density ratio 842.17}{\includegraphics[width=.48\linewidth]{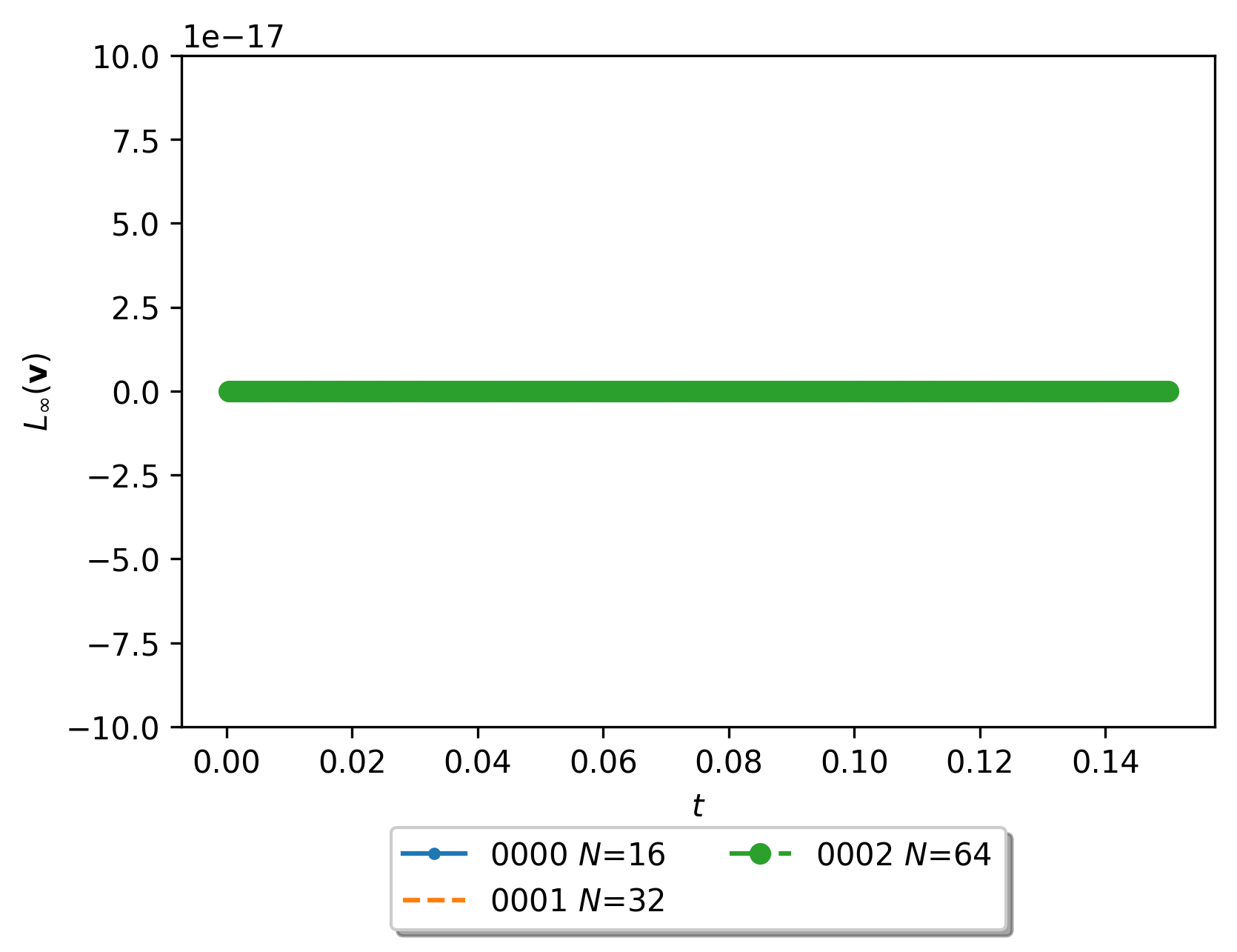}}    
    \subcaptionbox{Mercury droplet in air, density ratio 11431.37\label{fig:mercury_air}}{\includegraphics[width=.48\linewidth]{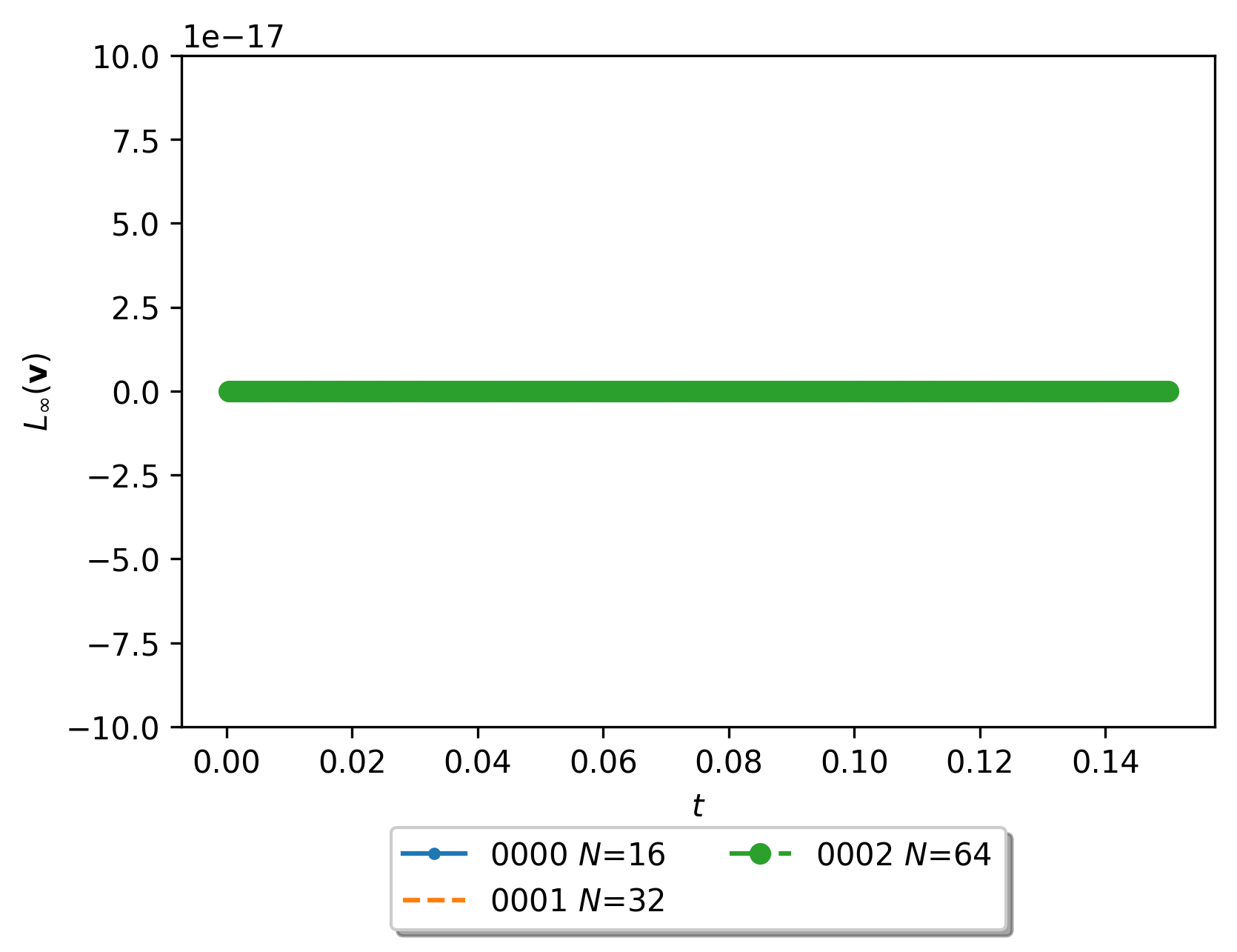}} 
    \caption{Temporal evolution of velocity error norm $L_{\infty}(\v)$ with pure advection: \rhoLENT{} method used in simulating two-phase flows with different density ratios, mesh resolution: $N\in(16,\ 32,\ 64)$.}
    \label{fig:different-fluids-rholent}
\end{figure}
\begin{figure}[!htb]
    \subcaptionbox{Silicone oil droplet in water, density ratio 0.96 }{\includegraphics[width=.48\linewidth]{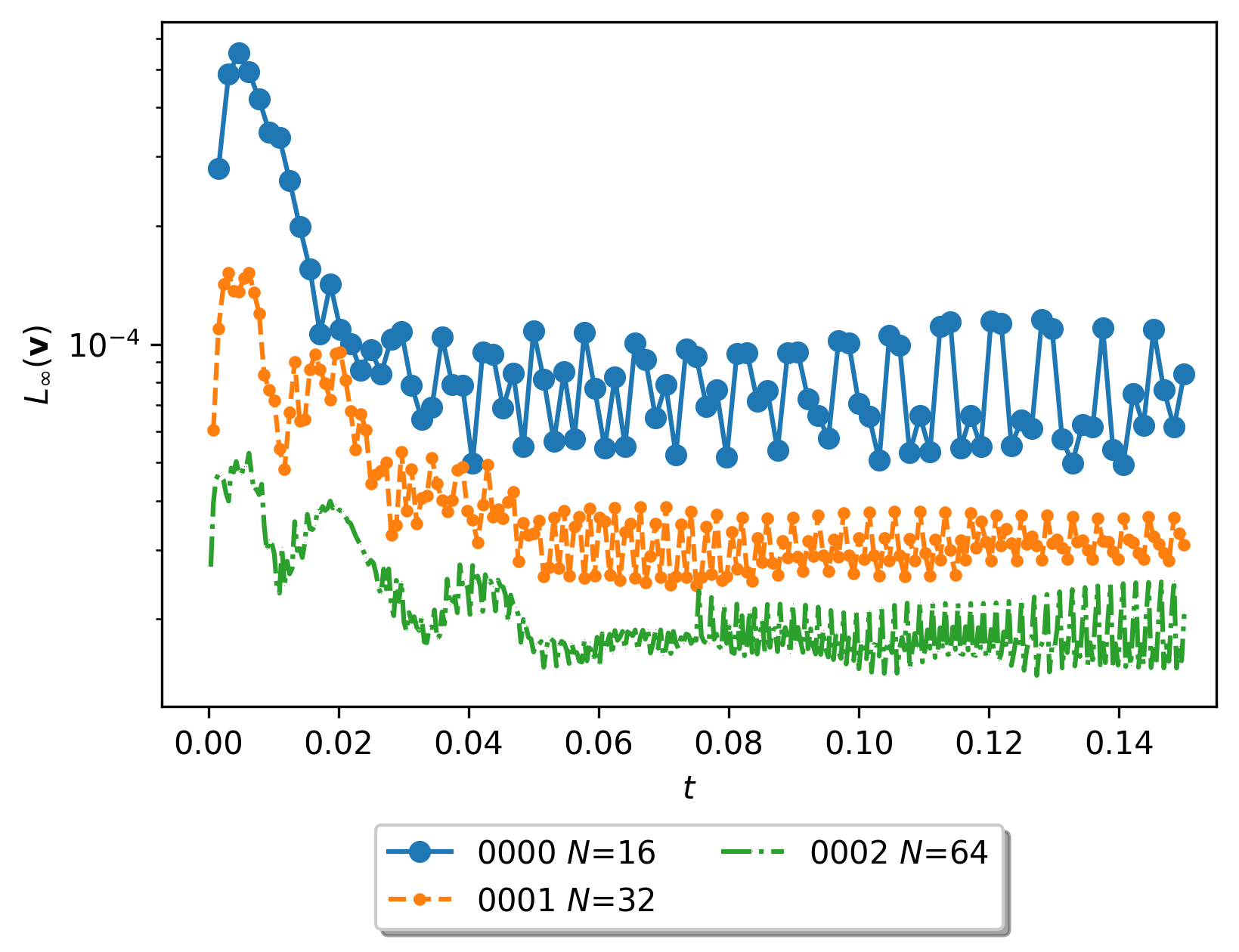}} 
    \subcaptionbox{Silicone oil droplet in air, density ratio 788.92}{\includegraphics[width=.48\linewidth]{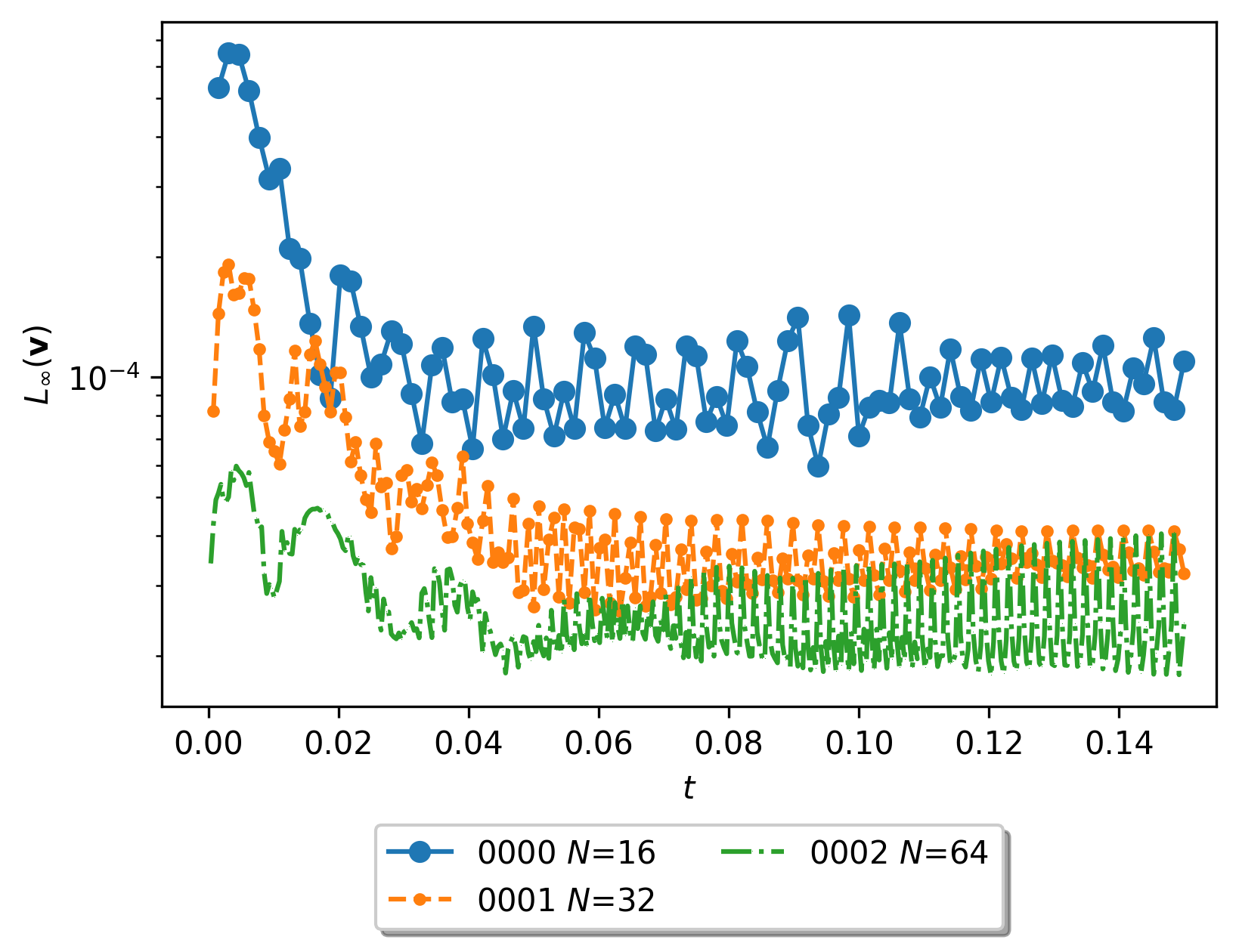}}
    \subcaptionbox{Water droplet in air, density ratio 842.17}{\includegraphics[width=.48\linewidth]{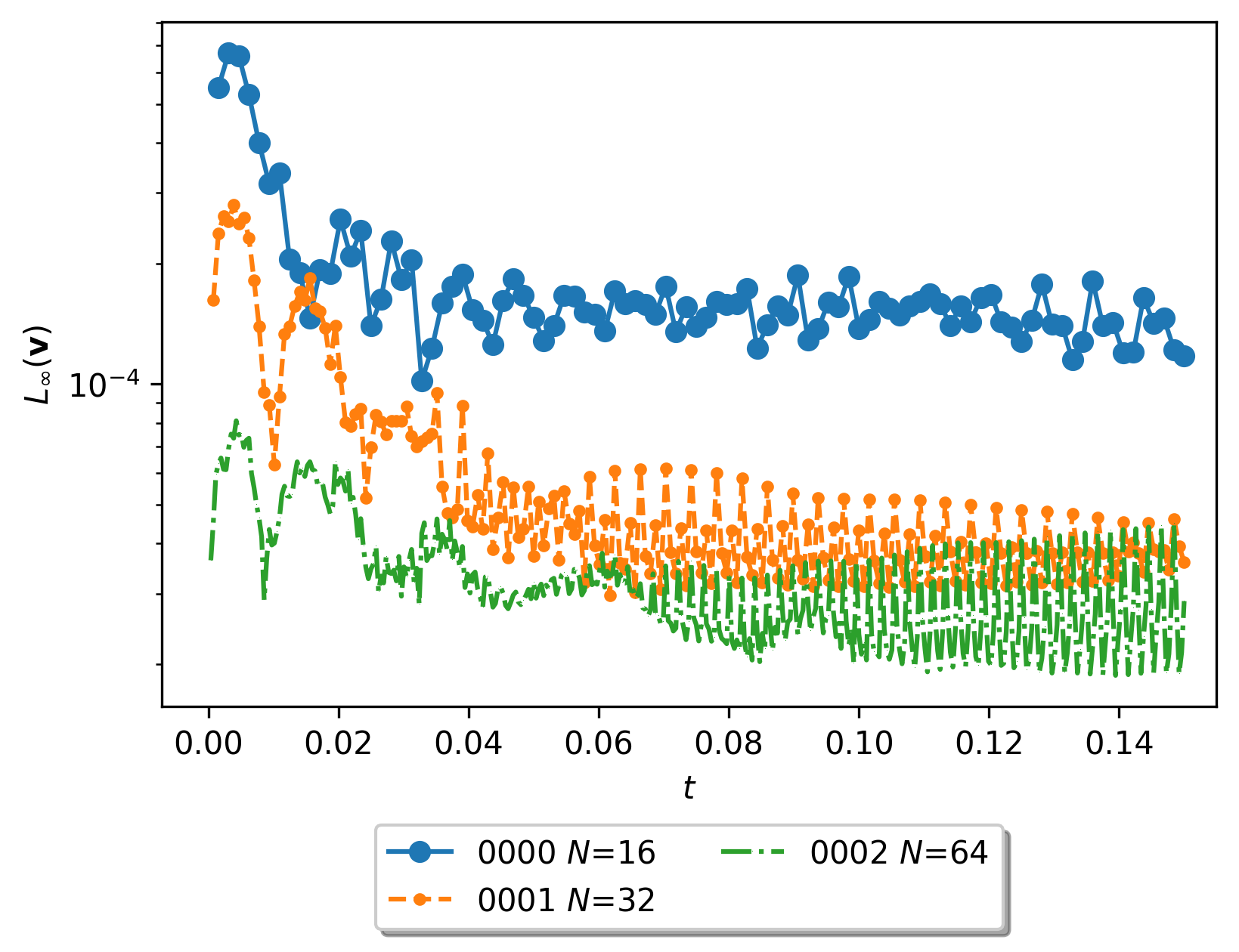}}    
    \subcaptionbox{Mercury droplet in air, density ratio 11431.37\label{fig:mercury_air1}}{\includegraphics[width=.48\linewidth]{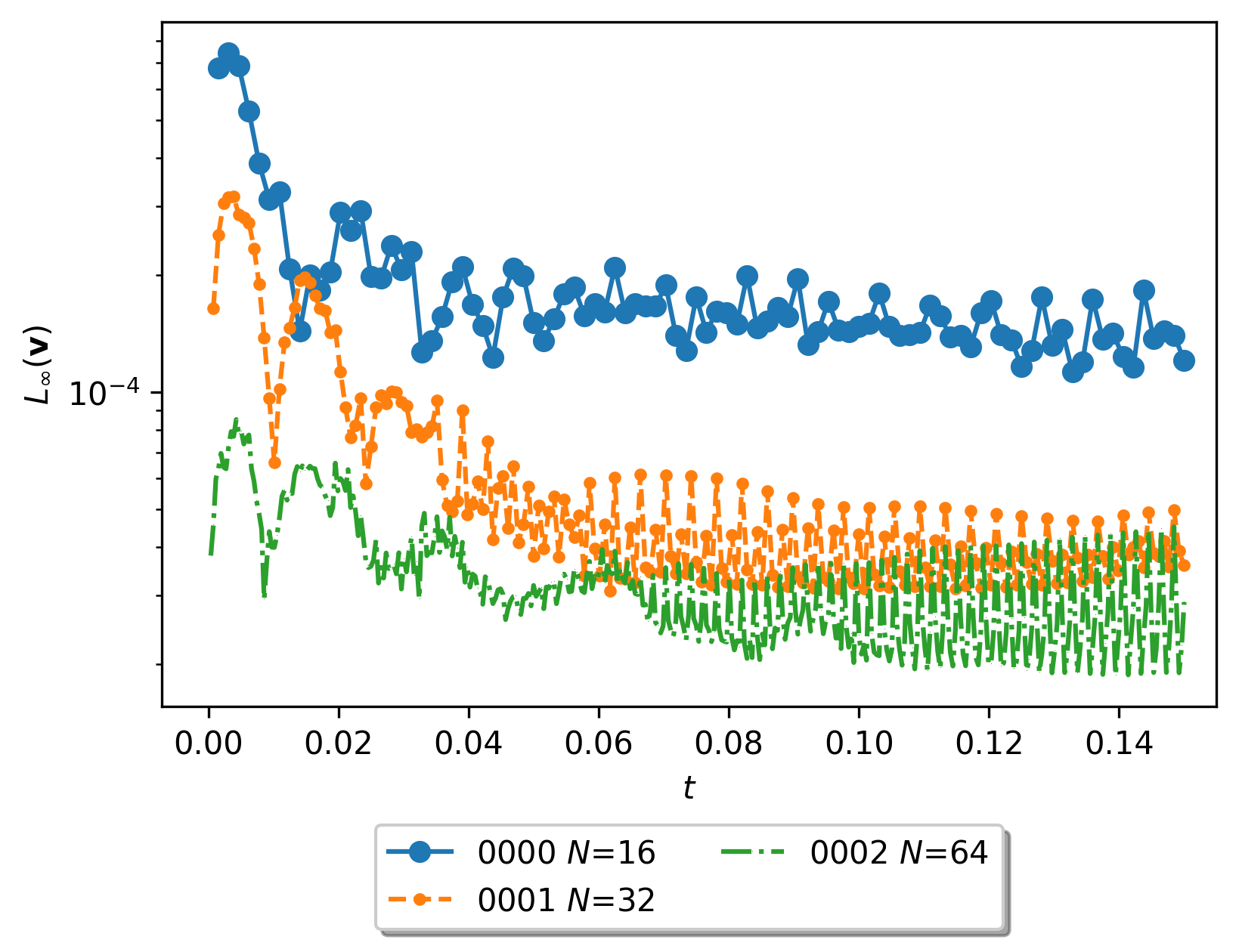}} 
    \caption{\textcolor{Reviewer24}{Temporal evolution of velocity error norm $L_{\infty}(\v)$ with the effect of viscosity and surface tension: \rhoLENT{} method used in simulating two-phase flows with different density ratios, mesh resolution: $N=16,32,64$.}}
    \label{fig:different-fluids-plusNuSigma-rholent}
\end{figure}
As depicted in \cref{fig:different-fluids-rholent},  it is obvious that $L_{\infty}(\v)$ remains stable  over time when the droplet translates. Even in the cases with a density ratio of over $10^4$, as shown in \cref{fig:mercury_air}, no matter how high the resolution is, the results from \rhoLENT{} the method can reach machine precision. 

\textcolor{Reviewer24}{The \cref{fig:different-fluids-plusNuSigma-rholent} illustrates the results of the same realistic droplets' cases, considering the influence of viscous forces and surface tension. It is observed that the errors decrease as the resolution increases, reaching magnitudes as low as $10^{-5}$ for all cases. This indicates the excellent capability of \rhoLENT{} to handle a wide range of density ratios in such cases.}

Apart from the observation mentioned above, \cref{table:performance-lent} \textcolor{Reviewer4}{reveals another advantage of \rhoLENT{} method - high computational efficiency. As shown in \cref{table:performance-lent}, the \rhoLENT{} method demonstrates very high computational efficiency in serial. Increasing the parallel computational efficiency requires further research, specifically, regarding a more efficient message-passing parallel implementation for unstructured Level Set / Front Tracking.}      
% Please add the following required packages to your document preamble:
% \usepackage{multirow}
% \usepackage[table,xcdraw]{xcolor}
% If you use beamer only pass "xcolor=table" option, i.e. \documentclass[xcolor=table]{beamer}
\begin{table}[!htb]
\begin{adjustbox}{width=0.6\textwidth}
\small
\begin{tabular}{cc|cc}
% \hline
% \rowcolor[HTML]{EFEFEF} 
%                                               &                                                   & \multicolumn{2}{c}{lentFoam}                                             \\
                                               \hline
                        \rowcolor[HTML]{EFEFEF} 
cases                                          & resolution & \textcolor{Reviewer4}{serial} execution time (s)  \\
\hline
\noalign{\vskip 1mm}
                                               & 16                                                                                & 6.62                                                                                          \\
                                               & 32                                                                         & 74.34                                                                                         \\ 
\multirow{-3}{*}{silicone oil droplet / water} & 64                                                                                 & 929.79                                                                                         \\ \hline
                                               & 16                                                                                 & 7.77                                                                                          \\
                                               & 32                                                                         & 91.64                                                                                         \\
\multirow{-3}{*}{water droplet / air}          & 64                                                                         & 1324.38                                                                                        \\ \hline
                                               & 16                                                                         & 7.71                                                                                          \\
                                               & 32                                                                         & 94.57                                                                                         \\
\multirow{-3}{*}{mercury droplet / air}        & 64                                                                         & 1334.36                                                                                        \\ \hline
                                               & 16                                                                         & 7.31                                                                                           \\ 
                                               & 32                                                                        & 83.43                                                                                        \\
\multirow{-3}{*}{silicone oil droplet / air}   & 64                                                                         & 1318.43                                                
\\ \hline
\end{tabular}
\end{adjustbox}
    \caption{\textcolor{Reviewer4}{Serial} execution time for the \rhoLENT{} method.}
\label{table:performance-lent}
\end{table}

\textcolor{Reviewer3}{\subsubsection{Oscillating droplet}}

\textcolor{Reviewer24}{An ellipsoidal droplet is submerged in an ambient fluid with \textcolor{Reviewer24}{an approximate axially symmetric solution} provided by \citet{Lamb1932hydrodynamics}.  The solution can be represented as a summation of a constant and a Legendre polynomial $\mathit{P}_n(\cos\theta)$, i.e.,
\begin{equation}
    \mathit{R}(\theta,t) \doteq \mathit{R}_0 +  a_n\mathit{P}_n(\cos\theta)\sin(\omega_nt),\ \ \theta \in [0,2\pi],
    \label{eq:ellipsoid_radius}
\end{equation}
where $R_0$ is the initial unperturbed radius, $a_n$ is the amplitude of the $n$-th oscillation mode, $\theta$ is the angle between the radius line of a droplet point and the symmetric axis, $\omega_n$ represents
the oscillation frequency. The latter $\omega_n$ has the form
\begin{equation}
    \omega^2_n = \frac{n(n+1)(n-1)(n+2)\sigma}{[(n+1)\rho_d+n\rho_a]\mathit{R}^3_0},
    \label{eq:oscillation_frequency}
\end{equation}
where $n$ is the mode number, $\rho_d$, $\rho_a$ represent the droplet and ambient flow density respectively, $\sigma$ indicates the surface tension coefficient. \citet{Lamb1932hydrodynamics} derived \cref{eq:ellipsoid_radius} neglecting the viscous effect, and used constant $a_n$. \citet{chandrasekhar1959oscillations,miller1968oscillations,prosperetti1980normal} extended the expression of $a_n$ to include the influence of the viscosity, where the amplitude $a_n$ decays exponentially over time by 
\begin{equation}
    a_n(t) = a_0e^{-\gamma t},\ \gamma = \frac{(n-1)(2n+1)\nu}{\mathit{R}^2_0},
\end{equation}
where $\nu$ is the kinematic viscosity. \citet{hiller1989experimental} conducted a series of experiments to validate the decay expression.} 

\begin{figure}[!htb]
      \centering
      \includegraphics[width=.7\textwidth]{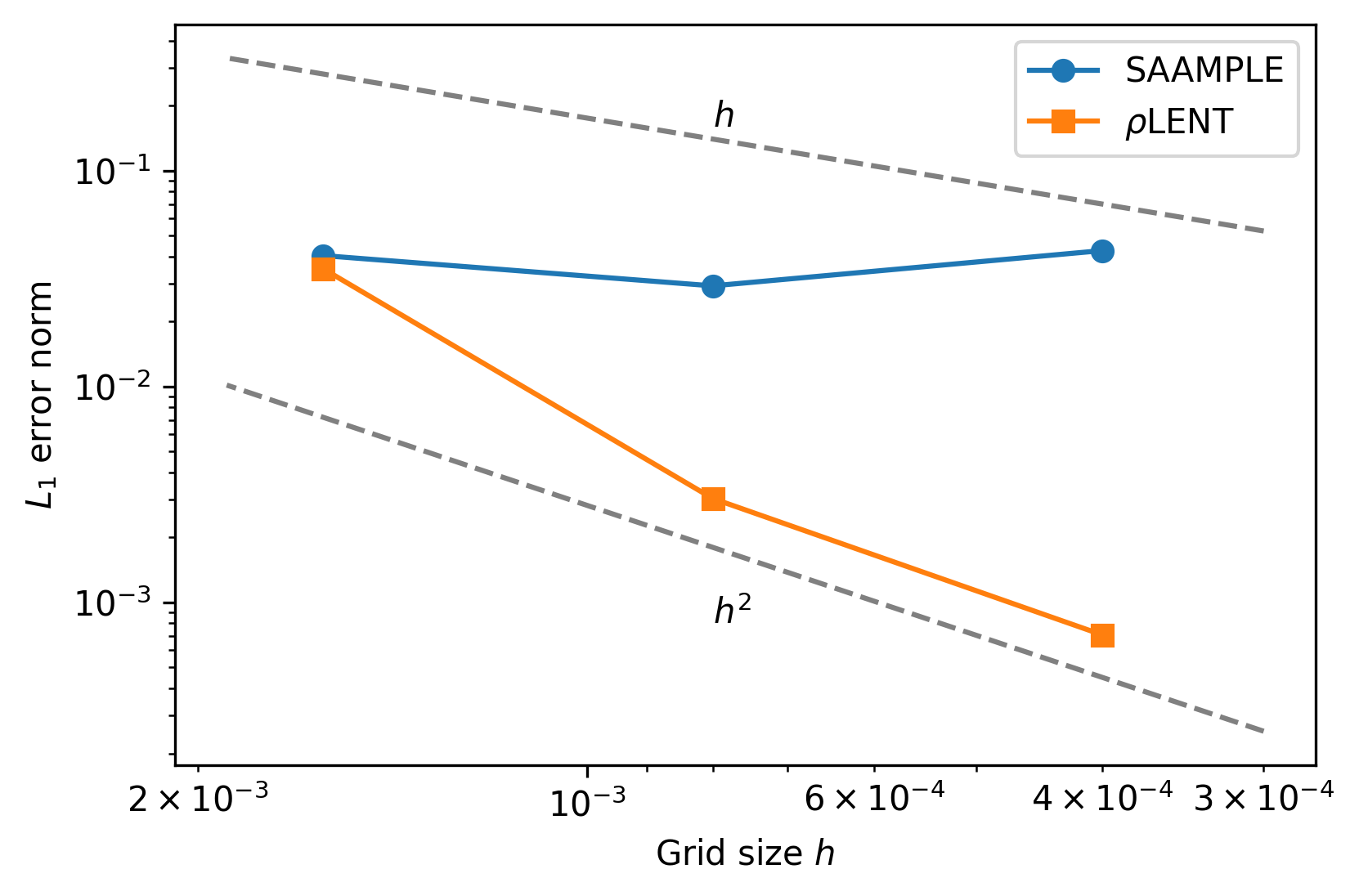}
    \caption{\textcolor{Reviewer24}{$L_1$ norms of the oscillation frequency errors.} }
    \label{fig:Oscillat_L1-norm}
\end{figure}

\textcolor{Reviewer3}{We have applied the physical properties of mercury and air from \ref{table:physical-properties} to the droplet and the ambient fluid. The droplet interface is initialized with the parameters: $R_0=0.01$, $n=2$, $\epsilon=0.00025$, $t = \pi/(2\omega_n)$ and the center $(0.0200001,0.0199999,0.020000341)$. The computational domain size is $(0,0,0)\times(0.04,0.04,0.04)$. \textcolor{Reviewer24}{The gravity is neglected in this case.} At the simulation's beginning, the droplet and the flow are still, i.e. $\v(t=0)=0$ holds for the whole field. Since the analytical frequency $\omega_a$ can be acquired from \cref{eq:oscillation_frequency}, the results are evaluated by an error norm
\begin{equation}
    L_1(\omega)= \frac{|\omega-\omega_a|}{\omega_a}.
\end{equation}
Three mesh resolutions are tested to verify the convergence of both methods. The results of the \textcolor{Reviewer24}{SAAMPLE\citep{tolle2020saample}} and \textcolor{Reviewer24}{$\rho$LENT} with increasing mesh resolutions are summarized in the \cref{table:oscillation_freq_comp}. \textcolor{Reviewer24}{As shown in  \cref{fig:Oscillat_L1-norm},} the oscillating frequencies calculated from \textcolor{Reviewer24}{SAAMPLE} do not converge. On the contrary, when deploying the new consistent method, \textcolor{Reviewer24}{$L_1$ norm exhibits second-order convergence.}%the frequency converges in $L_1(\omega)$.  
\begin{table}[H]
{
\begin{adjustbox}{width=1\textwidth}
\small
\begin{tabular}{lllllllllllll}
\hline
\rowcolor[HTML]{EFEFEF}
Grid size $h$ &  & \multicolumn{2}{c}{Background mesh} &  & \multicolumn{2}{c}{Front mesh} &  & \multicolumn{2}{c}{\textcolor{Reviewer24}{SAAMPLE}} &  & \multicolumn{2}{c}{\textcolor{Reviewer24}{$\rho$LENT}} \\  \cline{3-4} \cline{6-7} \cline{9-10} \cline{12-13} 
\rowcolor[HTML]{EFEFEF}
       &  & Points           & Cells            &  & Points         & Tris.         &  & Frequency          & $L_1$ norm         &  & Frequency         & $L_1$ norm        \\ \hline
0.0016 &  & 17576            & 15625            &  & 3150           & 6296          &  & 16.255923          & 0.040397           &  & 16.346002         & 0.035080          \\
0.0008 &  & 132651           & 125000           &  & 12662          & 25320         &  & 16.444685          & 0.029254           &  & 16.888921         & 0.003031          \\
0.0004 &  & 1030301          & 1000000          &  & 50732          & 101460        &  & 16.218606          & 0.042600           &  & 16.928274         & 0.000708          \\ \hline
\end{tabular}
\end{adjustbox}
\caption{The analysis of the oscillation frequency convergence and its comparison between the \textcolor{Reviewer24}{SAAMPLE} method and the consistent \textcolor{Reviewer24}{\rhoLENT{}} method.}
\label{table:oscillation_freq_comp}}
\end{table}}

\textcolor{Reviewer3}{ \subsection{Rising bubble}}

\textcolor{Reviewer3}{In this test case, we apply the proposed method to a single bubble rising in quiescent viscous liquid. We test the configuration from \citet{Anjos2014}, who simplified the rising bubble experiments previously conducted by \citet{Bhaga_weber_1981} and selected three different viscosity ratios to perform tests and comparisons. In this section, we focus on the most challenging case, i.e. the case with the smallest viscosity ratio, which corresponds Morton number $Mo=g\nu_l^4/\rho_l\sigma^3=1.31$, where $g$ is the gravitational acceleration value, and $\nu_l$, $\rho_l$, $\sigma$ indicate the viscosity, density of the ambient liquid and the surface tension. The initial state of the air bubble is idealized to be spherical with a diameter of $D=2.61\text{ cm}$. The physical properties of the air are characterized by a viscosity of $1.78\times10^{-5}\text{ kg/ms}$ and a density of $1.225\text{ kg/m}^3$, whereas the properties of the liquid are defined by a viscosity of $0.54\text{ kg/ms}$ and a density of $1350\text{ kg/m}^3$. Additionally, the surface tension between the air bubble and the liquid is $0.078\text{ N/m}$. The computational domain is defined as $(-4D,-4D,-2D)\times(4D,4D,6D)$, \textcolor{Reviewer24}{which show the positions of space diagonal vertices of the computational domain}, and the initial position of the bubble is set as the origin, $(0,0,0)$.}

% \subsubsection{Interpolation method}
% \textit{do we need this subsection to explain our displacement interpolation, which is distinguishable from the widespread velocity interpolation}

% \textcolor{Reviewer3}{\subsubsection{Results}}
%%
\begin{figure}[!htb]
      \centering
      \includegraphics[width=.4\textwidth]{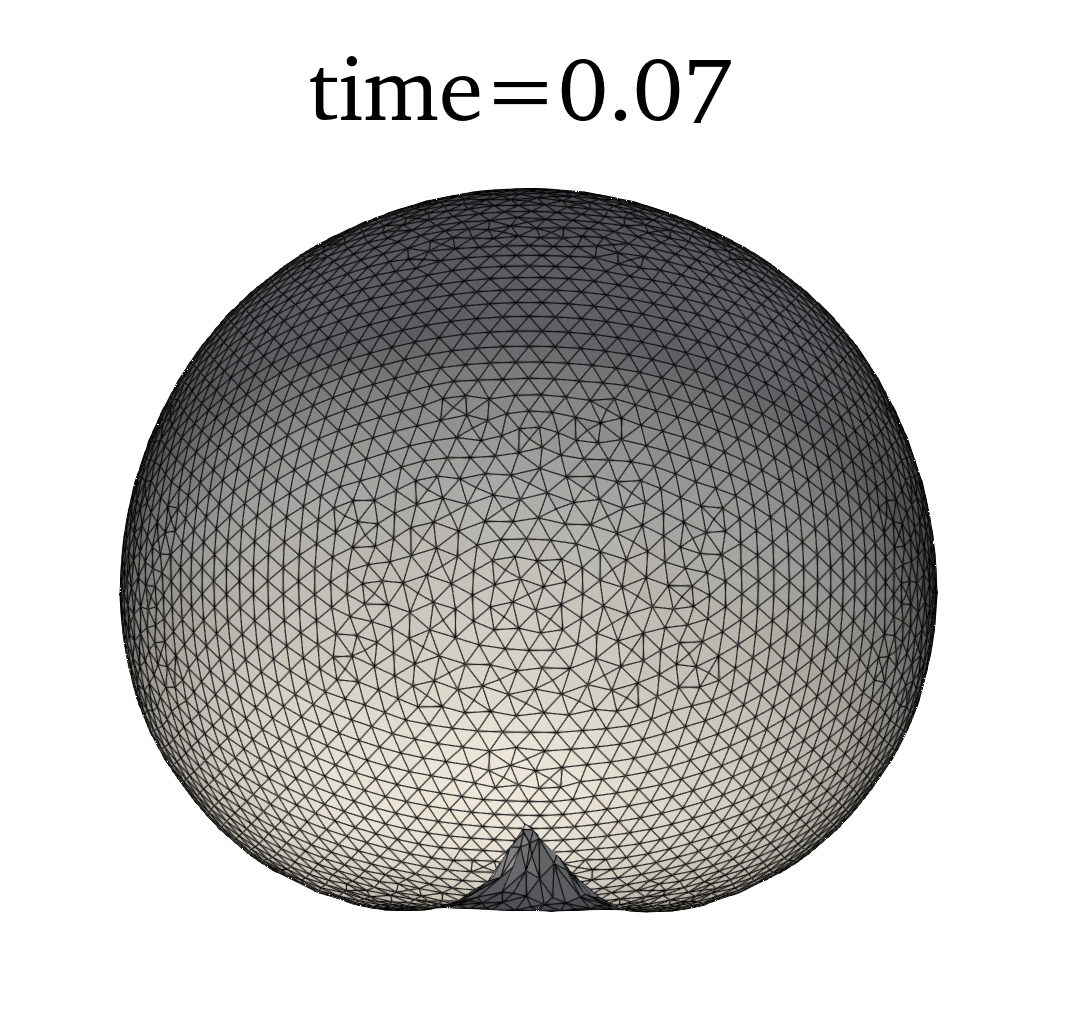}
    \caption{\textcolor{Reviewer24}{SAAMPLE} method: the collapsed bubble shape \textcolor{Reviewer24}{caused by numerical inconsistency}. }
    \label{fig:collapsed_risingBubble-shape}
\end{figure}
\begin{figure}[!htb]
      \centering
      \includegraphics[width=.8\textwidth]{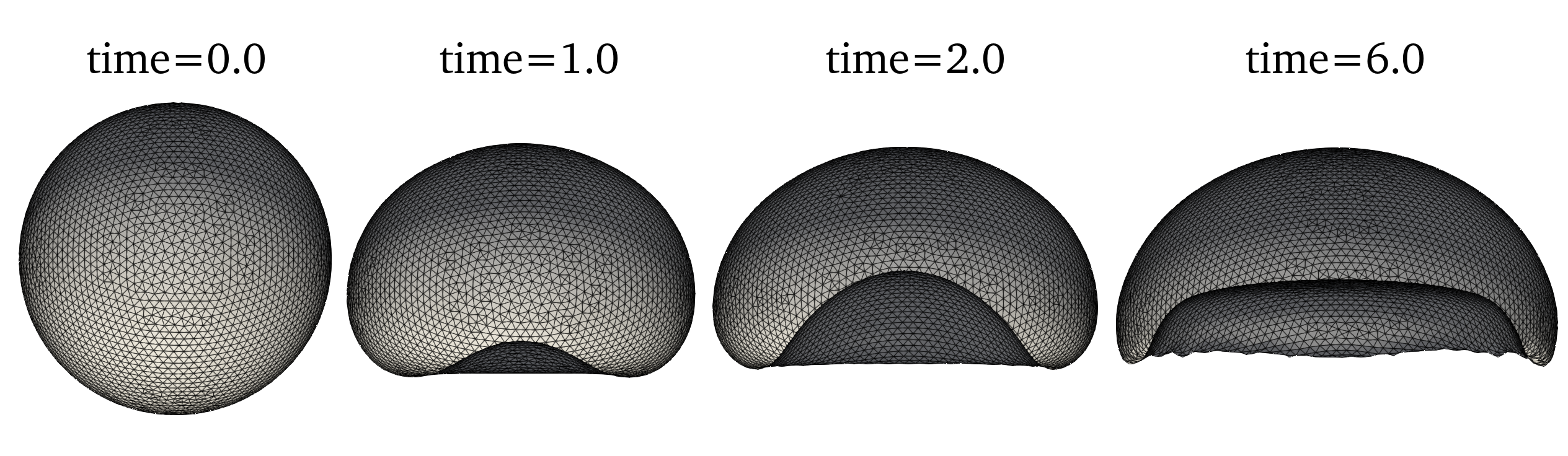}
    \caption{\textcolor{Reviewer24}{\rhoLENT{}}: the temporal evolution of bubble's shape \textcolor{Reviewer24}{with resolution $N=128$}.}
    \label{fig:risingBubble-shape}
\end{figure}
As the parallel computing module in the $\rho$LENT method is still in the developmental phase, the entire domain was resolved using a single core. Consequently, \textcolor{Reviewer24}{relatively coarse meshes were utilized to simulate the motion of the bubble, i.e. $N\in(64,\ 96,\ 128,\ 160)$, where $N$ indicates the grid numbers in all three directions of the computational domain.} To estimate the results, a set of dimensionless characteristic variables was introduced as follows:
\begin{equation}
    \mathbf{w}=\frac{\v}{\sqrt{gD}},\ t=\sqrt{\frac{g}{D}}\tau,
\end{equation}
where $\tau$ indicates the realistic time. When deploying the inconsistent method, the simulation crashed at an early stage, as shown in \ref{fig:collapsed_risingBubble-shape}. Inconsistent method deployment resulted in a simulation crash at an early stage. The velocity of the bubble's bottom region increased abruptly, causing the front's vertices in that region to have much higher velocity than the neighbor region, which explains the bottom sharp cone formation as shown in Figure \ref{fig:collapsed_risingBubble-shape}. Conversely, Figure \ref{fig:risingBubble-shape} depicts the temporal evolution of the bubble's shape using the consistent method \textcolor{Reviewer24}{with $N=128$}. The predicted bubble shapes show good agreement with the previous simulation results \citep{Hua_2007,Hua_2008,Anjos2014} and the experimental results \citep{Bhaga_weber_1981}. 
\begin{figure}[!htb]
      \centering
      \includegraphics[width=.8\textwidth]{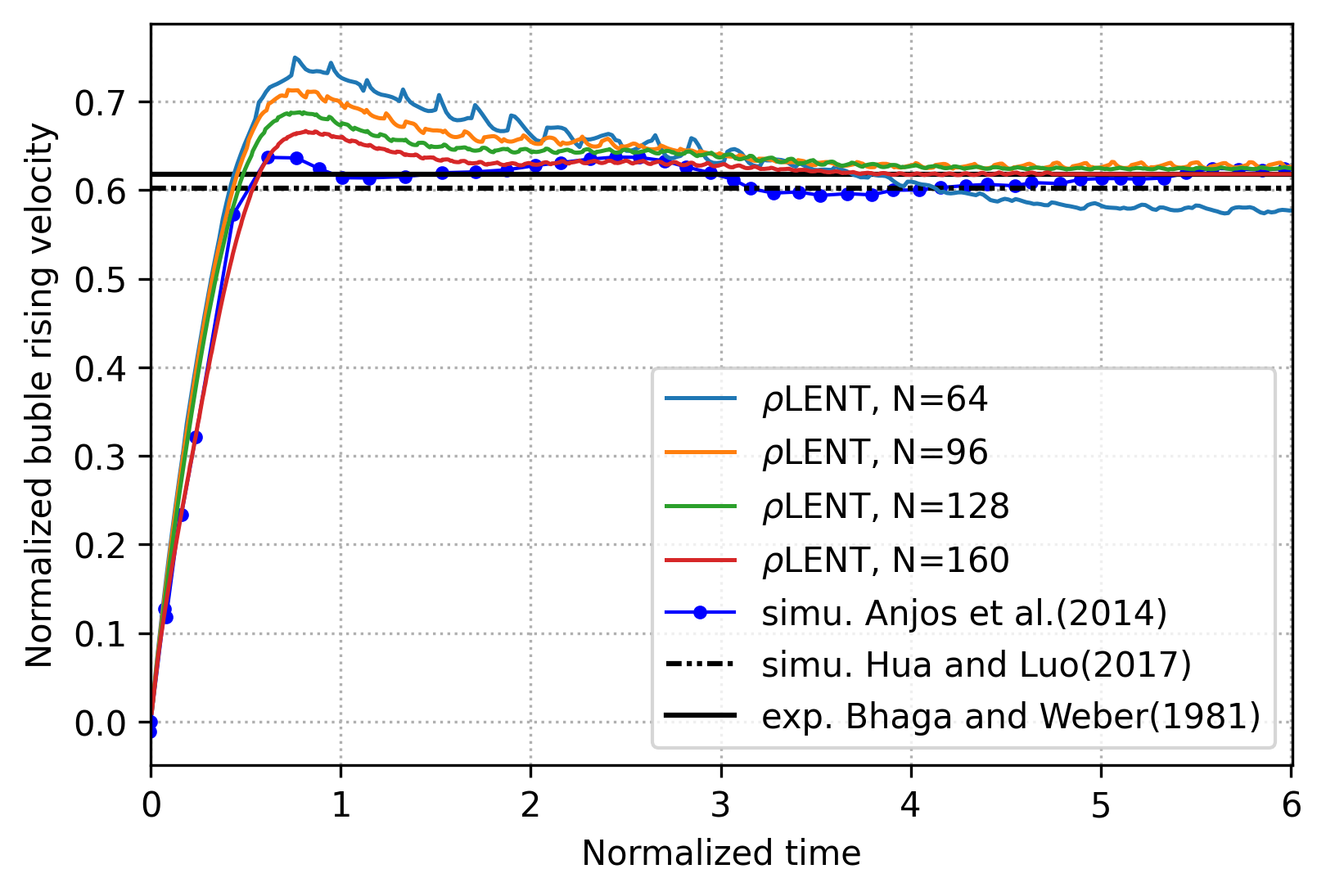}
    \caption{\textcolor{Reviewer24}{ The average bubble velocities from $\rho$LENT with four resolutions are compared with: the experimental results (black solid line) from \citet{Bhaga_weber_1981}, the simulation results (black dashed double-dotted line) from \citet{Hua_2007}, and the extracted simulation results (blue solid dotted line ) from \citet{Anjos2014}. Simu. and exp. are the abbreviation of simulation and experiment.}}
    \label{fig:risingBubble-velocity}
\end{figure}
Additionally, \ref{fig:risingBubble-velocity} shows a comparison of the bubble's rising velocities between our $\rho$LENT method and some previous works. \textcolor{Reviewer24}{At the acceleration stage, the \textcolor{Reviewer24}{predicted rising velocity} from $\rho$LENT method with the finest mesh $N=160$  agrees remarkably well with the results from \citet{Anjos2014}, whereas the velocities from the cases with coarser meshes are slightly higher than the results from \citet{Anjos2014}. The deceleration stage of the bubble can be observed for all cases with different resolutions, which also exists in the results of \citet{Anjos2014}. Except for in the case with the coarse mesh $N=64$, the rising velocities in cases with higher resolutions reach a stable state and converge to the experimental value from \citet{Bhaga_weber_1981}.}
\hfill\\
%Although a relatively coarse mesh was deployed, the predicted bubble rising velocity from the consistent $\rho$LENT method agrees remarkably well with the results from \citep{Anjos2014} during the early acceleration stage. In the steadily rising phase, our results show greater stability than those from \citep{Anjos2014} and a great correspondence with the results from \citep{Hua_2007}.}
\section{Conclusions}
\label{sec:concl}

The proposed \rhoLENT{} method exactly ensures numerical consistency of the single-field incompressible two-phase momentum convection, discretized by the unstructured collocated Finite Volume Method. The \rhoLENT{} method is straightforward and can be applied directly to any two-phase flow simulation method that relies on the collocated FV method for equation discretization of two-phase single-field Navier-Stokes equations by adding a geometrical computation of area fractions $\alpha_f^{n+1}$ from the approximated fluid interface $\tilde{\Sigma}^{n+1}$ and the auxiliary density equation to the solution algorithm. We \textcolor{Reviewer24}{provide an analysis that connects the mass conservation, phase indicator function conservation and momentum convection, theoretically justifying the numerical consistency of the cell-centered density $\rho_c^{n+1}$ computed by a mass flux identical to the one used in the two-phase momentum convective term. This provides the theoretical reasoning behind the auxiliary mass conservation equation, originally introduced by \citet{ghods2013}.} Following the importance of the face-centered (mass flux) density pointed out by \citet{zuzio2020}, we derive the expression for the mass flux density using the principle of mass conservation and connect the mass flux density with the phase indicator. We achieve this by avoiding the temporal integration of the conserved property as done very recently by \citet{Arrufat2021}, which allows us to express the mass fluxes using the phase indicator in a discrete setting. The consistent cell-centered density $\rho_c^{n+1}$ is used in the $p-\mathbf{v}$ coupling algorithm \cite{tolle2020saample} to obtain the velocity $\mathbf{v}_c^{n+1}$, necessary to evolve the fluid interface in the next step from $t^{n+1}$ to $t^{n+2}$. Once the velocity is obtained by $p-\mathbf{v}$ coupling, the cell-centered density $\rho_c^{n+1}$ is again made consistent with the fluid interface. 
Using the face-centered (mass-flux) density in the $p-\v$ coupling and advecting the interface first, enables \rhoLENT{} to discretize the momentum convection term implicitly, compared to the explicit convective term discretization that is used by \citet{bussmann2002, ghods2013} in the collocated Finite Volume setting. 
The consistency of the mass flux in the auxiliary density equation with the mass flux computed using the phase indicator, justifies theoretically the use of the same interpolation  schemes for these two fluxes by \citet{ghods2013,patel2017,manik2018}.

Results demonstrate the recovery of an exact solution, with the error in the $L_\infty$ norm exactly equaling $0$, for the canonical droplet translation verification case studies \cite{popinet2009accurate}.
Droplets with sub-millimeter diameters and with realistic fluid properties are also advected exactly. 
Validation cases with realistic surface tension forces and viscosity demonstrate numerical stability of \rhoLENT{}, resulting in the relative $L_\infty$ norm for the parasitic currents between $10^{-4}$ and $10^{-2}$ for  density ratios up to $10^4$. 
\textcolor{Reviewer3}{Our consistent method successfully recovers the accurate frequency of oscillation for the ellipsoidal droplet with a relative error of $10^{-4}$. The simulation also accurately captured the strong deformation of the rising bubble, in excellent agreement with experimental results.}

\section{Acknowledgments}

Funded by the German Research Foundation (DFG) – Project-ID 265191195 – SFB 1194. Calculations for this research were conducted on the Lichtenberg high performance computer of the TU Darmstadt. 

%% The Appendices part is started with the command \appendix;
%% appendix sections are then done as normal sections
%% \appendix

%% \section{}
%% \label{}

%%
\bibliographystyle{elsarticle-num-names} 
\bibliography{literature}

%% End of file `elsarticle-template-num-names.tex'.

\clearpage
\appendix
%\section{Connection between the phase indicator equation and mass conservation}
%\label{app:whole-figures}

\section{\textcolor{Reviewer25}{Full figures of the results from translating droplet cases}}%{Parameter study figures with legends}
\label{sec:appendix-a-imageWithLegend}
\begin{figure}[!htb]
    \centering
    \includegraphics[width=.8\textwidth]{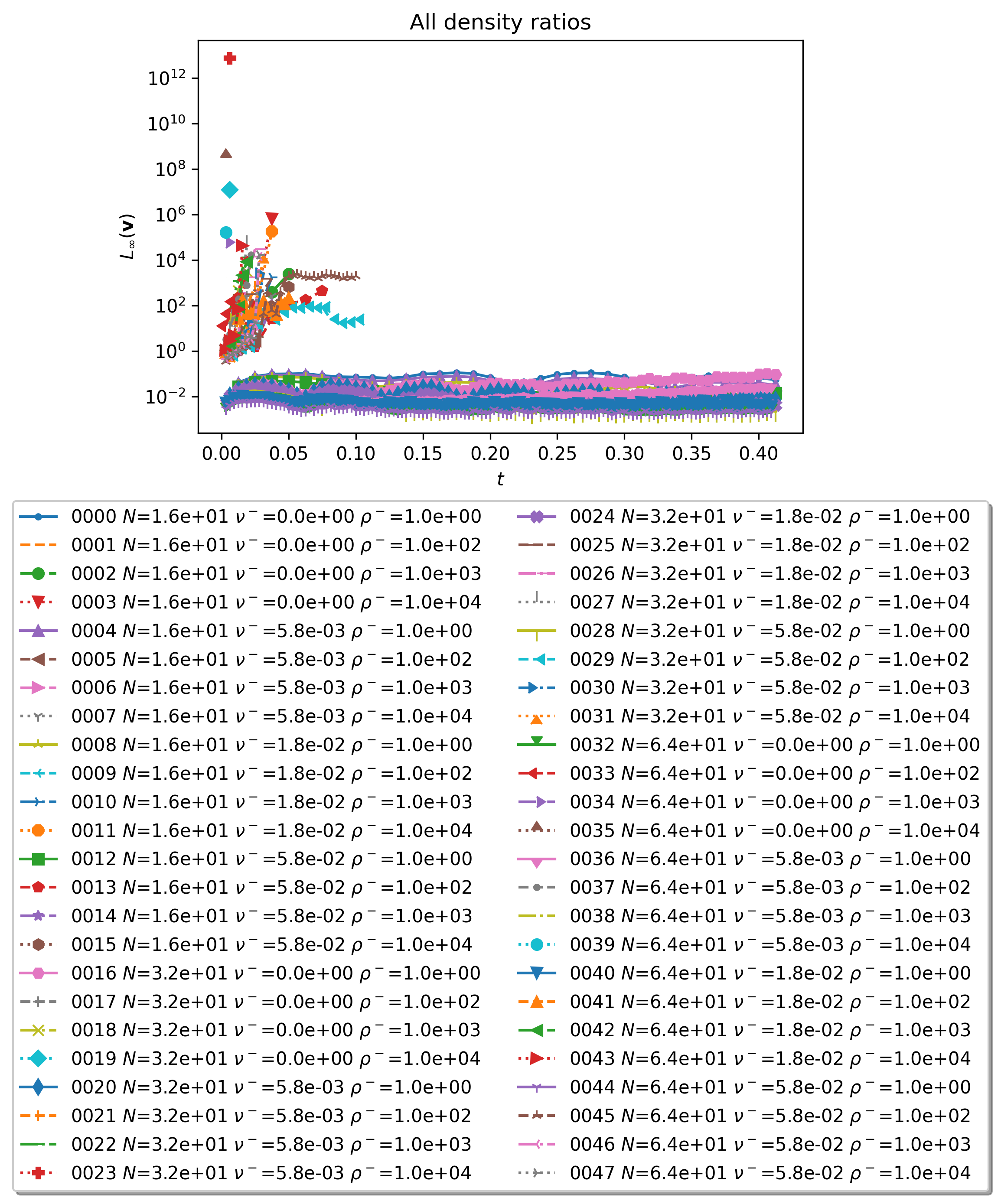}
    \caption{Full figure of \cref{fig:fullforces-old-inconsistent-method}}
    \label{legend:no_with_LV0_withlegend}
\end{figure}

\begin{figure}[!htb]
    \centering
    \includegraphics[width=.9\textwidth]{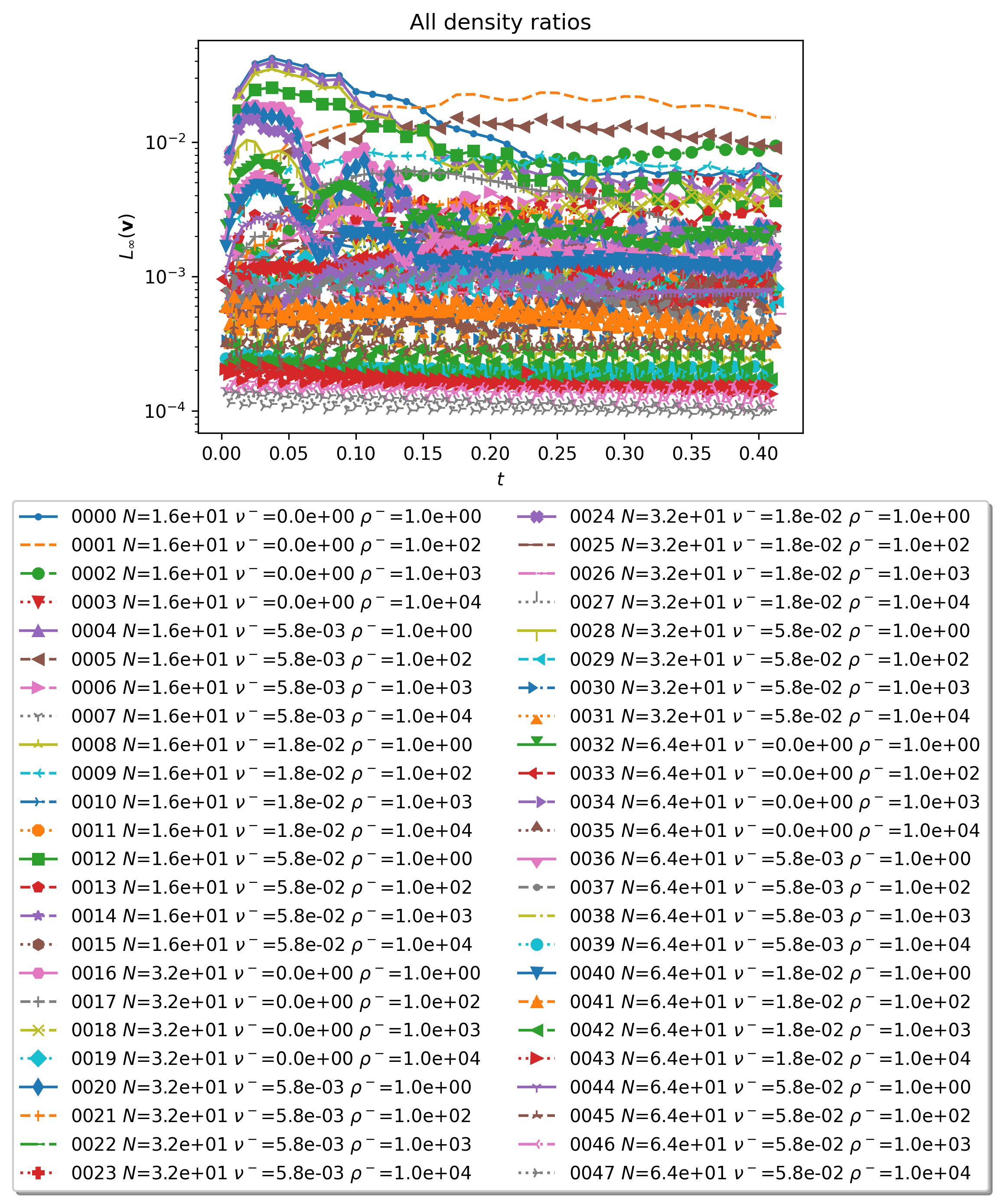}
    \caption{Full figure of \cref{fig:fullforces-new-consistent-method}}
    \label{legend:with_with_LV0_withlegend}
\end{figure}

\begin{figure}[!htb]
    \centering
    \includegraphics[width=.9\textwidth]{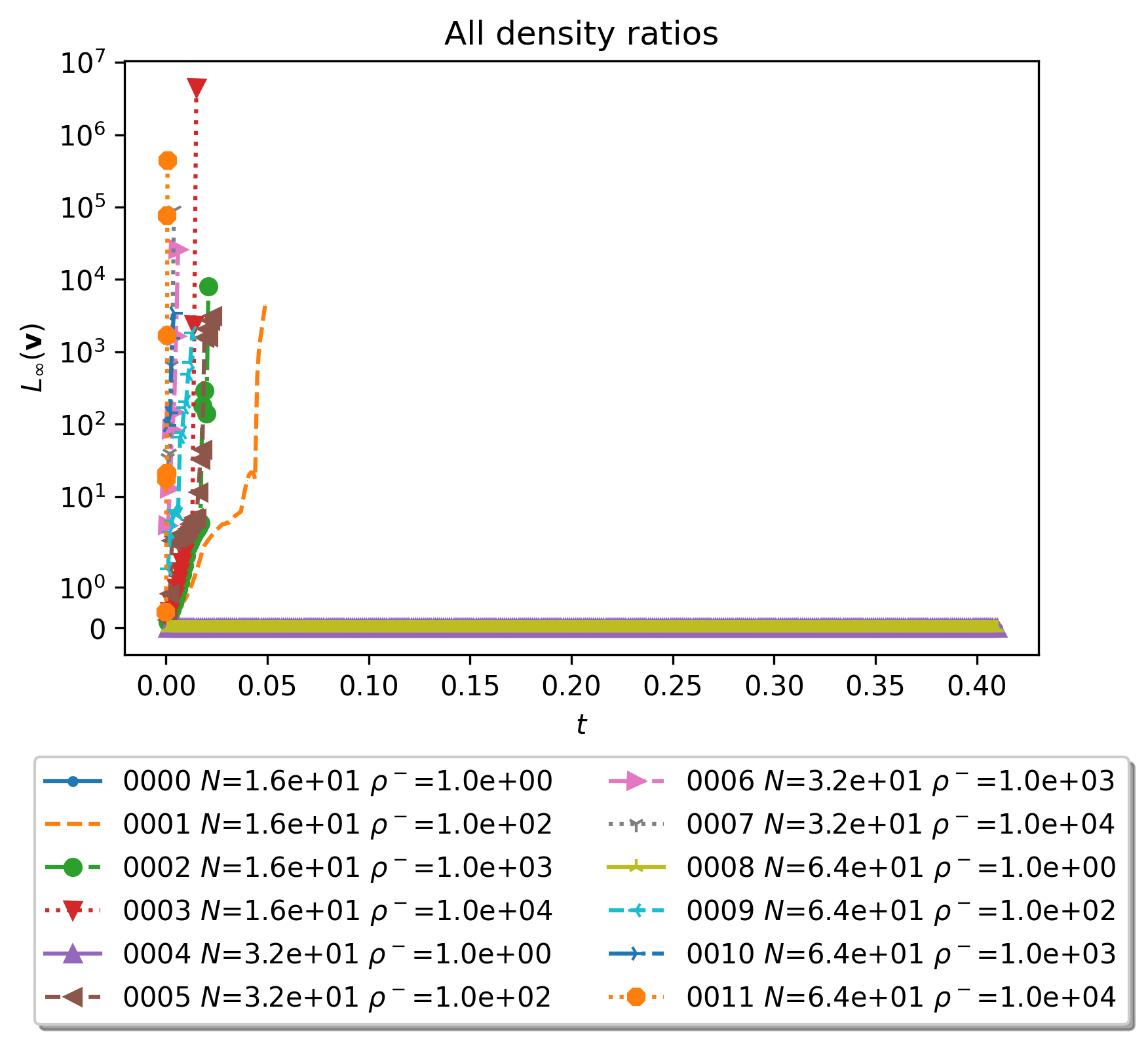}
    \caption{\textcolor{Reviewer25}{Full figure of \cref{fig:old-inconsistent-method}}}
    \label{legend:densityRatioInfluence_fullTime}
\end{figure}

% \begin{figure}[!htb]
%     \centering
%     \includegraphics[width=.9\textwidth]{figures/Trans_popinet2009_noRhoEquation_noMomentumPredictor_limitedLinearV0_withlegend.png}
%     \caption{Full figure of \cref{fig:inconsistent_limiter_0}}
%     \label{legend:inconsistent_limiter_0}
% \end{figure}

% \begin{figure}[!htb]
%     \centering
%     \includegraphics[width=.9\textwidth]{figures/Trans_popinet2009_noRhoEquation_noMomentumPredictor_limitedLinearV0.5_withlegend.png}
%     \caption{Full figure of \cref{fig:inconsistent_limiter_0.5}}
%     \label{legend:inconsistent_limiter_0.5}
% \end{figure}

% \begin{figure}[!htb]
%     \centering
%     \includegraphics[width=.9\textwidth]{figures/Trans_popinet2009_noRhoEquation_noMomentumPredictor_limitedLinearV1_withlegend.png}
%     \caption{Full figure of \cref{fig:inconsistent_limiter_1}}
%     \label{legend:inconsistent_limiter_1}
% \end{figure}

% \begin{figure}[!htb]
%     \centering
%     \includegraphics[width=.9\textwidth]{figures/Trans_popinet2009_noRhoEquation_withMomentumPredictor_limitedLinearV1_withlegend.png}
%     \caption{Full figure of \cref{fig:inconsistent_withMomentumPredictor}}
%     \label{legend:inconsistent_withMomentumPredictor}
% \end{figure}

% \begin{figure}[!htb]
%     \centering
%     \includegraphics[width=.9\textwidth]{figures/Trans_popinet2009_noRhoEquation_noMomentumPredictor_limitedLinearV1_withlegend.png}
%     \caption{Full figure of \cref{fig:inconsistent_noMomentumPredictor}}
%     \label{legend:inconsistent_noMomentumPredictor}
% \end{figure}

\end{document}